\documentclass[longauth]{aa}
\usepackage{graphicx}
\usepackage{txfonts}
\usepackage{hyperref}
\usepackage{url}
\usepackage{color}
\usepackage{comment}
\usepackage[switch]{lineno}
\begin{document}

    \title{Study of the variable broadband emission of Markarian 501 during the most extreme {\it Swift} X-ray activity}

   \titlerunning{Mrk\,501 during an extreme X-ray outburst in 2014}



   
%
\author{
MAGIC Collaboration: 
V.~A.~Acciari\inst{1} \and
S.~Ansoldi\inst{2,21} \and
L.~A.~Antonelli\inst{3} \and
A.~Babi\'c\inst{4} \and
B.~Banerjee\inst{5} \and
U.~Barres de Almeida\inst{6} \and
J.~A.~Barrio\inst{7} \and
J.~Becerra Gonz\'alez\inst{1}\thanks{Corresponding authors: J. Becerra Gonz\'alez (jbecerra@iac.es), D. Paneque (dpaneque@mppmu.mpg.de), C. Wendel (cwendel@astro.uni-wuerzburg.de)} \and
W.~Bednarek\inst{8} \and
E.~Bernardini\inst{9,14,23} \and
A.~Berti\inst{10,}\inst{24}
% ...
}

\institute {Inst. de Astrof\'isica de Canarias, E-38200 La Laguna, and Universidad de La Laguna, Dpto. Astrof\'isica, E-38206 La Laguna, Tenerife, Spain
\and Universit\`a di Udine, and INFN Trieste, I-33100 Udine, Italy
\and National Institute for Astrophysics (INAF), I-00136 Rome, Italy
\and Croatian MAGIC Consortium: University of Rijeka, 51000 Rijeka; University of Split - FESB, 21000 Split; University of Zagreb - FER, 10000 Zagreb; University of Osijek, 31000 Osijek; Rudjer Boskovic Institute, 10000 Zagreb, Croatia
\and Saha Institute of Nuclear Physics, HBNI, 1/AF Bidhannagar, Salt Lake, Sector-1, Kolkata 700064, India
\and Centro Brasileiro de Pesquisas F\'isicas (CBPF), 22290-180 URCA, Rio de Janeiro (RJ), Brasil
\and Unidad de Part\'iculas y Cosmolog\'ia (UPARCOS), Universidad Complutense, E-28040 Madrid, Spain
\and University of \L\'od\'z, Department of Astrophysics, PL-90236 \L\'od\'z, Poland
\and Deutsches Elektronen-Synchrotron (DESY), D-15738 Zeuthen, Germany
\and Istituto Nazionale Fisica Nucleare (INFN), 00044 Frascati (Roma) Italy
\and Max-Planck-Institut f\"ur Physik, D-80805 M\"unchen, Germany
\and Institut de F\'isica d'Altes Energies (IFAE), The Barcelona Institute of Science and Technology (BIST), E-08193 Bellaterra (Barcelona), Spain
\and Universit\`a di Siena and INFN Pisa, I-53100 Siena, Italy
\and Universit\`a di Padova and INFN, I-35131 Padova, Italy
\and Universit\`a di Pisa, and INFN Pisa, I-56126 Pisa, Italy
\and Finnish MAGIC Consortium: Finnish Centre of Astronomy with ESO (FINCA), University of Turku, FI-20014 Turku, Finland; Astronomy Research Unit, University of Oulu, FI-90014 Oulu, Finland
\and Technische Universit\"at Dortmund, D-44221 Dortmund, Germany
\and Departament de F\'isica, and CERES-IEEC, Universitat Aut\`onoma de Barcelona, E-08193 Bellaterra, Spain
\and Universitat de Barcelona, ICCUB, IEEC-UB, E-08028 Barcelona, Spain
\and ICRANet-Armenia at NAS RA, 0019 Yerevan, Armenia
\and Japanese MAGIC Consortium: ICRR, The University of Tokyo, 277-8582 Chiba, Japan; Department of Physics, Kyoto University, 606-8502 Kyoto, Japan; Tokai University, 259-1292 Kanagawa, Japan; RIKEN, 351-0198 Saitama, Japan
\and Inst. for Nucl. Research and Nucl. Energy, Bulgarian Academy of Sciences, BG-1784 Sofia, Bulgaria
\and Humboldt University of Berlin, Institut f\"ur Physik D-12489 Berlin Germany
\and also at Dipartimento di Fisica, Universit\`a di Trieste, I-34127 Trieste, Italy
% ...
}


 
  \abstract
   {Markarian~501 (Mrk~501) is a very high-energy (VHE) gamma-ray blazar located at $z=0.034$, which is regularly monitored by a wide range of multi-wavelength (MWL) instruments, from radio to VHE gamma rays. During a period of almost two weeks  in July 2014, the highest X-ray activity of Mrk~501 was observed in   $\sim$14 years of operation of the \textit{Neil Gehrels Swift Gamma-ray Burst Observatory}.}
   {We characterize the broadband variability of Mrk~501 from radio to VHE gamma rays during the most extreme X-ray activity measured in the last 14 years, and evaluate whether it can be interpreted within theoretical scenarios widely used to explain the broadband emission from blazars.}
   {The emission of Mrk~501 was measured at radio with Mets\"ahovi, at optical--UV with KVA and {\it Swift}/UVOT, at X-ray with {\it Swift}/XRT and {\it Swift}/BAT, at gamma ray with {\it Fermi}-LAT, and at VHE gamma rays with the FACT and MAGIC telescopes.  The multi-band variability and correlations were quantified, and the broadband spectral energy distributions (SEDs) were compared with predictions from theoretical models.}
   {The VHE emission of Mrk~501 was found to be elevated during the X-ray outburst, with a gamma-ray flux above 0.15 TeV varying from $\sim$0.5 to $\sim$2 times the Crab nebula flux (CU).  The X-ray and VHE emission  both varied on  timescales of 1 day and were found to be correlated. We measured a general increase in the fractional variability with energy, with the VHE variability being twice as large as the X-ray variability. The temporal evolution of the most prominent and variable segments of the SED, characterized on a day-by-day basis from 2014 July 16 to 2014 July 31, is described with a one-zone synchrotron self-Compton model with variations in the break energy of the electron energy distribution (EED), and with some adjustments in the magnetic field strength and spectral shape of the EED. These results suggest that the main flux variations during this extreme X-ray outburst are produced by the acceleration and the cooling of the high-energy electrons. A narrow feature at $\sim$3~TeV was observed in the VHE spectrum measured on 2014 July 19 (MJD~56857.98), which is the day with the highest X-ray flux ($>0.3$~keV)  measured during the entire {\it Swift} mission. This feature is inconsistent with the classical analytic functions to describe the measured VHE spectra (power law, log-parabola, and log-parabola with exponential cutoff) at more than 3\,$\sigma$. A fit with a log-parabola plus a narrow component is preferred over the fit with a single log-parabola at more than 4\,$\sigma$, and a dedicated Monte Carlo simulation estimated the significance of this extra component to be larger than 3~$\sigma$. Under the assumption that this VHE spectral feature is real, we show that it can be reproduced with three distinct theoretical scenarios: a) a pileup in the EED due to stochastic acceleration; b) a structured jet with two-SSC emitting regions, with one region dominated by an extremely narrow EED; and c) an emission from an IC pair cascade induced by electrons accelerated in a magnetospheric vacuum gap, in addition to the SSC emission from a more conventional region along the jet of Mrk~501.}
   {}


   \maketitle
%

\section{Introduction}
Markarian~501 (Mrk~501) is a well-known gamma-ray blazar located at $z=0.034$. It was first detected at very high-energy (VHE, E$>$100~GeV) gamma rays with the Whipple Observatory \citep{quinn}. It is classified as a BL~Lac object, whose optical spectra are dominated by the nonthermal continuum from the jet. In BL~Lac objects there are no signs of a strong broad-line region (BLR) or of a dusty IR torus, and therefore, in absence of any strong external photon field interacting with the jet, they are typically modeled by Synchrotron Self-Compton models \citep[SSC; see, e.g.,][]{ssc_maraschi}. 

Mrk~501 is one of the few VHE objects that can be detected with the current generation of Imaging Air Cherenkov Telescopes (IACTs) in relatively short integration times even during their low state emission periods. This  makes Mrk~501 an ideal blazar for long-term  multi-wavelength (MWL) monitoring with the aim of performing detailed studies that cannot be carried out for other blazars that are fainter, located farther away, or have more complicated structures. Motivated by this goal, an extensive multi-instrument program was organized to characterize and study the temporal evolution, over many years, of the broadband emission of Mrk~501 \citep[see, e.g.,][]{2011ApJ...727..129A,2015A&A...573A..50A,2015ApJ...812...65F,2017A&A...603A..31A,2018AhnenSubmitted}. This observational campaign was enhanced by the beginning of the \textit{Fermi} era, providing a continuous coverage over a wide range of gamma-ray energies. Thanks to the large amount of data already investigated in the past, the extensive time and energy coverage keep bringing new clues to better understand the emission mechanisms of this blazar.

During the MWL campaign performed in July 2014, we observed a $\sim$two-week  flaring activity in the X-ray and VHE bands. The X-ray activity was exceptionally high, yielding the largest fluxes detected with the X-ray Telescope (XRT) instrument on board the
\textit{Neil Gehrels Swift Gamma-ray Burst Observatory} \citep{2004ApJ...611.1005G} during its almost 14 years of operation after its launch in 2004. The X-ray activity during these two weeks appears to be similar to that observed during the large historical flare from 1997, when  the \textit{BeppoSAX} satellite reported a large increase in the X-ray flux of Mrk~501. During this 1997 flare, the peak of the synchrotron bump was located above 100\,keV, hence indicating a shift by more than two orders of magnitude of the peak position compared to that of the typical (nonflaring) state  
\citep[][]{1998ApJ...492L..17P,1999A&A...347...30V,historical_flare}. Within the framework of the planned multi-instrument observations, the  First G-APD Cherenkov Telescope
(FACT) was observing Mrk\,501 daily (provided atmospheric conditions
allow), but other facilities such as Mets\"{a}hovi, KVA, \textit{Swift}, and MAGIC
were observing only once every a few days (typically once every 3--4 days). Triggered by the
outstanding X-ray activity observed in the \textit{Swift} data collected during the campaign, we organized
multi-band observations every day, as shown in Fig.~\ref{fig:mwl_lc}.
These multi-instrument data allowed us to characterize  with a
wide energy coverage and fine temporal sampling the evolution of the
broadband spectral energy distribution (SED) during this period of
outstanding activity. This manuscript reports the results from these measurements together with a characterization of the variability and correlation among the various energy bands, and a physical interpretation of this remarkable behavior using theoretical leptonic scenarios that are commonly used in the literature.

This paper is structured as follows. In Sect.~\ref{sec:MWL} we briefly describe the instruments whose data are used, along with their data analyses. In Sect.~\ref{sec:Results} we report all the observational results, namely the multi-instrument light curves, the quantification of the variability and correlations among energy bands, and a detailed study of the X-ray and VHE gamma-ray spectra. Sections ~\ref{sec:Model} and ~\ref{sec:ModelNarrow} characterize the broadband SED and its temporal evolution within standard leptonic scenarios, and provide a theoretical interpretation of the obtained results.  Finally, Sect.~\ref{sec:Conclusions} provides a short summary and concluding remarks. Additional details on the analysis are given in the Appendix. Throughout this work we adopt the cosmological parameters $H_0=70\,\mathrm{km}\,\mathrm{s}^{-1}\, \mathrm{Mpc}^{-1}$, $\Omega_{\Lambda}=0.7$, and $\Omega_{M}=0.3$.

\section{Multi-wavelength observations}
\label{sec:MWL}
Many different observatories, from radio to VHE gamma rays, participated in the MWL campaign on Mrk~501 performed between March and September 2014. The extensive dataset collected during the 2014 campaign will be reported in a future study. In this paper, we only report measurements from the  $\sim$two-week interval in July 2014 when an extremely high X-ray activity was observed. This paper focuses mainly on the X-ray and VHE gamma-ray bands, which are the two segments of the broadband SED with the highest energy flux, and that show the highest variability (see Sec.~\ref{sec:Results}). In the following sections the observations and data analyses for each instrument used in this work will be briefly described.

\subsection{MAGIC}\label{magic_analysis}

The MAGIC stereoscopic telescope system is composed of two IACTs located at the Roque de los Muchachos Observatory on La Palma, one of the Canary Islands ($28.7^{\circ}$ N, $17.9^{\circ}$ W), at a height of 2200 m above sea level \citep{magic_upgrade1}. Each telescope has a large mirror dish  17\,m in diameter. MAGIC can detect air Cherenkov showers initiated by gamma rays in the energy range from $\sim$50\,GeV to $\sim$50\,TeV.

This paper reports MAGIC data taken from 2014 July 16 to 2014 July 31 (MJD~56854-56869). The observations were performed within a zenith angle range from $10.0^{\circ}$ to $41.2^{\circ}$. The energy threshold of the analysis, calculated as the peak of the number of events for these observation conditions and a spectral index of $-2.2$, is located at approximately 130 GeV. Since the data sample covers a wide range of zenith angles, the low zenith angle data allow us to characterize the spectrum at energies below the average energy threshold. The data analysis was carried out using the standard analysis package developed by the MAGIC collaboration named MAGIC Analysis and Reconstruction 
Software \citep[MARS,][]{zanin,magic_upgrade2}. Mrk~501 was observed with the MAGIC telescopes for a total of 13.5 h under dark and good quality conditions. Detailed information about the data collection can be found in Table~\ref{tab:magic_obs}.

\begin{table}
   \centering
   \begin{tabular}{c c c c c} 
        Date & MJD & Obs. time [h] & Zd [$^{\circ}$] & Significance  \\
        \hline
        20140716 & 56854.91 & 0.45 & 10-15 & 13.1~$\sigma$ \\
        20140717 & 56855.91 & 0.37 & 11-14 & 15.9~$\sigma$ \\
        20140718     & 56856.91 & 0.48 & 11-14 & 22.5~$\sigma$ \\
        20140719$^a$ & 56857.98 & 1.54 & 10-40 & 36.5~$\sigma$ \\
        20140720 & 56858.98 & 0.63 & 10-41 & 27.9~$\sigma$ \\
        20140721 & 56859.97 & 1.48 & 10-38 & 40.0~$\sigma$ \\
        20140723 & 56861.01 & 0.49 &  24-32 & 16.6~$\sigma$ \\
        20140724 & 56862.02 & 1.28 &  25-42 & 21.9~$\sigma$ \\
        20140725 & 56863.00 & 0.49 &  25-32 & 17.8~$\sigma$ \\
        20140726 & 56864.02 & 1.26  & 24-41 & 26.6~$\sigma$ \\
        20140727 & 56865.00 & 0.44 & 25-32 & 17.4~$\sigma$ \\
        20140728 & 56866.00 & 2.13 & 17-41 & 57.4~$\sigma$ \\
        20140729 & 56867.00 & 0.49 & 27-33 & 21.1~$\sigma$ \\
        20140730 & 56868.01 & 1.29 & 26-43 & 42.7~$\sigma$ \\
        20140731 & 56869.93 & 0.66 & 11-18  & 19.2~$\sigma$ \\
        \hline
   \end{tabular}
   \caption{Summary of the Mrk~501 VHE observations performed with the MAGIC telescopes during the flaring activity that occurred in July 2014. The center of the observation time bin is given in MJD. The significance is calculated according to equation~17 in \citet{1983ApJ...272..317L}. $^a$ Observation showing a hint of a narrow feature at $\sim$3~TeV, see section~\ref{bump}.}
   \label{tab:magic_obs}
\end{table}

\subsection{FACT}

The First G-APD Cherenkov Telescope  is located next to the two
MAGIC telescopes. With its 9.5\,$\mathrm{m}^2$ mirror and its camera consisting of 1440 pixels with
silicon-based photosensors (G-APDs, also known as SiPM), it has been designed to
perform an intense monitoring of bright TeV blazars \citep{2013JInst...8P6008A,2014JInst...9P0012B}. FACT has been operating since October 2011, and it has already collected  more than 11000 hours of data. 

This manuscript reports FACT observations of Mrk\,501 from 2014 July 14 until 2014 August 5, amounting to 51.8\,hours,  of which 47.2\,hours passed the quality selection based on the cosmic-ray rate described in \citet{HildebrandICRC2017}. The FACT analysis was performed as described in 
\citet{2015arXiv150202582D}. The excess rate was corrected
for the effect of changing zenith distance and changing trigger
threshold as described in \citet{2013arXiv1311.0478D} and \citet{MahlkeICRC2017}.
The identically corrected excess rate measured from the Crab nebula 
during the same season was used to convert the observed excess rates
into photon fluxes. The resulting light curve, with an energy threshold of about 0.83~TeV, is shown in Fig.~\ref{fig:mwl_lc}. The analysis pipeline does not permit reliable spectral measurements for the observations reported here. Variability in the spectral shape of Mrk\,501 would introduce an additional uncertainty in the FACT fluxes reported in Fig.~\ref{fig:mwl_lc}. However, given the spectral variability measured with MAGIC (see Table~\ref{tab:vhe_fits}), this systematic error would be  only $\sim$5\,\%, which is much smaller than the statistical uncertainties on the VHE fluxes measured by FACT reported in Fig.~\ref{fig:mwl_lc}.

\subsection{\it{Fermi}-LAT}

We analyzed the data collected by the Large Area Telescope (LAT) on board the {\it Fermi} Gamma-ray Space Telescope from 200 MeV to 800 GeV. The data selection was centered at the position of Mrk~501 and a circular region of $10^{\circ}$ was chosen. The analysis used Pass 8 source class events. A first unbinned likelihood analysis was carried out for 8 months of data from 2014 April 1 to 2014 December 1 (MJD 56748-56992) in order to discard  nonvariable weak sources that cannot be detected by {\it Fermi}-LAT on short timescales. In this first step, all the sources present in the 3FGL catalog \citep{3fgl} within $20^{\circ}$ of Mrk~501 were included in the analysis (85 point-like sources). The sources located within $10^{\circ}$ were left free to vary both in flux and spectral shape. On the contrary, for the sources beyond $10^{\circ}$, only the flux normalization was left free while the spectral shapes where fixed to their 3FGL catalog values. The analysis was performed using the Science-Tools software package version v11-07-00, the instrument response function \verb|P8R2_SOURCE_V6| and the diffuse background models \verb|gll_iem_v06| and \verb|iso_P8R2_SOURCE_V6_v06|\footnote{https://fermi.gsfc.nasa.gov/ssc/data/access/lat/BackgroundModels.html}. After the first unbinned likelihood fit, the sources with test statistics \citep[TS,][]{mattox} value of TS$<$5 were removed from the model. The resulting simplified model was used to analyze the two-week period covered by the flare detected at VHE gamma rays and X-rays. The preliminary FL8Y point source list\footnote{https://fermi.gsfc.nasa.gov/ssc/data/access/lat/fl8y/} was checked to search for any additional source not previously included in the 3FGL catalog which might have an impact on the analysis. No new sources with TS$>$25 were found within $20^{\circ}$ from Mrk~501.

The light curve (LC) was calculated with a  three-day binning because Mrk~501 is a relatively weak source for {\it Fermi}-LAT. For the LC analysis, the shape of the source spectrum was described with a power-law function, with the flux normalization as a free parameter, and the spectral index fixed to $\Gamma=1.78$, which is the value found for the almost $\sim$two-week time period considered in this manuscript\footnote{The spectral analysis of LAT data does not show significant variability during the time period considered in this manuscript, and hence it is reasonable to assume that the spectral shape is constant throughout this period.}. The normalizations of the diffuse background models were allowed to vary during the likelihood fit. Additionally, we also performed a spectral analysis of {\it Fermi}-LAT data for two time intervals, 4 days and 10 days, centered at the time of the observations performed daily with MAGIC. Due to the low photon statistics, it is not possible to derive constraining spectral parameters in shorter observation windows for {\it Fermi}-LAT observations of Mrk\,501. Owing to the variability on one-day timescales measured at X-ray and VHE energies, the {\it Fermi}-LAT spectral points are considered  an estimate of the HE spectra at the time of the X-ray and VHE observations, and used only as a guide in the theoretical modeling  of the broadband SEDs.

\subsection{Swift}

This study reports observations performed with the three instruments on
board the \textit{Neil Gehrels Swift Gamma-ray Burst Observatory}
\citep{2004ApJ...611.1005G}; namely the 
Burst Alert Telescope \citep[BAT,][]{2005ApJ...633L..77M}, the X-ray Telescope \citep[XRT,][]{2005SSRv..120..165B}, and the Ultraviolet/Optical 
Telescope \citep[UVOT,][]{2005SSRv..120...95R}.

\subsubsection{BAT}
\label{SwiftBAT}

We analyzed the BAT data available from Mrk~501 during the period of  high activity in July 2014. We use BAT survey data in this analysis, which contain 80 energy channels and are pre-binned by the onboard software in $\sim 300$\,s \citep[for details, see Sect. 3.3.1 in][]{Markwardt07}. These data are processed using the standard BAT pipeline, \textit{batsurvey}\footnote{https://heasarc.gsfc.nasa.gov/ftools/caldb/help/batsurvey.html}. We adopted eight energy bands in our analysis  from 14 to 195 keV. When computing the BAT fluxes in one-day time intervals, these eight channels had to be  combined into a single energy bin because of the low signal event count. For each observation, the {\it batsurvey} pipeline produced the mask-weighted counts (i.e., background-subtracted counts) in these eight energy bands at the source location. We added up the resulting counts in each day and calculated the corresponding uncertainties through error propagation, and then used this information to create an eight energy band spectrum. As we only use the survey data when {\it Swift} was pointing at Mrk\,501, the counts could be added up without adjusting for different source incident angles and partial coding fractions. We then used the BAT tool {\it batdrmgen} to generate the corresponding BAT detector response file. 

The analysis was performed using two different timescales: daily analysis integrating the observations within one day centered at the MAGIC observations, and a stacked analysis over the time interval considered in this manuscript, namely MJD~56854.5--MJD~56872.5. In the stacked analysis (60.2 ks of exposure), the source is detected with a signal-to-noise ratio of 19.2 and is well described by a power-law function ($\chi^2$/df=2.4/6) with spectral index of $2.3\pm0.1$ and a flux of $(4.1\pm0.3)\times 10^{-10} \mathrm{erg}\,\mathrm{cm}^{-2}\mathrm{s}^{-1}$ in the 14--195 keV energy band.

The BAT flux for each day during this observation period was found by fitting the eight-bin spectra using the commonly adopted X-ray Spectral Fitting Package, Xspec\footnote{https://heasarc.gsfc.nasa.gov/xanadu/xspec/}. Because the source is only detected at a relatively low significance on a daily timescale, we only allowed the flux normalization to vary in the fitting procedure. Two different spectral shapes were used: a) the power-law function from the 18-day stacked analysis of the BAT data and b) the spectral parameters from the XRT spectral analysis reported in Table~\ref{tab:xrt_fits}. The calculation of the flux and uncertainty range were carried out with Xspec, using the {\emph cflux} command. In the spectral analysis for MJD~56862, the counts are too low and Xspec did not find any solution. Consequently, we calculated the $2 \sigma$ flux upper limit based on the exposure time using Eq. 9 in \citet{Baumgartner13}, which gives an approximation of the BAT sensitivity. The results are reported in Table~\ref{tab:bat}.

\subsubsection{XRT}

The XRT data were taken in the framework of the planned extensive multi-instrument campaign. The high activity of Mrk~501 in the X-ray band motivated the increase in the number of observations from one pointing every $\sim$4 days, to one per day between MJD~56855 and MJD~56870.  All observations were carried out in
the Windowed Timing (WT) readout mode, with an exposure of $\sim$1~ks per pointing. The data were  
processed using the XRTDAS software package (v.3.4.0), which was developed by the ASI Science Data 
Center and released by HEASARC in the HEASoft package (v.6.2.2).  The data were  
calibrated and cleaned with standard filtering criteria using the \textit{xrtpipeline} task and 
the calibration files available from the \textit{Swift}/XRT CALDB (version 20140709). For the spectral analysis, events in the energy channels between 0.3 keV and 10 keV were selected within a 20-pixel ($\sim$46~arcsecond) radius, which contains 90\% of the 
point spread function (PSF). 
The background was estimated from a nearby circular region with a radius of 20~pixels. Corrections for 
the PSF and CCD defects were applied from response files generated using the \textit{xrtmkarf} task 
and the cumulative exposure map. The spectra were binned to ensure a minimum of 20 counts per bin, fitted in the band 0.3--10~keV, and corrected for  
absorption with a neutral-hydrogen column density fixed to the Galactic 21 cm value in the 
direction of Mrk\,501, namely \mbox{1.55 $\times~10^{20}~\mathrm{cm}^{-2}$} \citep{2005A&A...440..775K}.
The spectral results are reported in Table~\ref{tab:xrt_fits}.

\subsubsection{UVOT}

We also used the \textit{Swift}/UVOT observations performed with the UV lenticular filters (W1, M2, and W2) that were taken within the same observations acquired by XRT. The emission from these bands is not affected by the host galaxy emission.  We evaluated the photometry of the source according to the
recipe in \citet{2008MNRAS.383..627P}, extracting source counts with an aperture
of 5~arcsecond radius and an annular background aperture with inner and outer radii
of 20~arcsecond and 30~arcsecond. The count rates were converted to fluxes using the
updated calibrations \citep{2011AIPC.1358..373B}. Flux values were then corrected for mean Galactic extinction using an $E (B - V )$
value of 0.017 \citep{2011ApJ...737..103S} using the UVOT filter effective
wavelength and the mean Galactic interstellar extinction curve in \citet{1999PASP..111...63F}.

\subsection{Optical and radio}

The optical data in the R band were obtained with the KVA telescope, at the Roque de los Muchachos (La Palma, Spain).
The data analysis was performed as described in \cite{2018A&A...620A.185N}. The calibration was performed using the stars reported by \citet{1998A&AS..130..305V} and the Galactic extinction was corrected using the coefficients given in \citet{2011ApJ...737..103S}. The contribution from the host galaxy in the R band, which is about 2/3 of the measured flux, was determined using 
\citet{2007A&A...475..199N}, and subtracted from the values reported in Fig.~\ref{fig:mwl_lc}.

The radio fluxes at 37~GHz were obtained with the  14~m Mets\"{a}hovi Radio telescope at the  Mets\"{a}hovi Radio Observatory. 
Details of the observation and analysis strategies are given in \citet{1998A&AS..132..305T}.

\section{Results}
\label{sec:Results}

\subsection{Multi-wavelength flux evolution and quantification of the variability}

\begin{figure*}
   \centering
   \includegraphics[scale=0.70, angle=0]{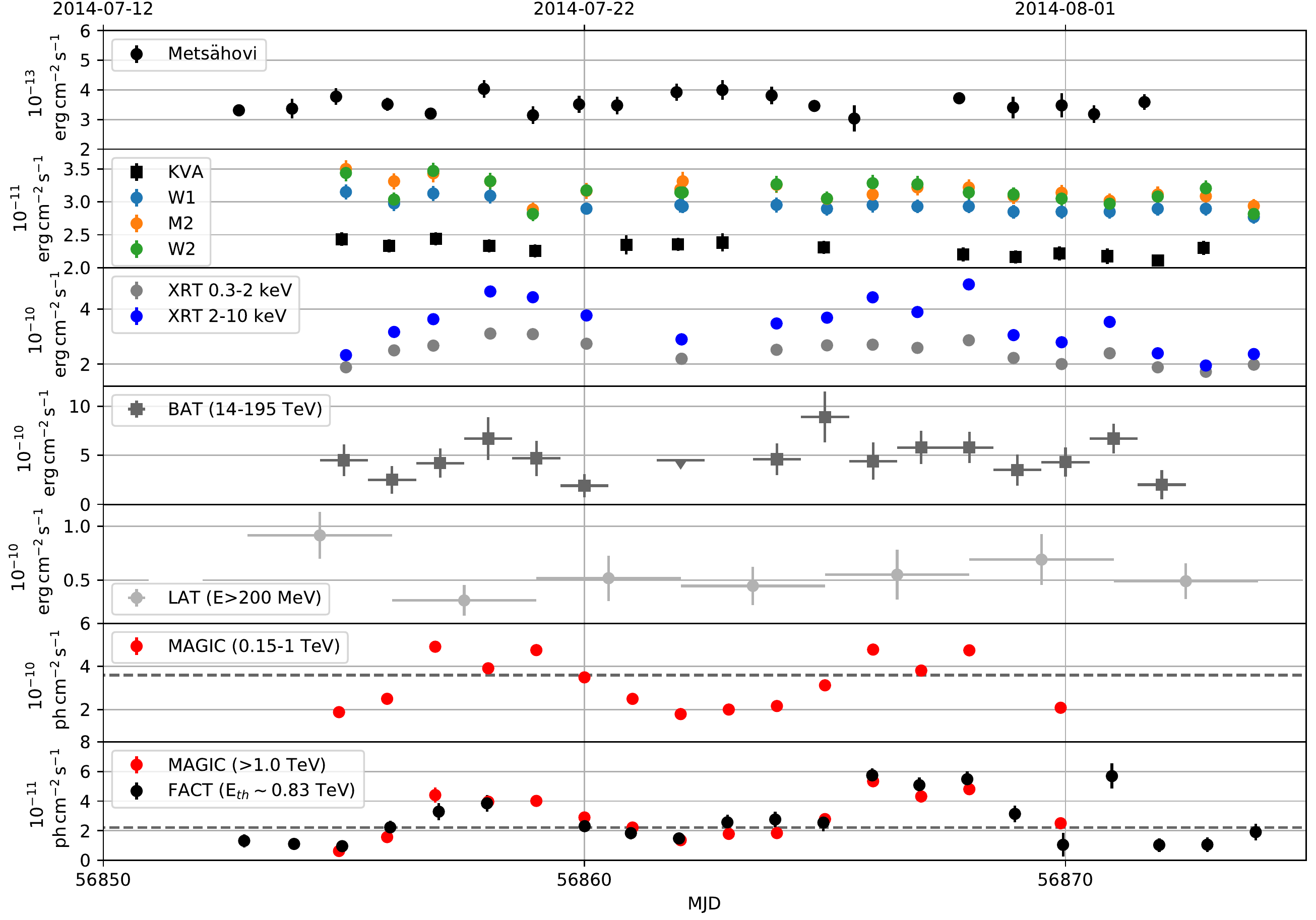} 
   \caption{Multi-wavelength light curve for Mrk 501 during the highest X-ray activity measured with {\it Swift}-XRT to date. The correspondence between the instruments and the measured quantities is given in the legends. The horizontal dashed lines in the VHE light curves depict the flux of the Crab nebula reported in \citet{magic_upgrade2}. For BAT, the daily fluxes were computed using the spectral shape from the time interval MJD~56854.5--MJD~56872.5 (see section \ref{SwiftBAT}).} 
   \label{fig:mwl_lc}
\end{figure*}

The MWL LC from radio to the VHE band is reported in Fig.~\ref{fig:mwl_lc}. Only marginal variability is detected in radio, optical-UV, and low-energy gamma rays above 200\,MeV observed by {\it Fermi}-LAT. On the contrary, there are large flux variations in X-rays and VHE gamma rays. In both energy bands, the flux evolution during the two-week high activity shows a two-peak structure of similar amplitude with respect to each other. Variability on one-day timescales is  significantly detected, but no intra-night variability is observed in any of the energy bands studied in this work.

To quantify and compare the variability observed at different energy bands, the fractional variability ($F_{var}$) is calculated. Following the prescription from \cite{vaughan}, $F_{var}$ is defined as

\begin{equation}
F_{var}=\sqrt{\frac{S^2-<\sigma^2_{err}>}{<F_{\gamma}>^2}}
,\end{equation}
where $<\mathrm{F}_{\gamma}>$ denotes the average photon flux, S the standard deviation of the different flux measurements, and $<\sigma^2_{err}>$ represents the mean squared error of the flux measurements. The uncertainty of $F_{var}$ is calculated following the prescription in \citet{poutanen}, as described in \citet{2015A&A...573A..50A}, such that these uncertainties are also valid in the case when $\Delta F_{var} \sim F_{var}$~,

\begin{equation}
\Delta F_{var} = \sqrt{F^2_{var} + err(\sigma^2_{NXS})} - F_{var}
,\end{equation}
where $\sigma^2_{NXS}$ is calculated following equation 11 from \cite{vaughan}. This prescription to determine the multi-band variability has some caveats related to the different sensitivity and observing sampling among the various instruments used \citep[see, e.g.,][]{2014A&A...572A.121A,2015A&A...576A.126A}. However, it provides a relatively simple way of quantifying and comparing the flux variability in the different energy bands. 

\begin{figure}
   \centering
   \includegraphics[scale=0.62, angle=0]{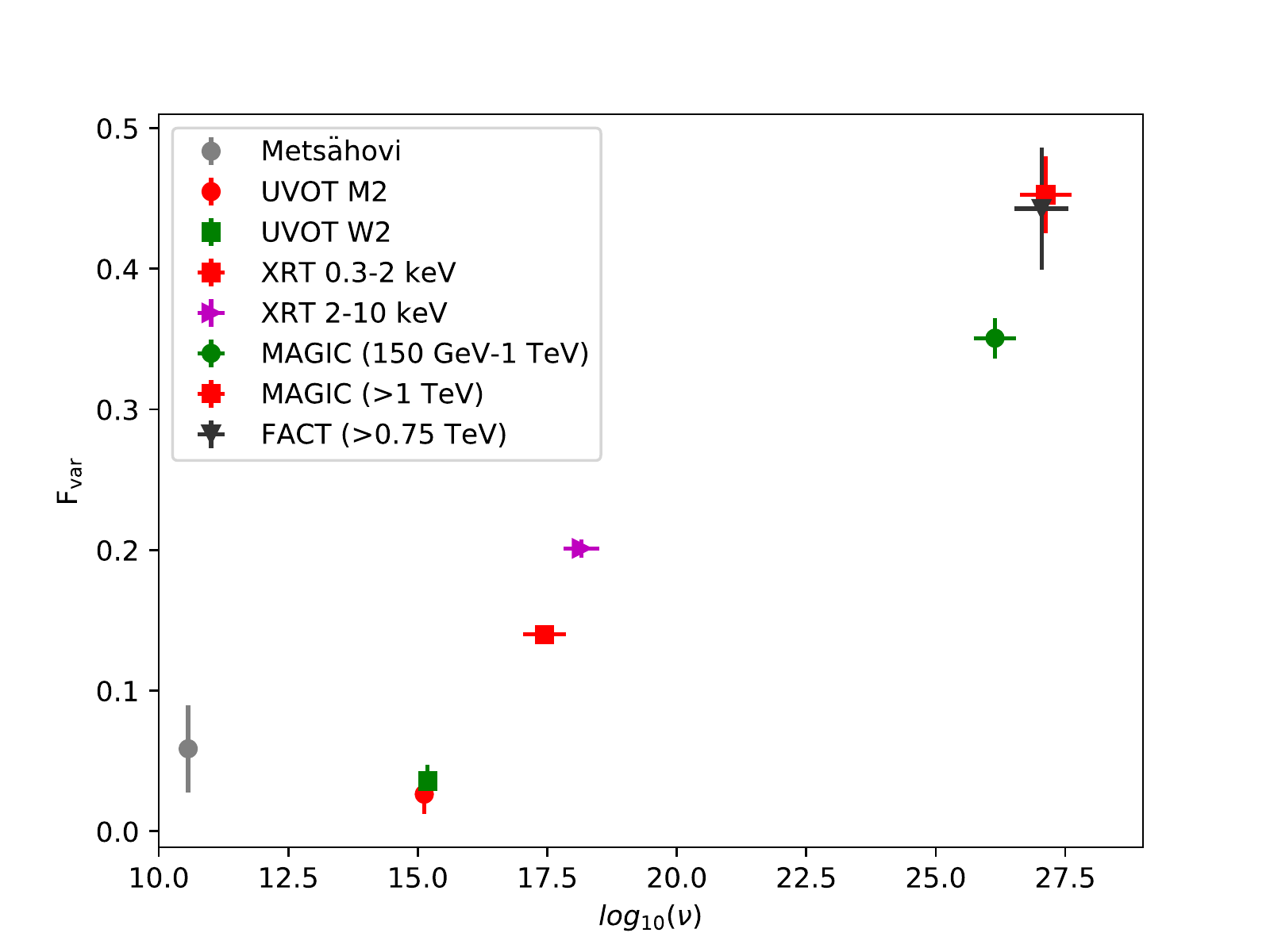} 
   \caption{Fractional variability $F_{var}$ as a function of frequency.}
   \label{fig:fvar}
\end{figure}

The results of the $F_{var}$ calculation for each energy band, as reported in Fig.~\ref{fig:mwl_lc}, are shown in Fig.~\ref{fig:fvar}. The fractional variability   is not defined in the case of radio and HE gamma rays observed with {\it Fermi}-LAT, as the excess variance is negative ($S^2$ is smaller than \mbox{$<\sigma^2_{err}>$}). A negative excess variance implies that either there is no variability or that the instruments are not sensitive enough to detect it.

There is a general increase in  the fractional variability with increasing energy of the emission, showing the highest variability in the VHE band.  At optical and UV bands the fractional variability is about 0.05, at the X-ray bands it is about 0.2, and at the VHE gamma-ray bands it is about 0.4. A comparable variability pattern in the broadband emission of Mrk~501 has been observed in most of the previous extensive campaigns
\citep[see, e.g.,][]{2015A&A...573A..50A,2017A&A...603A..31A,2018AhnenSubmitted}, indicating that it is a typical characteristic of Mrk501, during low and high activity. In contrast, for the other classical TeV blazar, Mrk421,  a well-defined double-peak structure is observed in the plot of $F_{var}$ versus energy, where the variability in the X-ray band is comparable (and even greater) than that at VHE gamma-ray energies 
\citep[see, e.g.,][]{2015A&A...576A.126A,2015A&A...578A..22A,2016ApJ...819..156B}.

\subsection{Correlation between the X-ray and VHE gamma-ray bands. }
\label{sec:XrayVHECorr}

\begin{table}
   \centering
   \begin{tabular}{c c c} 
        &0.15-1 TeV &   $>1$ TeV \\
    \hline
    \hline
    ~ & Pearson ($\sigma$) | DCF  & Pearson ($\sigma$) | DCF \\
    \hline
    0.3-2 keV& 0.75 (2.9$\sigma$) ~|~ 0.7$\pm$0.2 & 0.59 (2.0$\sigma$) ~|~ 0.6$\pm$0.2 \\
    2-10 keV & 0.85 (3.6$\sigma$) ~|~ 0.8$\pm$0.2 & 0.81 (3.4$\sigma$) ~|~ 0.8$\pm$0.2\\
    \hline
    \hline
   \end{tabular}
   \caption{Quantification of the correlation: VHE vs X-ray flux at different energy bands. The Pearson correlation and its
significance (in brackets) are calculated following \citet{2002nrca.book.....P}. The discrete correlation function (DCF) and errors are calculated as prescribed in \cite{1988ApJ...333..646E}. }
   \label{tab:vhe_vs_xrays}
\end{table}

\begin{figure}[htbp]
   \centering
   \includegraphics[scale=0.5, angle=0]{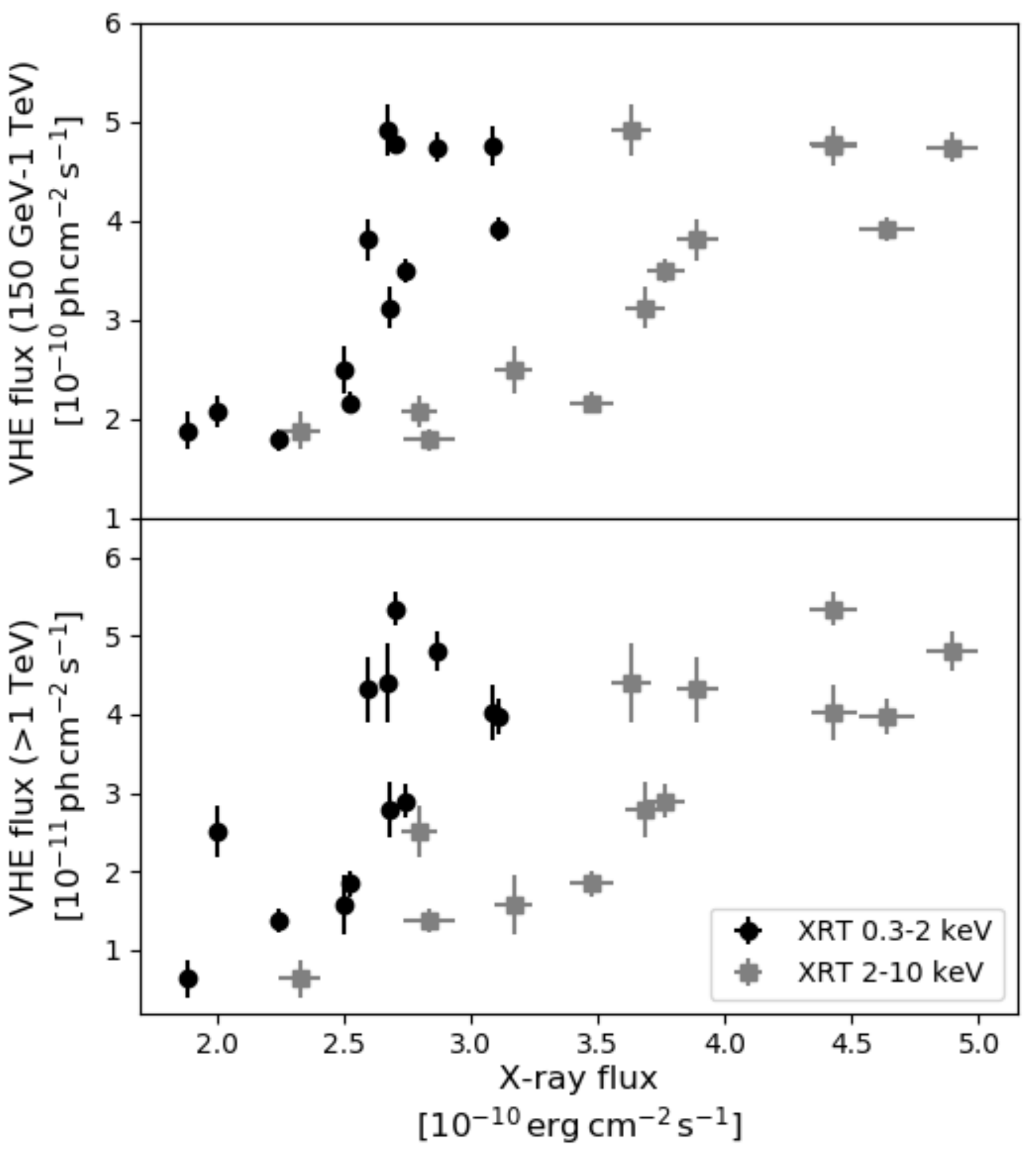} 
   \caption{VHE flux in two energy bands (0.15--1~TeV and $>$1~TeV)  as a function of the {\it Swift}/XRT flux in two energy bands (0.3--2~keV and 2--10~keV). } 
   \label{fig:corr_xrays_vhe}
\end{figure}

This section investigates the cross-correlation between the two segments of the electromagnetic spectrum with the highest variability, namely the X-rays and the VHE gamma rays (see Fig.~\ref{fig:fvar}). Figure~\ref{fig:corr_xrays_vhe} shows the integral VHE gamma-ray flux from two energy bands (0.15--1~TeV and $>$1~TeV) measured by MAGIC, plotted against the X-ray flux in two energy bands (\mbox{0.3--2~keV} and \mbox{2--10~keV)} observed by \textit{Swift}/XRT. The 13 X-ray and VHE fluxes depicted in this figure are taken within a maximum difference of 3\,hours from each other\footnote{The time difference is computed using the center of the time interval of the observations.}. Given that we did not find any significant intra-night variability (neither in the \textit{Swift}/XRT nor in the MAGIC and FACT data), the used X-ray and VHE data can be safely considered simultaneous. The correlation between these two bands is quantified using two methods: the Pearson correlation coefficient (and the significance of this correlation) and the discrete correlation
function \citep[DCF,][]{1988ApJ...333..646E}. The DCF has an advantage over the Pearson correlation in that it also uses  the uncertainties in the individual flux measurements, which naturally contribute to the dispersion in the flux values. The results are shown in Table~\ref{tab:vhe_vs_xrays}. Despite the relatively short time interval considered in this study, and  that Mrk~501 was in an elevated state at X-rays and VHE gamma-rays during the entire period,  we observe a significant correlation between the X-ray and VHE gamma-ray bands. This correlation increases slightly with the increasing energy in X-rays: it is $\sim3 \sigma$ for the 0.3--2~keV band and $\sim4 \sigma$ for the 2--10~keV band. A stronger correlation with increasing X-ray energy was also reported for Mrk501 \citep[see Tables 1 and 4 of][]{2018AhnenSubmitted}, but in that case for a much longer time interval (three months instead of two weeks). These observations indicate that, within the one-zone SSC theoretical framework, the electrons that dominate the emission at 2--10~keV make a larger contribution to the emission at VHE gamma rays than those that dominate the emission at 0.3--2~keV \citep[see][for further details]{2018AhnenSubmitted}.

It  is interesting to note that during periods of low activity the correlation between the X-ray and VHE bands has been shown to be only marginally significant or even nonexistent \citep[see, e.g.,][]{2015A&A...573A..50A,2017A&A...603A..31A,2018AhnenSubmitted}. On the other hand, this correlation is very strong for well-sampled and long-term light curves covering periods of low activity together with periods of very high activity \citep[see, e.g.,][]{2006ApJ...646...61G}. Naturally, our ability to detect significant correlations improves when considering accurate flux measurements and periods with large flux changes. The study reported here shows, for the first time for Mrk~501, a significant ($>3 \sigma$) correlated behavior between  X-rays and VHE gamma rays during a short period of time (two weeks) of persistent elevated activity. A correlation on weekly  timescales  was also claimed for Mrk~501 in \cite{1997ApJ...487L.143C} and \cite{Sambruna2000}, but the significance of this correlation was not computed in either of these two previous studies. On the other hand, a significant correlation between the X-ray and the VHE gamma-ray band during a $\sim$two-week elevated state has also been reported for Mrk~421 \citep{2015A&A...578A..22A}.
Such a X-ray--VHE correlation is actually expected within the framework of the synchrotron self-Compton (SSC) emission scenario \citep[see, e.g.,][]{ssc_maraschi}, which  predicts that the X-ray and the VHE gamma-ray emission are produced by the same population of electrons and positrons. This is the most widely used theoretical scenario for describing the emission of high-peaked BL Lac-type objects such as Mrk~501, and will be also used to model the broadband SEDs of these two weeks of remarkably high X-ray activity (see Sec.~\ref{sec:Model}).

\subsection{X-ray and VHE gamma-ray spectral variability}
\label{sec:spectra}
Most of the X-ray spectra measured with {\it Swift}/XRT are well characterized by a power-law function (PL), as reported in Appendix A (see Table~\ref{tab:xrt_fits}). A hint of harder-when-brighter evolution is observed in X-rays at $\sim2\sigma$ and $\sim4\sigma$ for soft \mbox{(0.3--2 keV)} and hard \mbox{(2--10 keV)} X-rays, respectively, as reported in Appendix~\ref{App:SpectrumVSFlux}.


The VHE gamma-ray spectra from MAGIC are characterized on one-day timescales because we did not find any significant intra-night variability during the observation campaign reported in this paper. The gamma-ray spectra are absorbed and distorted due to the interaction with the extragalactic background light (EBL) via pair production of an electron and a positron \citep[see, e.g.,][and references therein]{dominguez}. Both the observed and EBL-corrected \citep[assuming the EBL model from][]{dominguez} VHE spectra can  typically be well fitted by a simple power-law function (PL, eq.~\ref{eq:power-law}), except for two or three cases out of the 15 nights  which show curvature, and a log-parabola fit (LP, eq.~\ref{eq:logparabola}) is preferred over a PL fit with a significance higher than $3~\sigma$ (see Table~\ref{tab:vhe_fits}). The PL function is defined as

\begin{equation}
\frac{dF}{dE}=f_0 \left(\frac{E}{500 \, \mathrm{GeV}}\right)^{\Gamma}
\label{eq:power-law}
,\end{equation}
where $f_0$ represents the normalization constant and $\Gamma$ the spectral index. The LP function is given by 

\begin{equation}
\frac{dF}{dE}=f_0 \left(\frac{E}{500 \,\mathrm{GeV}}\right)^{\Gamma-b \cdot log \frac{E}{500 \, \mathrm{GeV}}}
\label{eq:logparabola}
,\end{equation}
which uses the $b$ parameter in addition to eq.~\ref{eq:power-law} to parameterize  the spectral curvature. 

The flux and spectral evolution in the VHE band, as observed by MAGIC, does not show a harder-when-brighter trend, as reported in  Appendix~\ref{App:SpectrumVSFlux}. During the observations taken on 2014 July 19 (MJD 56857.98), which is the day with the highest X-ray flux above 0.3~keV measured by {\it Swift} during its entire operation, a hint of a narrow spectral feature is observed. The investigation of this feature is discussed in Sec.~\ref{bump}.   

\subsection{Investigation of a feature in the VHE spectrum from \mbox{2014 July 19}}
\label{bump}

\begin{figure}
   \centering
   \includegraphics[scale=0.52]{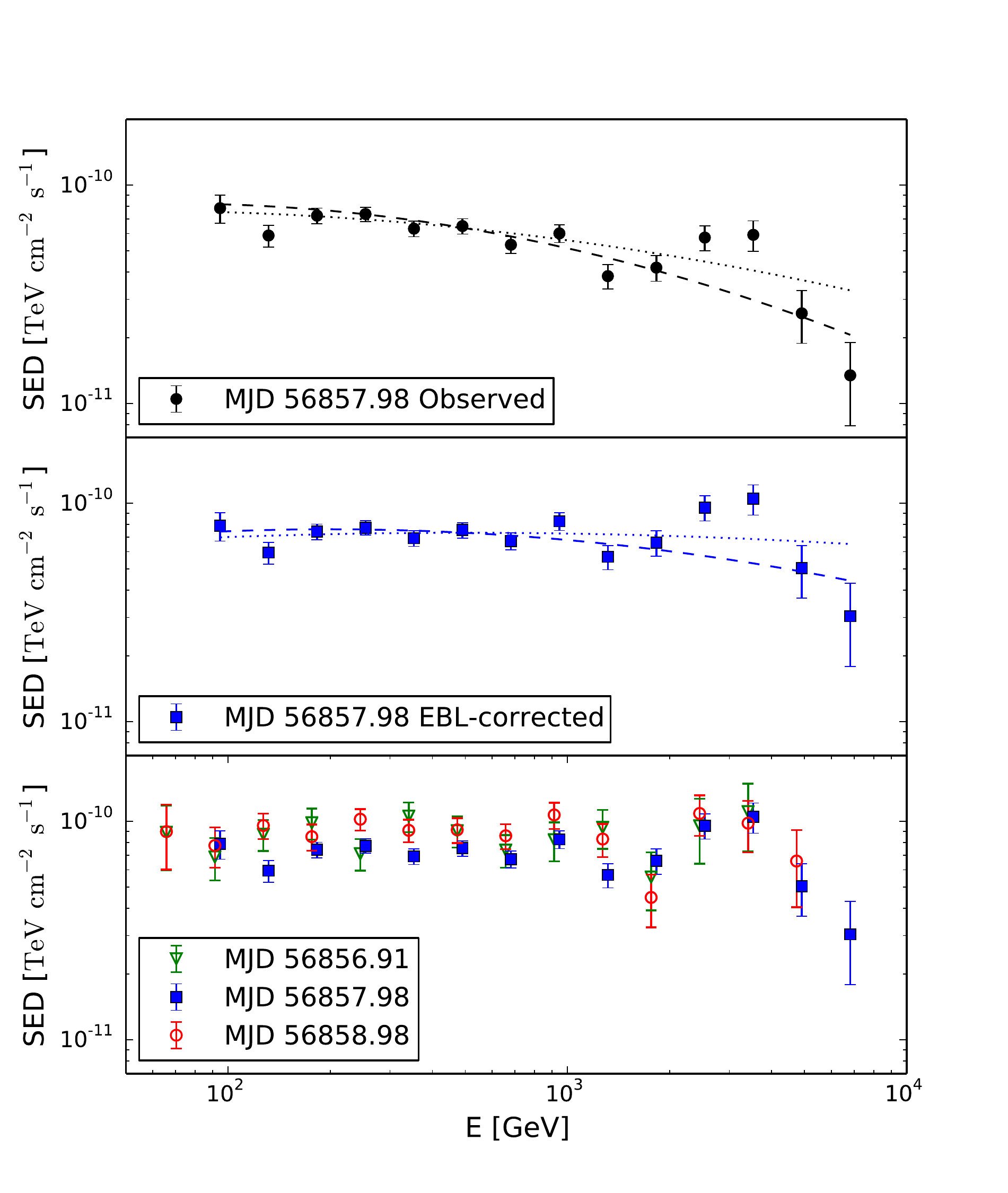} 
   \caption{VHE SEDs from the MAGIC telescopes during the highest \mbox{X-ray} flux measured with {\it Swift}-XRT. {\it Top and Middle panels:} Black circles represent the observed SED from 2014 July 19  (MJD 56857.98), while the blue squares denote the same spectrum corrected for EBL absorption  \citep[using the model from][]{dominguez}. In both panels the dotted lines depict the best LP fits (reported in Table~\ref{tab:bump_fit}), while the dashed lines show the best fits using data up to 1.5 TeV, and extrapolated beyond that energy (from the test reported in Table \ref{tab:vhe_partialfit}).  {\it Bottom panel:} VHE SEDs after EBL correction during three consecutive nights around 2014 July 19  (MJD 56857.98).}
   \label{fig:VHE_bump}
\end{figure}

The VHE spectrum observed by MAGIC on 2014 July 19 shows a hint of a narrow spectral feature, as depicted in the upper panel of Fig.~\ref{fig:VHE_bump}. To test the significance of this feature, the goodness of the fit to the spectrum was evaluated by means of a $\chi^2$ test using different functions: a PL (see eq.~\ref{eq:power-law}), an LP (eq.~\ref{eq:logparabola}), and an exponential log-parabola (ELP) defined as

\begin{equation}
\frac{dF}{dE}=f_0 \left(\frac{E}{500 \,\mathrm{GeV}}\right)^{\Gamma-b \cdot log \frac{E}{500 , \mathrm{GeV}}} e^{-E/E_{c}}
\label{eq:epl}
,\end{equation}
where in addition to the parameters used in eq.~\ref{eq:logparabola}, the parameter $E_{c}$ sets the exponential cutoff energy. These three spectral functions have been widely used to successfully parameterize the spectra of VHE gamma-ray sources.

\begin{table*}
   \centering
   \begin{tabular}{c c c c c c c c} 
        &Fit & $f_0$& $\Gamma$ & b & $E_{c}$ &$\mathrm{\chi}^2$/df & p-value \\
        &&$[10^{-10}\mathrm{TeV}^{-1}\mathrm{cm}^{-2}\mathrm{s}^{-1}]$& & & [TeV] & \\
        \hline
        \hline
        Observed&PL&2.32$\pm$0.07& -2.20$\pm$0.03 &- & - &52.1/15&$5.5\times10^{-6}$ (4.6$\sigma$) \\
        EBL--corr&PL&2.81$\pm$0.08& -2.02$\pm$0.03 &- & - &36.0/15&$1.8\times10^{-3}$ (3.1$\sigma$) \\
        \hline
        Observed&LP&2.54$\pm$0.09& -2.16$\pm$0.03 &0.08$\pm$0.02 & - &37.9/14&$5.4\times10^{-4}$ (3.5$\sigma$) \\
        EBL--corr&LP&2.93$\pm$0.10& -2.00$\pm$0.03 &0.04$\pm$0.02 & - &33.0/14&$2.9\times10^{-3}$ (3.0$\sigma$) \\
        \hline
    Observed&ELP&2.69$\pm$0.12& -2.02$\pm$0.07 & -0.02$\pm$0.05 & 5.7$\pm$2.9 &34.8/13&$9.0\times10^{-4}$ (3.3$\sigma$) \\
        EBL--corr&ELP&3.11$\pm$0.15& -1.87$\pm$0.08 &-0.05$\pm$0.05 & 5.8$\pm$3.2 &31.0/13&$3.3\times10^{-3}$ (2.9$\sigma$) \\
        \hline
        \hline
   \end{tabular}
   \caption{Results from the forward-folding fits with three different functions (PL, LP, and ELP) to the MAGIC VHE spectra observed and EBL-corrected~\citep[using the EBL model from ][]{dominguez} from 2014 July 19 (MJD 56857.98).}
   \label{tab:bump_fit}
\end{table*}

The parameters and the goodness of the spectral fits for both the observed and EBL-corrected spectra are reported for the three functions (PL, LP, ELP) in Table~\ref{tab:bump_fit}. We note that the reported spectral fits were obtained with a forward-folding procedure \citep[procedure details given in][]{ebl_magic}, where the number of degrees of freedom is related to the bins in estimated energy and not to the bins in true energy, which is what is shown in the broadband SEDs (Fig.~\ref{fig:VHE_bump}). As shown in Table~\ref{tab:bump_fit}, neither the observed nor the EBL-corrected spectrum can be fitted successfully with any of the three functions. The fits to the observed VHE spectra can be rejected at significance values ranging from $3.3\sigma$ to $4.6\sigma$, depending on the function. For the EBL-corrected spectrum, the rejection occurs at significance values from  $2.9\sigma$ to $3.1\sigma$.


A further test is performed fitting with  an LP (to allow possible curvature) all the single-night spectra up to 1.5 TeV, and evaluating the model-data agreement when extending the resulting fit function to energies higher than 1.5 TeV. This approach allows us to quantify how much the spectra change at high energies with respect to  the low energies, and hence investigate the potential existence of additional spectral components. This test is carried out only for the spectra with at least three spectral points beyond 1.5 TeV. The table with the results is found in Appendix C. As shown in Table~\ref{tab:vhe_partialfit}, the only extended fit beyond 1.5 TeV that can be rejected with a high confidence level is the one for the night of MJD 56857.98, with a significance of 6.0$\sigma$ for the observed spectrum and 5.0$\sigma$ for the EBL-corrected spectrum. 

\begin{figure}
   \centering
   \includegraphics[scale=0.6]{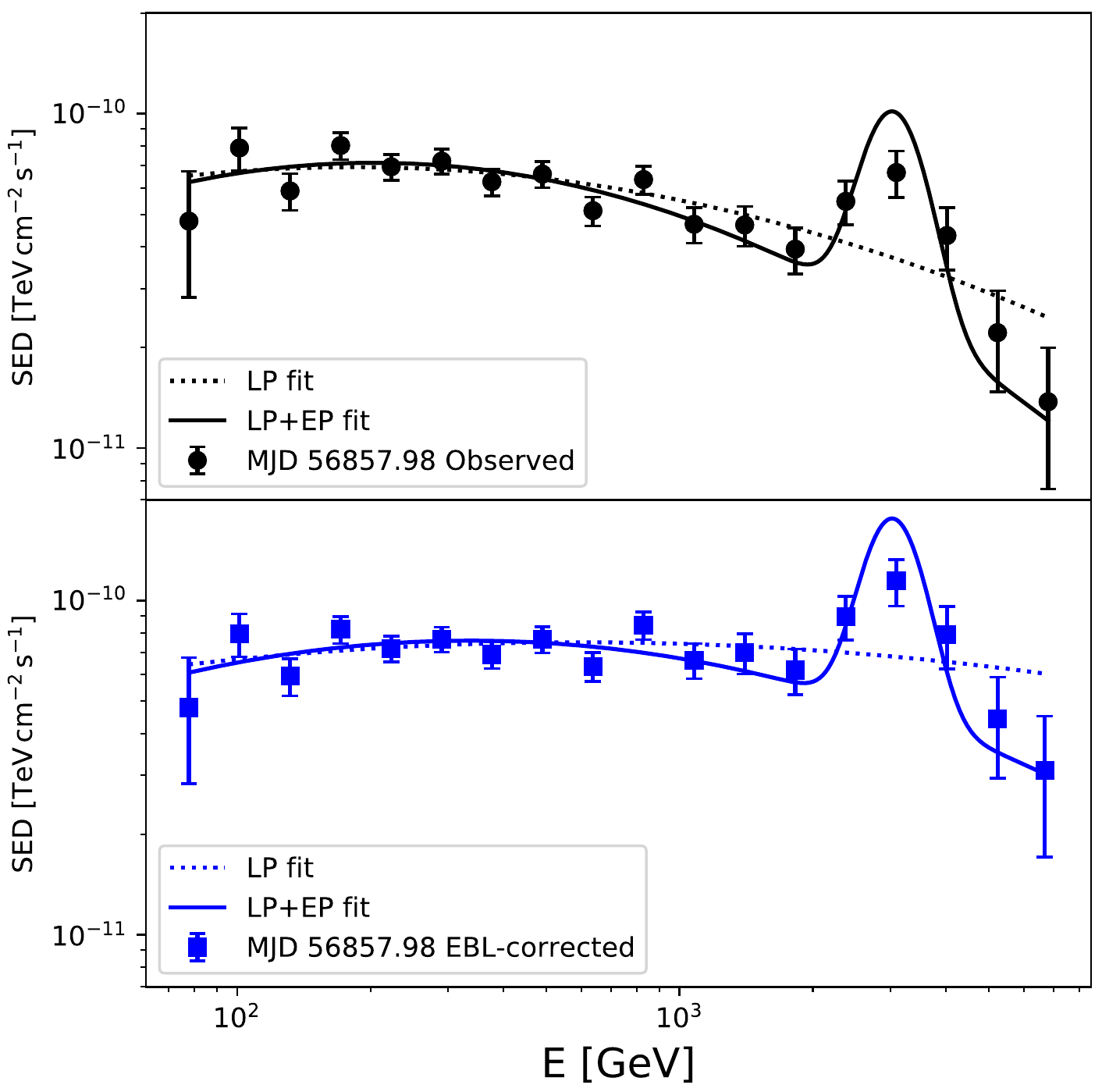} 
   \caption{VHE SED from 2014 July 19 (MJD 56857.98) measured with the MAGIC telescopes with an analysis that uses 25\% more bins in estimated energy with respect to that shown in Fig.~\ref{fig:VHE_bump}.  Black circles represent the observed SED, while the blue squares denote the same spectrum corrected for EBL absorption  \citep[using the model from][]{dominguez}. In both panels the dotted lines depict fits with an LP function, while the solid lines depict the fits with an LP+EP function. The parameter values resulting from the spectral fits are reported in Table~\ref{tab:bump_fit_extracomponent}.}
   \label{fig:VHE_bump_extracomponent}
\end{figure}

Motivated by the difficulty of fitting the spectrum from 2014 July 19 with the typical analytic functions used to describe the VHE spectra of blazars, we compare the goodness of the fit for an LP function with respect to an LP plus a strongly curved LP, described as an eplogpar model \citep[EP,][]{Tramacere2007} described in eq.~\ref{eq:eplogpar}, using a likelihood ratio test (LRT, where $\chi^2_{LRT}=\chi^2_{LP}-\chi^2_{LP+EP}$ with degrees of freedom $df=df_{LP}-df_{LP+EP}$).


\begin{equation}
\frac{dF}{dE}= \frac{K}{E^2} 10^{-\beta \log^2(E/Ep)}
\label{eq:eplogpar}
,\end{equation}where $K$ is a constant, $Ep$ represents the energy peak, and $\beta$ is the curvature term. The resulting spectral fits are depicted in Fig.~\ref{fig:VHE_bump_extracomponent} and the fit parameter values are reported in Table~\ref{tab:bump_fit_extracomponent}. In order to better characterize the relatively narrow spectral feature, we increased by 25\% the number of bins in estimated energy  with respect to those used in the spectral fits performed on all the single-night VHE spectra reported in this manuscript (see Table~\ref{tab:bump_fit} and Table~\ref{tab:vhe_fits}). This also increased the number of bins in true energy (i.e., the number of data points in Fig.~\ref{fig:VHE_bump_extracomponent} is larger than that of Fig.~\ref{fig:VHE_bump}).
This fine energy binning used to derive the spectral fitting results for 2014 July 19, as reported in Fig.~\ref{fig:VHE_bump_extracomponent} and Table~\ref{tab:bump_fit_extracomponent}, would not work on other days with lower gamma-ray activity and/or shorter observation times, due to the lower photon statistics. The LRT shows that the LP with the additional narrow component is preferred over the single LP function at 4.5$\sigma$ when using the observed spectrum and 3.9$\sigma$ when using the EBL-corrected spectrum.

\begin{table*}
   \centering
   \begin{tabular}{c c c c c c c c c c} %
   \hline
   \hline
   & Fit & $f_0 \cdot 10^{10}$ & $\Gamma$ & b & $K\cdot 10^{5}$ & $\beta$& Ep & $\mathrm{\chi}^2$/df & LRT\\
   &  & [$\mathrm{TeV}^{-1}\mathrm{cm}^{-2}\mathrm{s}^{-1}$] & &  & [ $\mathrm{TeV}^{-1}\mathrm{cm}^{-2}\mathrm{s}^{-1}$] & & [TeV] & & \\
   \hline
   \hline
   Observed & LP & $2.56\pm0.09$ & $-2.16\pm0.03$ & $0.08\pm0.02$ & - & - & - & 39.8/19 &  \\
   Observed & LP+EP & $2.54\pm0.10$ & $-2.26\pm0.04$ & $0.14\pm0.03$ & $7.7\pm1.7$ & $9.1\pm3.2$ & $3.04\pm0.10$ & 13.5/16 & $4.5\sigma$\\
   \hline
   \hline
   EBL-corr & LP& $3.00\pm0.11$ & $-1.99\pm0.03$ & $0.04\pm0.02$ & - & - & - &   35.4/19 & \\
   EBL-corr & LP+EP & $2.99\pm0.11$ & $-2.08\pm0.04$ & $0.10\pm0.03$ & $13.0\pm3.0$ & $10.0\pm3.6$ & $3.03\pm0.10$ & 14.6/16 & $3.9\sigma$\\
   \hline
   \hline
   \end{tabular}
   \caption{Results from the forward-folding fits with an LP and an LP+EP to the MAGIC VHE spectra observed and EBL-corrected~\citep[using the EBL model from ][]{dominguez} from 2014 July 19 (MJD 56857.98). The likelihood ratio test (LRT) of the second function with respect to the first  is reported in the last column. The spectral fits given in this table are depicted in Fig.~\ref{fig:VHE_bump_extracomponent}.}
   \label{tab:bump_fit_extracomponent}
\end{table*}

\begin{figure}
   \centering
   \includegraphics[scale=0.44]{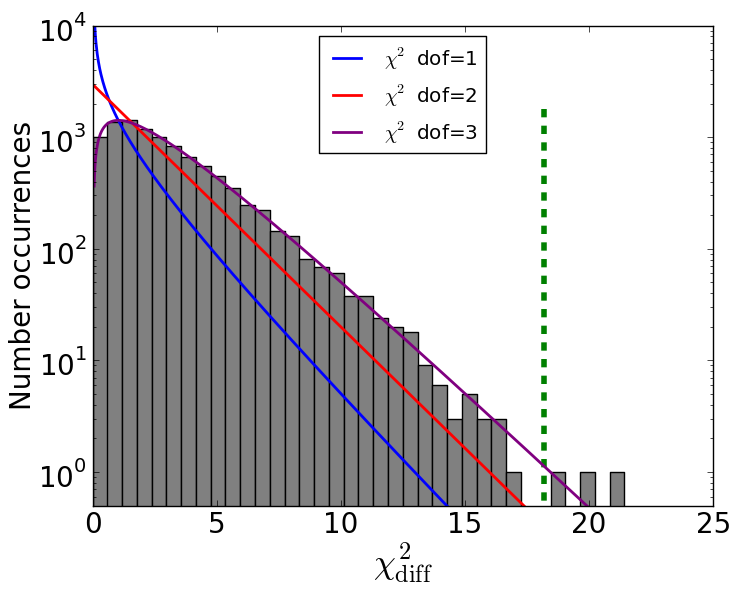}
   \includegraphics[scale=0.44]{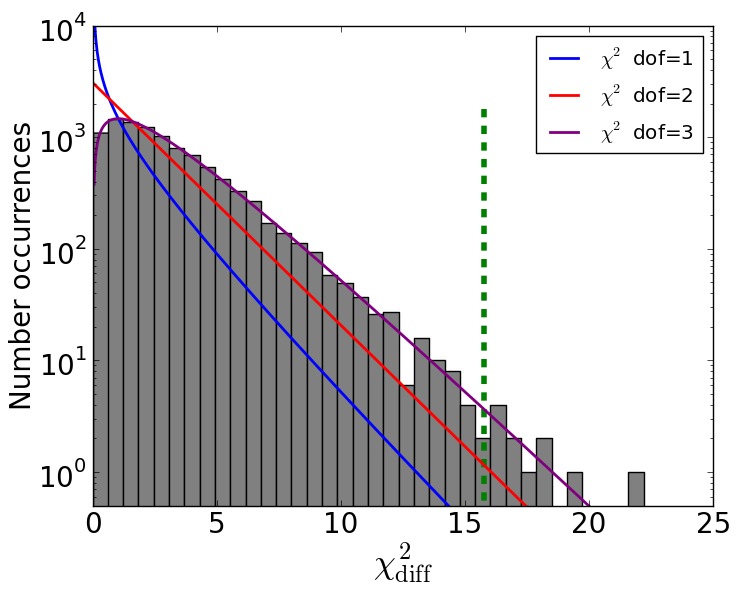}
   \caption{Distributions of $\chi^2_{diff}$ values (null distribution of the LRT statistic) obtained from a Monte Carlo test that uses 10000 simulated spectra to compare a baseline model (null hypothesis) parameterized with an LP function, and a narrow-feature model parameterized with an LP+EP function. The top panel shows results derived with the observed VHE spectrum, and the bottom panel shows the results obtained with the EBL-corrected VHE spectrum. The green dashed line indicates the $\chi^2_{diff-data}$ (LRT$_{data}$) obtained when comparing the LP and LP+EP fit results on the spectral data from Fig.\ref{fig:VHE_bump_extracomponent}. The blue, red, and purple solid lines depict the nominal $\chi^2$ distribution for 1, 2, and 3 degrees of freedom. See text in section~\ref{bump} for further details. }
   \label{fig:likelihood_mc_test_LP}
\end{figure}

It has been shown, in certain situations, that the LRT applied on a measured spectrum may overestimate or underestimate the significance of a narrow feature at an arbitrary location  \citep[][]{2002ApJ...571..545P}. In order to complement what is shown above,  we performed a dedicated Monte Carlo simulation to better quantify the significance of the narrow feature observed in the VHE spectrum from 2014 July 19. This test is performed on the VHE spectra in the plane of true energy, using the spectral data points reported in Fig. 5. This makes the test simpler and more transparent than performing the test on the plane of estimated energy, which would require using the  forward-folding methods specifically developed for the MAGIC software. While the forward-folding procedure might slightly affect the spectral index estimation, it cannot introduce narrow spectral features. Therefore, the use of the spectra in the plane of estimated energy (instead of true energy) should not have any impact on the test to validate the LRT methodology, while improving the repeatability of the test without the need of instrument dependent software. In this test, we first fit the spectral data points from Fig.5 \citep[calculated using the \texttt{flute} routine within MARS, as described in][]{zanin} with an LP function, which is used to describe the continuum model and represents the null hypothesis. Then we fit the spectral data points with an LP+EP function, which describes the hypothesis of the narrow feature.  The LP+EP hypothesis has three additional free parameters in comparison to the LP function: the normalization parameter $K$; the location of $E_p$, which can go from the energy of 0.08~TeV (first data point in the spectrum) to the energy 6.80~TeV (last data point in the spectrum); and the curvature parameter $\beta$, which can vary from 1 to 20. The difference between the $\chi^2$ from the two hypotheses ($\chi^2_{diff}=\chi^2_{LP} - \chi^2_{LP+EP}$) is $\chi^2_{diff-data}=18.1$ for the observed spectrum and $\chi^2_{diff-data}=15.8$ for the EBL-corrected one. These $\chi^2_{diff-data}$ values are somewhat lower than the difference of $\chi^2$ values reported in Table~\ref{tab:bump_fit_extracomponent} (e.g., for the observed spectrum $\chi^2_{diff}=\chi^2_{LP} - \chi^2_{LP+EP}$= \mbox{39.8-13.5= 26.3}),  where the LP and LP+EP spectral fits were performed in the plane of reconstructed energy. Apart from statistical fluctuations, the slightly higher LRT values reported in Table~\ref{tab:bump_fit_extracomponent} may occur because of the slightly higher resolution when performing the spectral fits in estimated energy, where the number of bins is larger than the number of energy bins in the VHE gamma-ray spectrum reported in Fig.\ref{fig:VHE_bump_extracomponent}. Then we use the LP function derived from the spectral fit (the null hypothesis) to generate 10000 realizations of this spectrum with data points that have the same statistical uncertainty as the spectral data points from Fig.\ref{fig:VHE_bump_extracomponent}. 
In order to account for the uncertainty in the  null hypothesis, following the prescription from \citet{2006ApJ...646..783M}, we fit  each of these simulated
spectra with an LP function and generated another simulated spectrum using the new LP values as input. This new simulated spectrum is then fit with an LP function, and the resultant $\chi^2$ is the one used to describe the goodness of the fit for the baseline (LP) model\footnote{Each spectral simulation is initiated with a slightly different realization of the null hypothesis model. This extra step does not change the overall results, but provides a more realistic simulation because the baseline model is known only within some statistical uncertainty.}. The distributions of $\chi^2_{diff}$ (= $\chi^2_{LP} - \chi^2_{LP+EP}$) values obtained from the 10000 simulated spectra (i.e., the null distributions of the LRT statistic) are shown in Fig.\ref{fig:likelihood_mc_test_LP}, and the summary of the resulting numbers are reported in Table~\ref{tab:MC_Results_LRT}. The distributions of $\chi^2_{diff}$ values follow closely a distribution of $\chi^2$ for three degrees of freedom, which is what one would expect when comparing, for a large number of simulated spectra, a hypothesis that has three additional degrees of freedom with respect to the baseline model. Therefore, the Monte Carlo test confirms the reliability of the LRT applied to the spectral data.

\begin{table*}
   \centering
   \begin{tabular}{c c c} %
   \hline
   \hline
        & Experimental data
        & MC data: 10$^4$ simulated spectra \\
    &  $\chi^2_{diff-data}$ ~~|~~ $p_{value}$~ (significance)
        & N$> \chi^2_{diff-data}$ ~~|~~ $p_{value}$~ (significance)\\
        \hline 
Observed  ~~~~
& 18.1 ~~~ | ~~~ $4.2\times10^{-4}$ (3.5$\sigma$)
& ~3 ~~~|~~~ $3.0\times10^{-4}$ (3.6$\sigma$) \\ 
EBL-corr ~~~~
& 15.8 ~~~|~~~ $1.2\times10^{-3}$ (3.2$\sigma$)
& 11 ~~~|~~~ $1.1\times10^{-3}$ (3.3$\sigma$) \\ 
\hline 
   \hline
   \end{tabular}
   \caption{Results from the Monte Carlo tests used to quantify the chance probability (and related significance) of observing a spectral feature (parameterized with an LP+EP function) on  top of the measured VHE gamma-ray spectrum described by an LP. The $p_{value}$ (and related significance) values in the   ``experimental data'' column are derived from the nominal $\chi^2$ distribution with 3 degrees of freedom, while the numbers reported in the column  ``MC data'' column are directly derived from the $10^4$ simulated spectra. See text in section~\ref{bump} for further details.}
   \label{tab:MC_Results_LRT}
\end{table*}

Additionally, we also performed a Monte Carlo test similar to the one reported in \citet{2010A&A...521A..57T}, which had been used to quantify the significance of line features obtained from a dedicated search over a large number of measured X-ray spectra. The context of this test is different from the one described above, which relates to the investigation of a feature observed in a single spectrum, but provides an alternative perspective to the evaluation of the random chance probability for the occurrence of narrow features in continuum spectra. The details of this test and the results obtained are given in Appendix~\ref{App:toyMC}. In addition to the EP function of arbitrary curvature used in the Monte Carlo test described above, we also used  an EP function with fixed shape (as in \citealt{2010A&A...521A..57T}) and a Gaussian function of arbitrary width. The results obtained for the three hypotheses are similar, and comparable within 0.5$\sigma$, to the results reported in Table~\ref{tab:MC_Results_LRT}.

The above-mentioned tests aim to quantify the statistical significance of the deviation of this narrow feature at $\sim$3~TeV with respect to (smooth) functions typically used to fit the spectra from gamma-ray sources, but they do not account for potential instrumental or analysis problems in the dataset. We performed several tests to search for these instrumental or analysis artifacts that may mimic similar spectral features to the one reported here. Specifically,   a) we performed three different analyses, all them yielding the same results; b) we inspected the effective area after gamma--hadron separation cuts \citep[see Sec. 3.4 and 4.2 in][]{magic_upgrade2}, which   did not show any discontinuity or feature; c) we varied the gamma--hadron separation cuts  (through the random forest hadronness parameter), and several VHE spectra (with different gamma efficiencies) were produced for 2014 July 19, all them showing the feature at $\sim$3~TeV; d) we produced spectra with and without the LIDAR atmospheric corrections \citep{2014arXiv1403.3591F,2015ApJ...812...65F}, both yielding spectra with the same spectral feature; and e) we applied the exact same data analysis procedures  to data from the Crab nebula taken under similar conditions, yielding a spectrum without features. Therefore, the Mrk~501 VHE spectrum from 2014 July 19, derived in different ways, always showed  the narrow TeV feature (at somewhat different magnitudes), deviating from an LP function at a significance varying from $\sim$2$\sigma$ to $\sim$5$\sigma$. Therefore, while the narrow spectral feature is statistically only marginally significant ($\sim$3-4$\sigma$), we are confident that it is not produced by any instrumental or analysis  artifact.

The shape of the VHE spectra from 2014 July 18 and 2014 July 20, above an energy of 1.5 TeV, appears to be compatible with the VHE spectrum from 2014 July 19, as  is shown in the bottom panel of Fig.~\ref{fig:VHE_bump}. While there is clear variability at energies below 1.5~TeV during these three consecutive nights, the spectral points appear to be similar at energies above 1.5~TeV. Nevertheless, as shown in Table~\ref{tab:vhe_fits}, the spectra obtained from the nights before and after that of July 19 are nicely described  with PL functions, and hence the deviations from the PL functions above 1.5~TeV are not significant. Therefore, the sensitivity of these observations with MAGIC is insufficient to constrain the duration of the $\sim$3~TeV feature to only one day: it may have lasted for three nights, which would correspond to the first of the two bumps in the VHE emission reported in the LC from Fig.~\ref{fig:mwl_lc}. We did not find any evidence of narrow spectral features in the VHE spectra during the second bump of the flare (MJD 56865-56867) when a similar X-ray and VHE flux is reached, as shown in Fig.~\ref{fig:mwl_lc}.

This is the first time that a narrow VHE spectral feature, inconsistent with a smooth function (PL, LP, and ELP) at more than 3$\sigma$, is found in the spectrum of Mrk~501 or any other blazar (see Table~\ref{tab:bump_fit}). With the caveat of doing the test a posteriori, the addition of a narrow component (EP, see eq.~\ref{eq:eplogpar}) to the VHE spectral fit is preferred at more than $\sim$3-4$\sigma$, depending on the method used for the test. This additional spectral component peaks at $\sim$3\,~TeV with a FWHM of $\sim$1.4\,~TeV, and, as we discuss in Sec.~\ref{sec:ModelNarrow}, it may be interpreted as an indication of additional physics in the theoretical framework aiming to explain the broadband emission of Mrk\,501.


\section{Characterization of the temporal evolution of the broadband spectral energy distribution}
\label{sec:Model}

Broadband SEDs were built with MWL simultaneous observations performed within hour timescales: out of the 15 SEDs considered here, the temporal difference between the X-ray and VHE measurements is less than 1 hour for six of them, between 1 and 2 hours for five of them, and 3 hours for two of them. The remaining two SEDs do not have X-ray data taken simultaneously with the VHE data observations, and we used the spectra from the night before and after as a guide. Given that we did not detect significant intra-night variability, we can assume that the variability timescales are longer than the time difference between observations. Therefore, all the observations used here can be considered  simultaneous. Each individual MWL SED is modeled using a one-zone SSC model from \cite{model}. The emitting region is assumed to be a sphere filled with relativistic electrons whose radius is compatible with the section of the jet. The electron energy distribution (EED) is described by a smoothed broken power law function as
\begin{equation}
N(\gamma)=K \gamma^{-n1} \left(1+\frac{\gamma}{\gamma_b}\right)^{n1-n2}
,\end{equation}
where $K$ represents the normalization factor, and the spectral indices before and after the break are given by $n1$ and $n2$, respectively. The energy (Lorentz factor) break is denoted by $\gamma_b$, and the function is defined between a minimum and maximum Lorentz factor $\gamma_{min}$ and $\gamma_{max}$. The synchrotron emission is produced by the interaction of this relativistic electron distribution with the tangled magnetic field ($B$). The synchrotron photons can interact with the same population of relativistic electrons via inverse Compton (IC) scattering, being responsible for the high-energy emission within the SSC scenario. In addition, the model also takes into account the bulk Lorentz factor and the viewing angle of the jet, included within a single parameter as the Doppler factor ($\delta$). The emitting region size is constrained by the causality relation: R$<(c \cdot t \cdot \delta)/(1+z)$. Assuming a $\delta=20$, which is often used to model the broadband SED of Mrk\,501 within SSC scenarios \citep[e.g.,][]{2011ApJ...727..129A,2015A&A...573A..50A}, and given that the shortest variability found within the MWL data sample is  on the order of one~day, the emitting region size can be constrained to R$<5\cdot 10^{16}$~cm. 

During the two-week time interval considered in this work, the X-ray emission observed by XRT display very hard spectra, compatible with the historical Mrk~501 flare from 1997 \citep[see, e.g.,][]{historical_flare}. Such hard X-ray spectra cannot be properly described together with the optical-UV emission with a single component. A similar situation occurred with the data collected from the extensive campaigns in 2009 and 2012 \citep{2017A&A...603A..31A,2018AhnenSubmitted}. Moreover, as shown in Fig.~\ref{fig:mwl_lc} and Fig.~\ref{fig:fvar}, the variability observed in the optical-UV band is much lower than in X-rays and VHE gamma rays, which also suggests that the emission at the optical and X-ray frequencies is dominated by different components, possibly located at different parts of the jet. 

A study using multi-year radio and optical light curves reported in \cite{lindfors} shows only a marginally significant  (2$\sigma$ confidence level) correlation between these two bands.  This suggests that a fraction of the optical emission might be produced co-spatially with  the radio emission. Due to self-absorption at radio frequencies, the radio emission is assumed to be produced in the outer regions of the jet. The radio emission is likely produced by a superposition of multiple self-absorbed jet components \cite{radio_absortion}. Emitting regions at radio wavelengths are typically larger and more complex. In particular, for Mrk~501 the radio observations reveal a complex jet with multiple components and a jet limb re-brightness \citep{giroletti2008}. Therefore, the simple one-zone SSC models are not the best approach to model the radio and optical-UV emission. In any case, just as an example, we tried and successfully managed to model the radio to optical--UV emission with an additional SSC component with a larger size. The details are given in Appendix D. 

\begin{table}
   \centering
   \begin{tabular}{c c c c c} 
        MJD&$\gamma_b$&$n1$&$n2$&$B$\\
        &[$10^{5}$]&&&[G]\\
        \hline
        \hline
        56854.91 & 1.4&2.018&3.1&0.140\\
        56855.91 & 2.0&2.00&3.1&0.127\\
        56856.91 & 8.5&1.99&3.1&0.087\\
        56857.98 & 4.0&2.00&3.1&0.120\\
        56858.98 & 9.0&2.00&3.1&0.105\\
        56859.97 & 4.0&2.00&3.1&0.110\\
        56861.01 & 3.5&2.015&3.1&0.115\\
        56862.02 & 1.9&2.015&3.1&0.134\\
        56863.00 & 1.9&2.01&3.05&0.130\\
        56864.02 & 2.5&2.03&3.05&0.149\\
        56865.00 & 4.0&2.00&3.1&0.110\\
        56866.00 & 20.0&1.99&3.1&0.078\\
        56867.00 & 9.5&1.99&3.1&0.084\\
        56868.01 & 11.0&1.99&3.1&0.090\\
        56869.92 & 3.0&2.016&3.1&0.115\\
        \hline
        \hline
   \end{tabular}
   \caption{One-zone SSC model results. The following parameters were fixed: $\gamma_{min}= 10^3$, $\gamma_{max}= 3\times10^6$, electron density$=2.1\times 10^4[\mathrm{cm}^{-3}]$, $R=2.9\times10^{15}$[cm], and $\delta=20$.}  
   \label{tab:modeling}
\end{table}

One of the goals of this work is to describe the evolution trend of the MWL SEDs observed during this two-week period of outstanding X-ray activity. Owing to the degeneracy in the parameter values from theoretical models used for blazars, we do not intend to produce model curves that describe perfectly the SEDs, but rather to evaluate how to reproduce the observed broadband behavior with simple variations in the model parameters.  For this purpose we attempted to model the data modifying only a few parameters. Given that the overall behavior observed during this period of extreme X-ray activity in 2014 is quite similar to that observed during the outstanding flaring activity in 1997 \citep{1998ApJ...492L..17P,1999A&A...350...17D,1999ApJ...518..693Q}, we decided to follow \cite{historical_flare}, and relate the overall changes in the broadband SED to variations in the parameter $\gamma_b$. In the canonical one-zone SSC scenario, the break in the electron energy distribution is related to the cooling of the electrons and hence inversely related to the size of the emitting region $R$ and the square of the magnetic field $B^2$ (see Appendix~\ref{App:RelationBreakAndB}), and hence any modification of $\gamma_b$ will come with changes in the parameters $R$ and/or $B$. For simplicity, we fixed the size of the emitting region, as well as the edges $\gamma_{min}$ and $\gamma_{max}$ and the electron number density and Doppler factor $\delta$, and allowed  the magnetic field strength $B$, the indices $n1$ and $n2$, and the break $\gamma_b$ of the EED to vary. Following the canonical one-zone SSC framework, we also kept the expected difference in the spectral indices $n2-n1 \sim 1$.

\begin{figure*}[h]
   \centering
   \includegraphics[scale=0.4, angle=0]{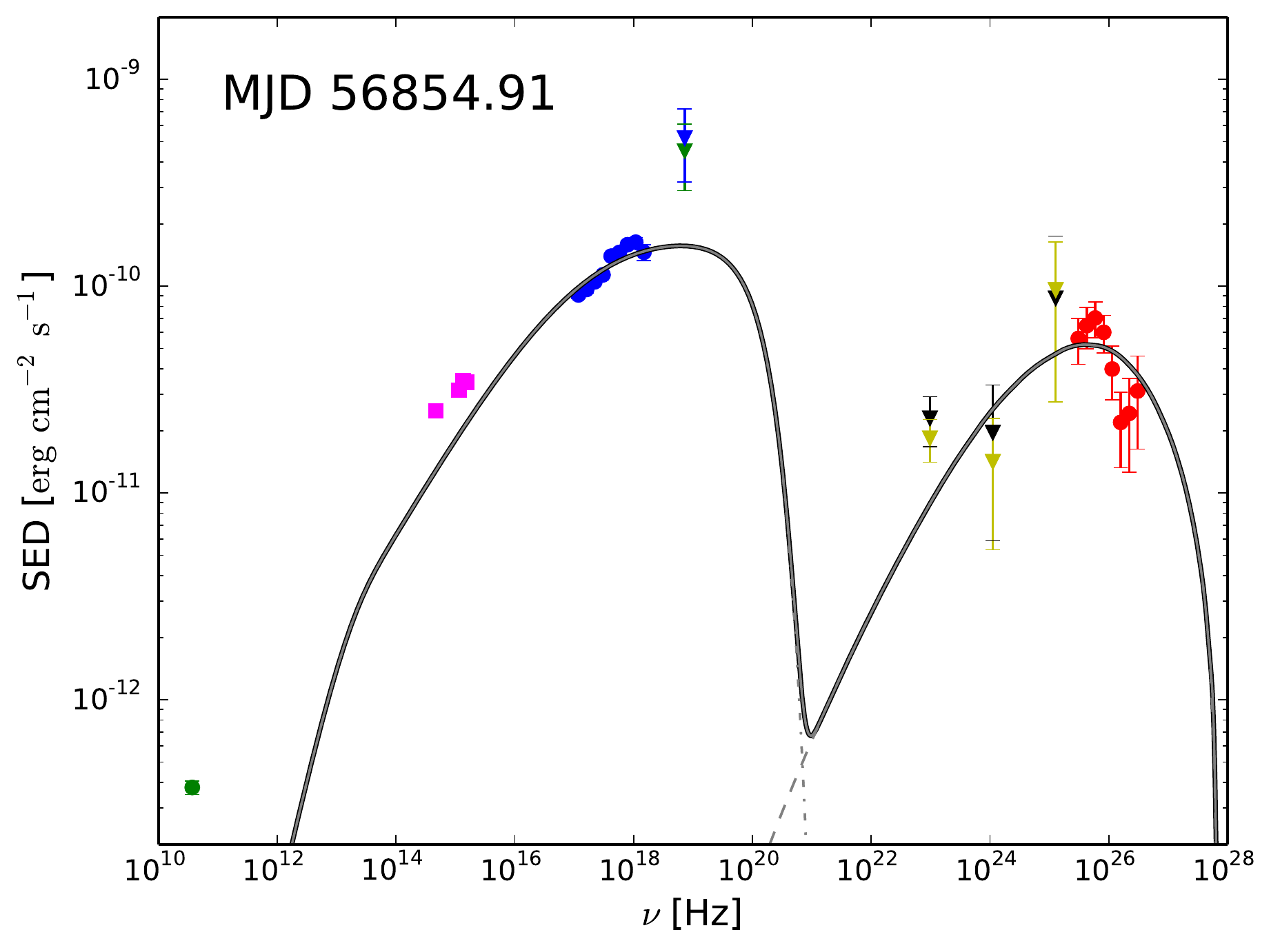}
   \includegraphics[scale=0.4, angle=0]{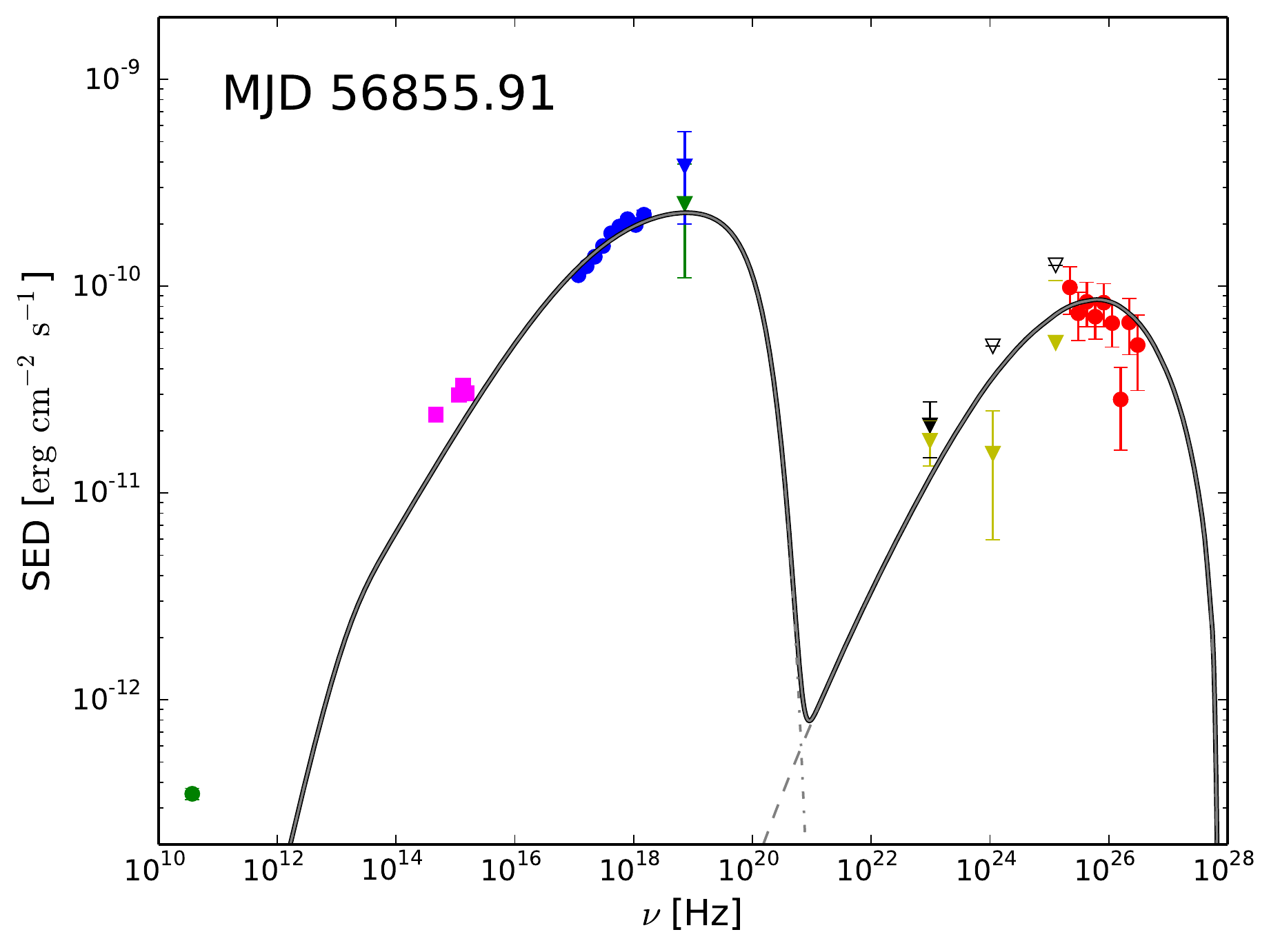}
   \includegraphics[scale=0.4, angle=0]{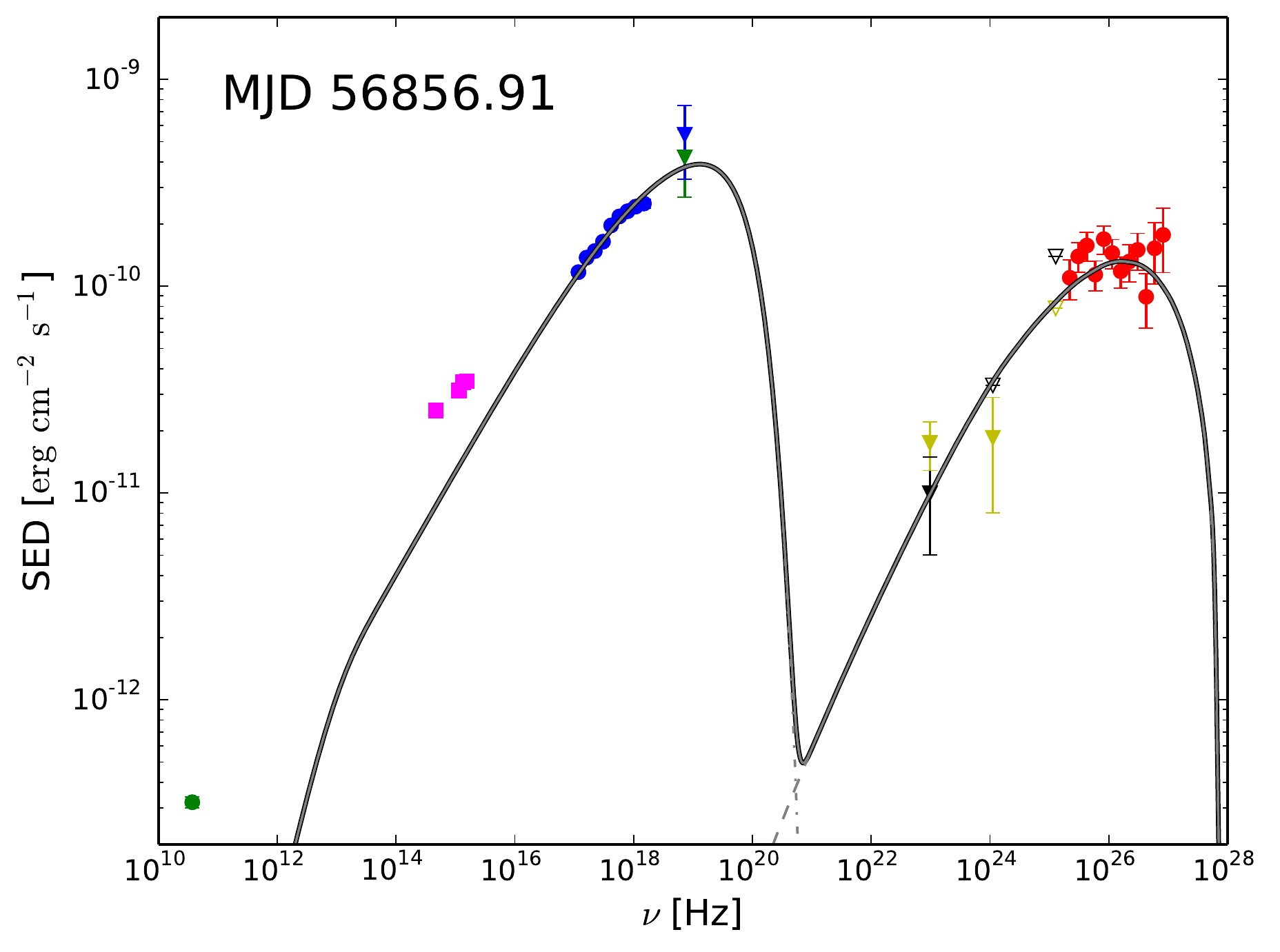}
   \includegraphics[scale=0.4, angle=0]{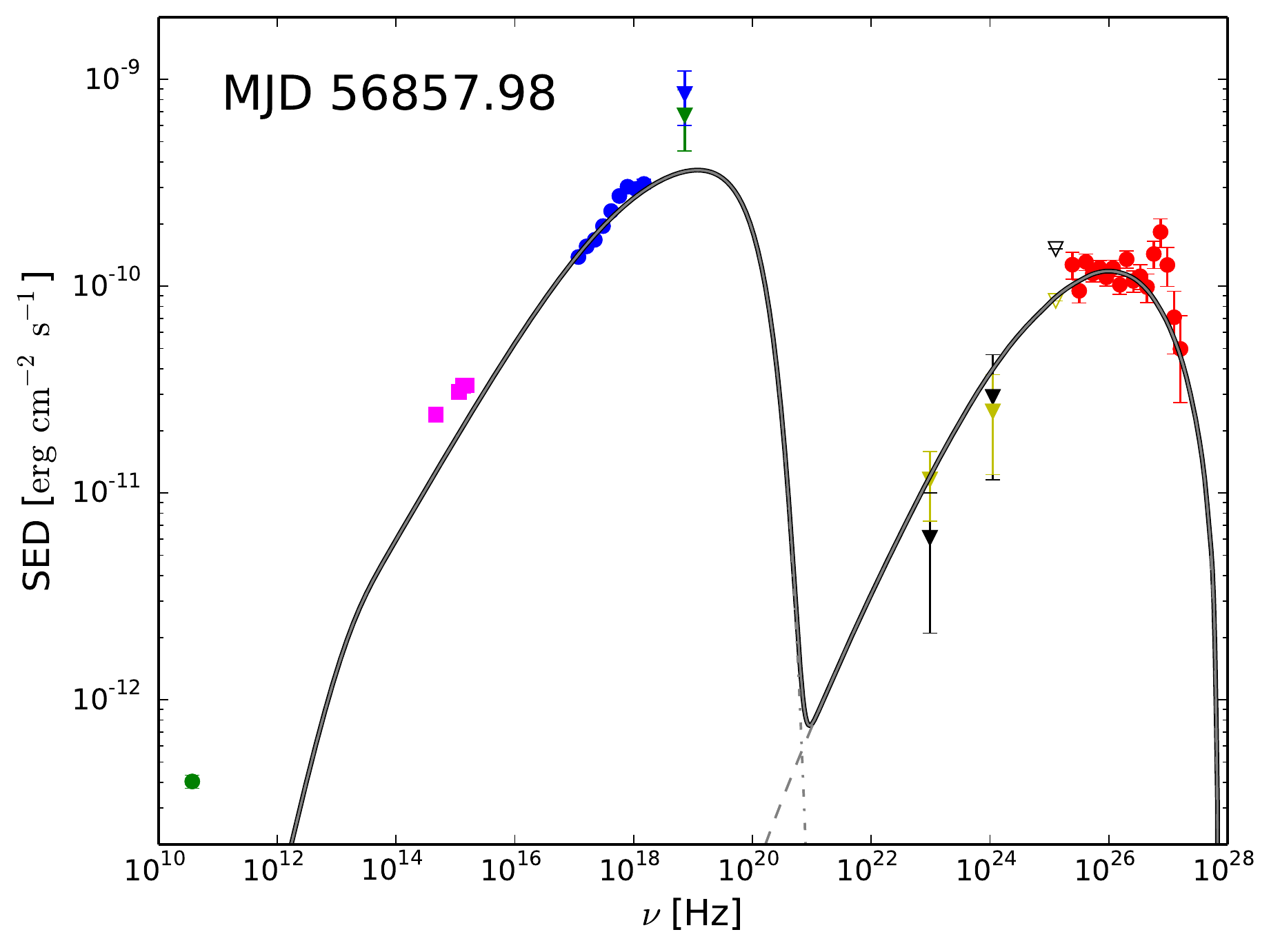}
   \includegraphics[scale=0.4, angle=0]{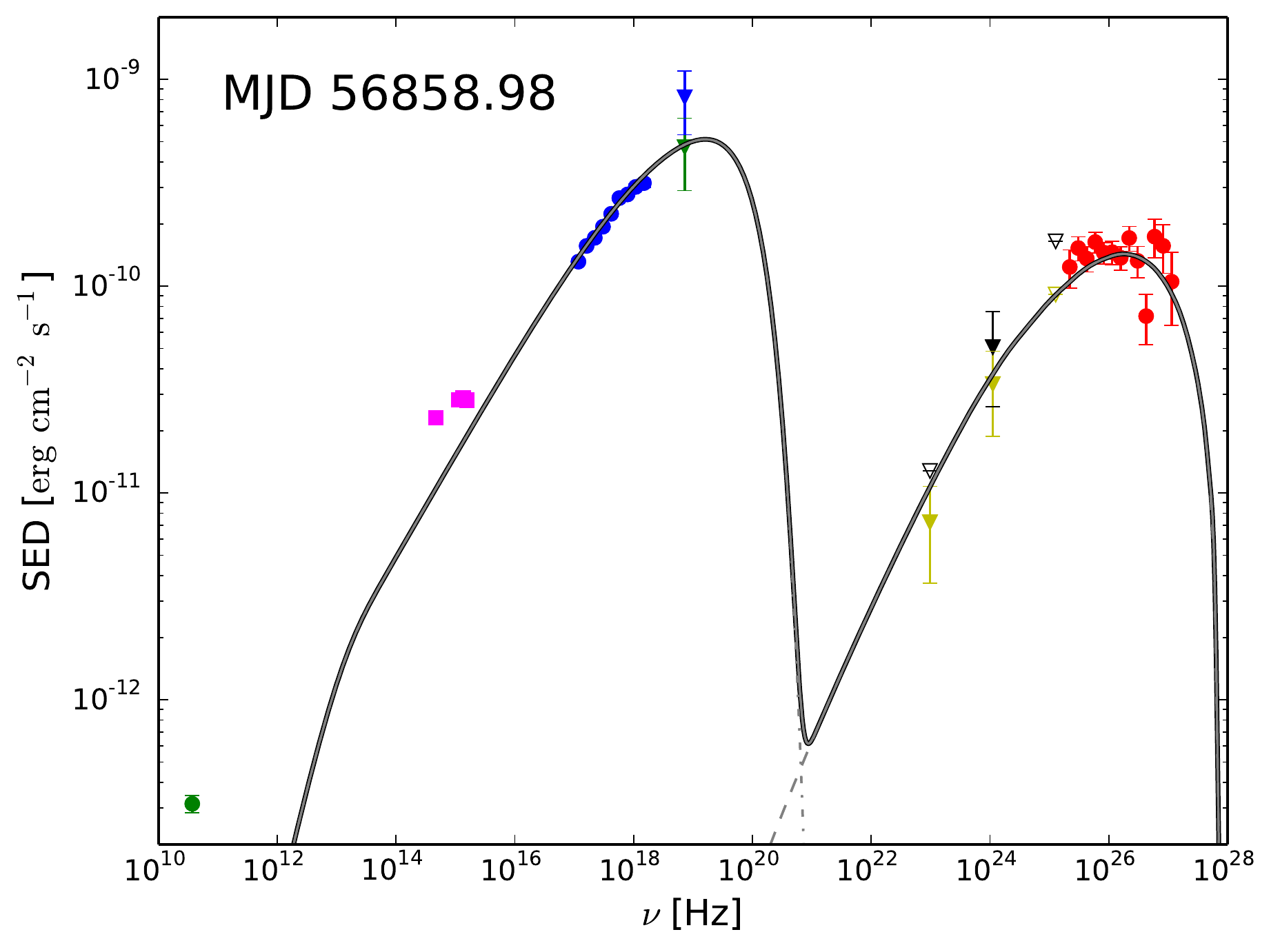}
   \includegraphics[scale=0.4, angle=0]{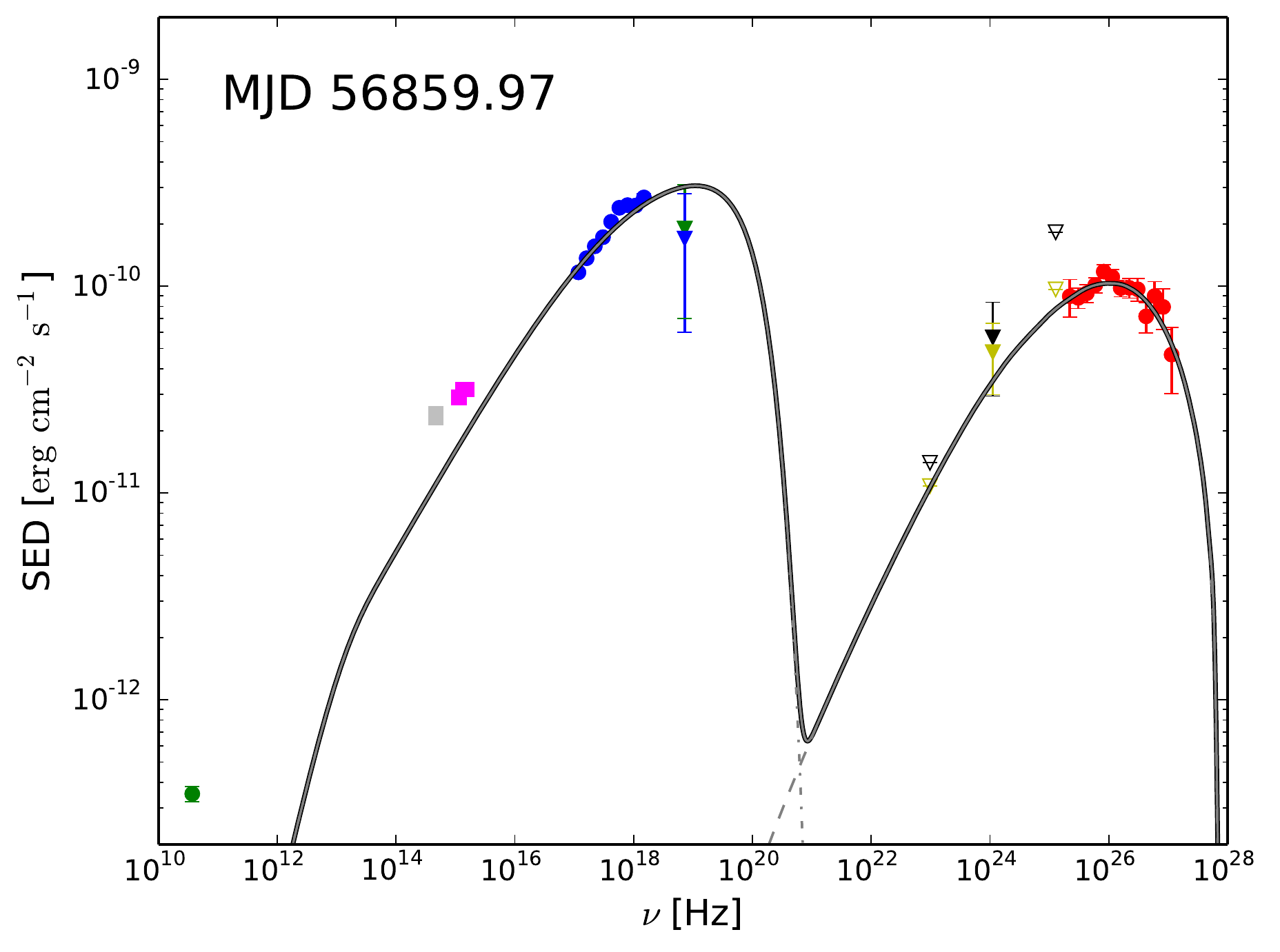}
   \includegraphics[scale=0.4, angle=0]{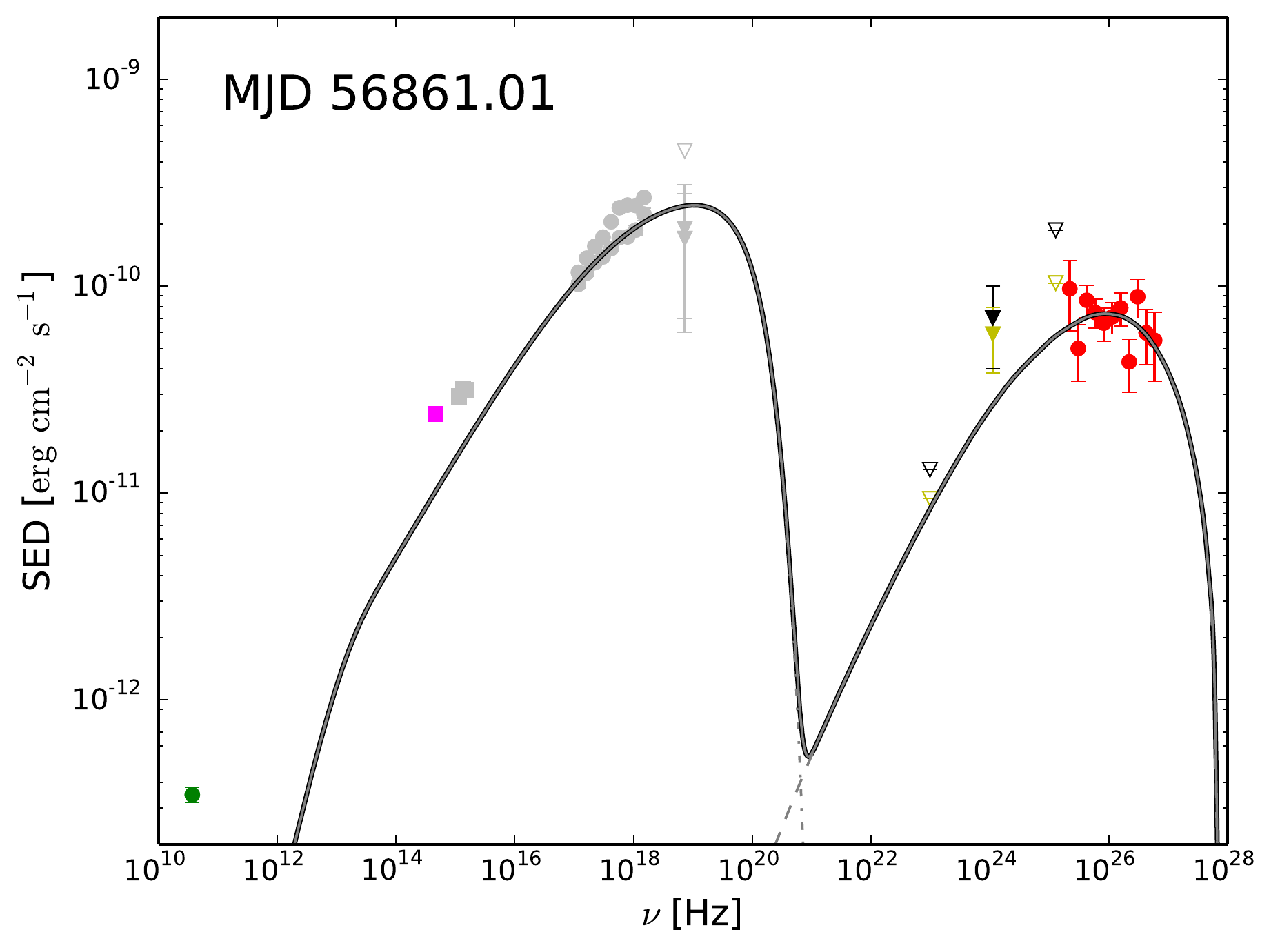}
   \includegraphics[scale=0.4, angle=0]{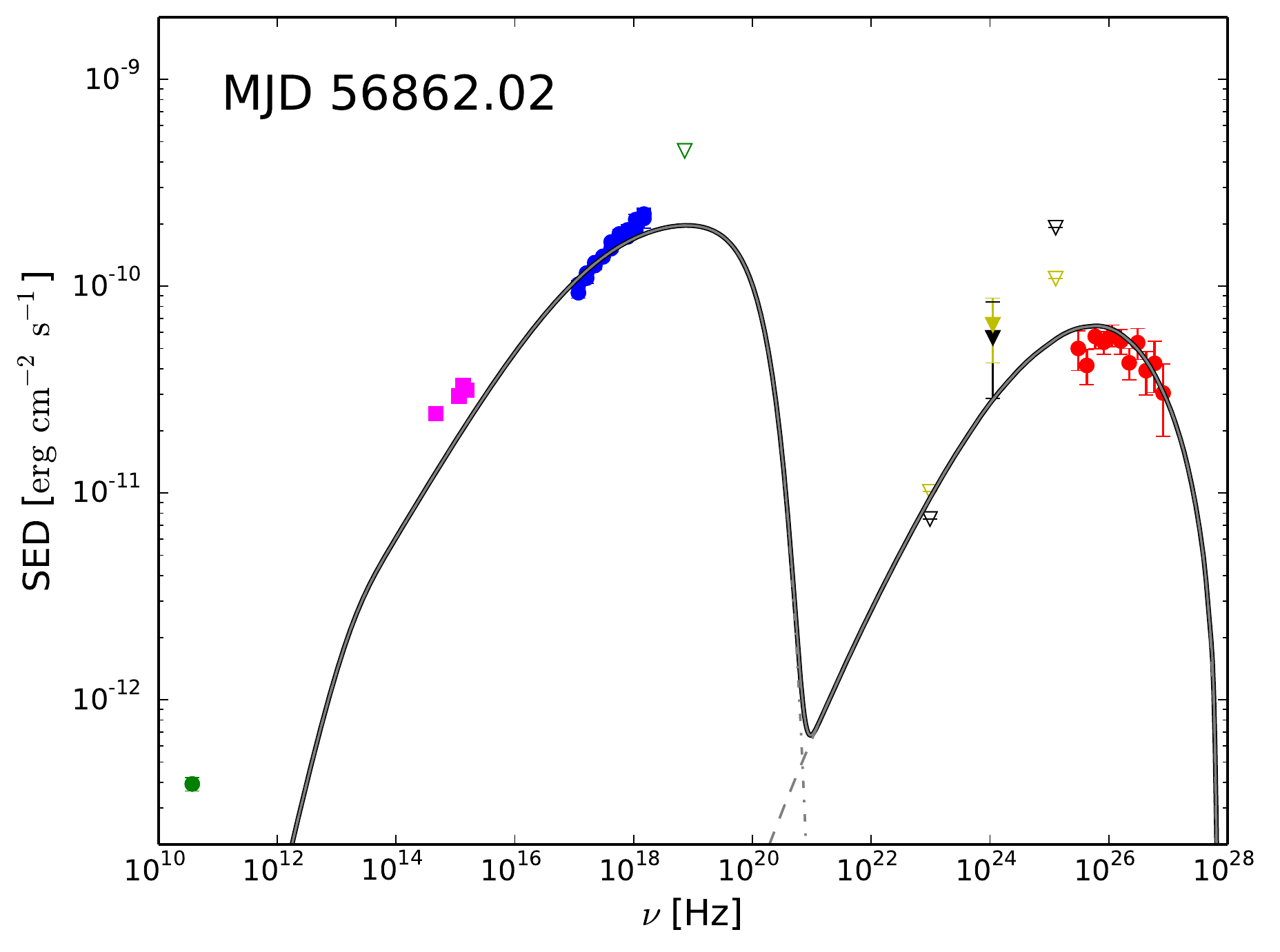}
   \caption{Single-night broadband SEDs described with a one-zone SSC model. The VHE gamma-ray spectra from MAGIC are represented by the red dots,  the {\it Fermi}-LAT spectra by the
black (4 days) and yellow (10 days) triangles, the BAT emission  by the blue triangles (using the spectral shape from XRT) and green triangles (using the spectral shape from the stacked BAT analysis over the time interval MJD~56854.5--MJD ~56872.5),  the binned X-ray spectra from XRT by the blue circles, the optical-UV observations from KVA and UVOT by the pink
squares, and  the radio observations from Mets\"ahovi by the green squares. Most of the data samples were selected from observations taken within 3 h of each other. For MJD~56861 and MJD~56863 there were no {\it Swift} observations taken within the same night of the MAGIC observations, and we depicted the spectra (UVOT, XRT, and BAT) from the night before and the night after with gray symbols. Upper limits are shown as open symbols. See text in Sect.~\ref{sec:Model} for further details.}
   \label{fig:SED_modeling}
\end{figure*}

The broadband SEDs for 15 consecutive days, together with the one-zone SSC models adjusted to describe the data points,  are shown in Fig.~\ref{fig:SED_modeling}. The model parameters are reported in Table~\ref{tab:modeling}.
The agreement between the SED data and the model curves is good, indicating that the adopted strategy to ascribe most of the broadband variations to $\gamma_b$ (with adjustments in the parameters $B$ and  $n1$ and  $n2$), as already done in \cite{historical_flare}, also works well for the extreme X-ray activity observed in July 2014. Within this framework, the variations in the broadband emission of Mrk~501 may be interpreted as being due to changes in the acceleration and cooling of the electrons in the shock in jet model \citep[see, e.g.,][]{1998A&A...333..452K}, which would produce substantial variations in the parameter $\gamma_b$, while many of the other model parameters characterizing the emitting region would remain almost stationary. This would naturally explain the existence of large variations close to the peaks of the two SED bumps (X-ray and VHE), while at lower energies (optical and below), where the emission is dominated by a large number of low-energy electrons, the magnitude of the flux variations would be small. 

\renewcommand{\thefigure}{\arabic{figure} (Cont.)}
\addtocounter{figure}{-1}

\begin{figure*}[h]
   \centering
   \includegraphics[scale=0.4, angle=0]{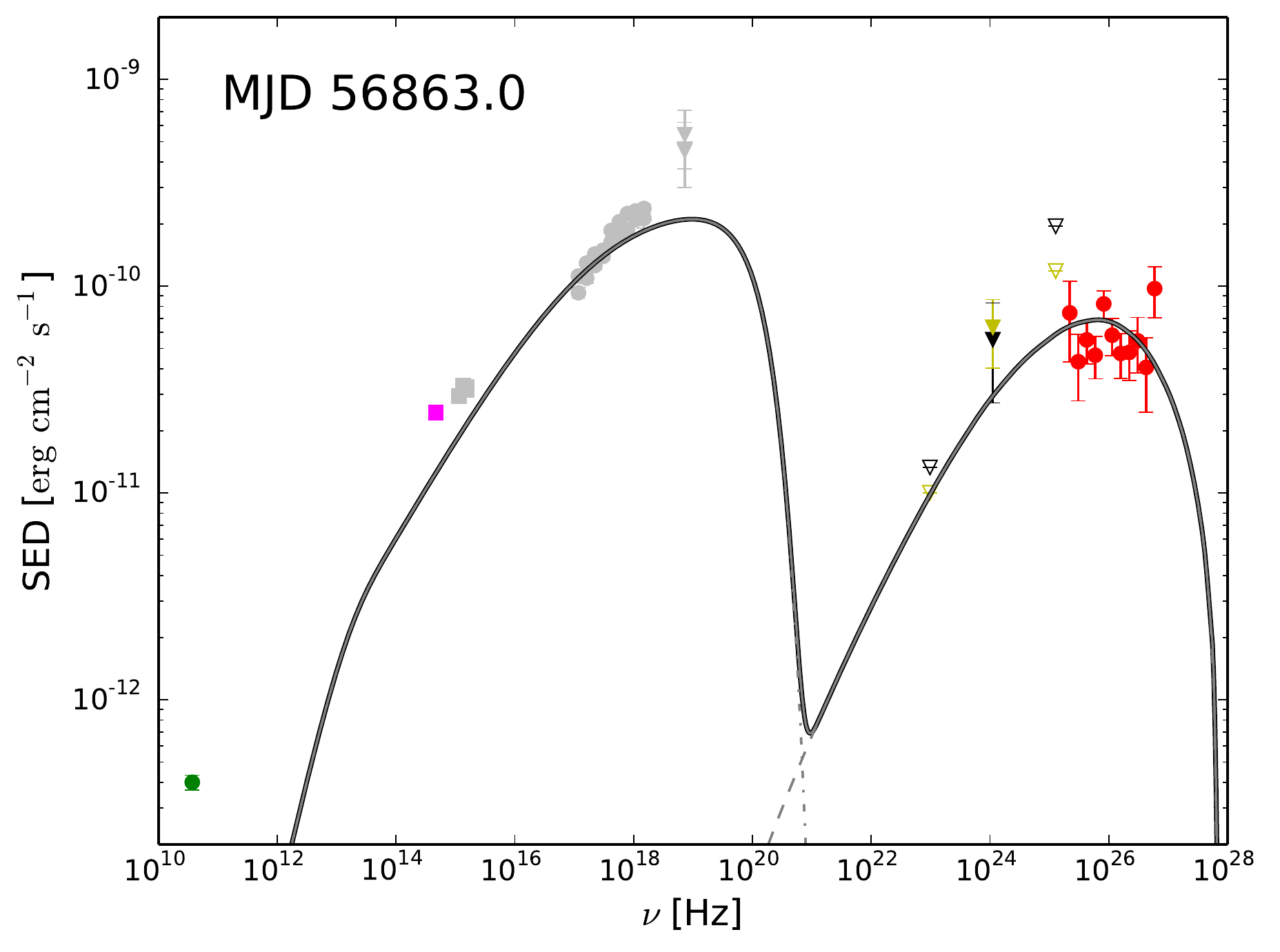}
   \includegraphics[scale=0.4, angle=0]{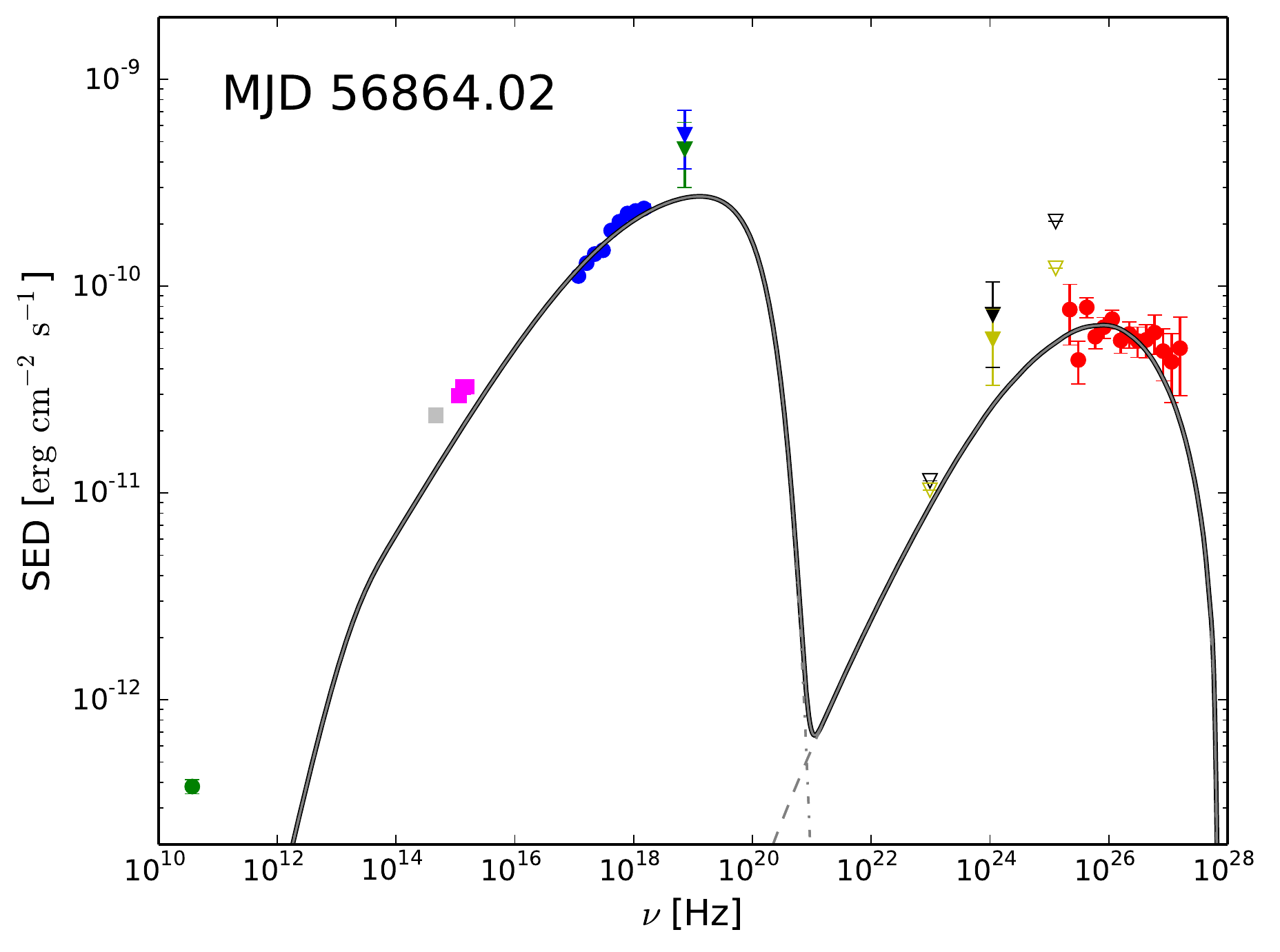}
   \includegraphics[scale=0.4, angle=0]{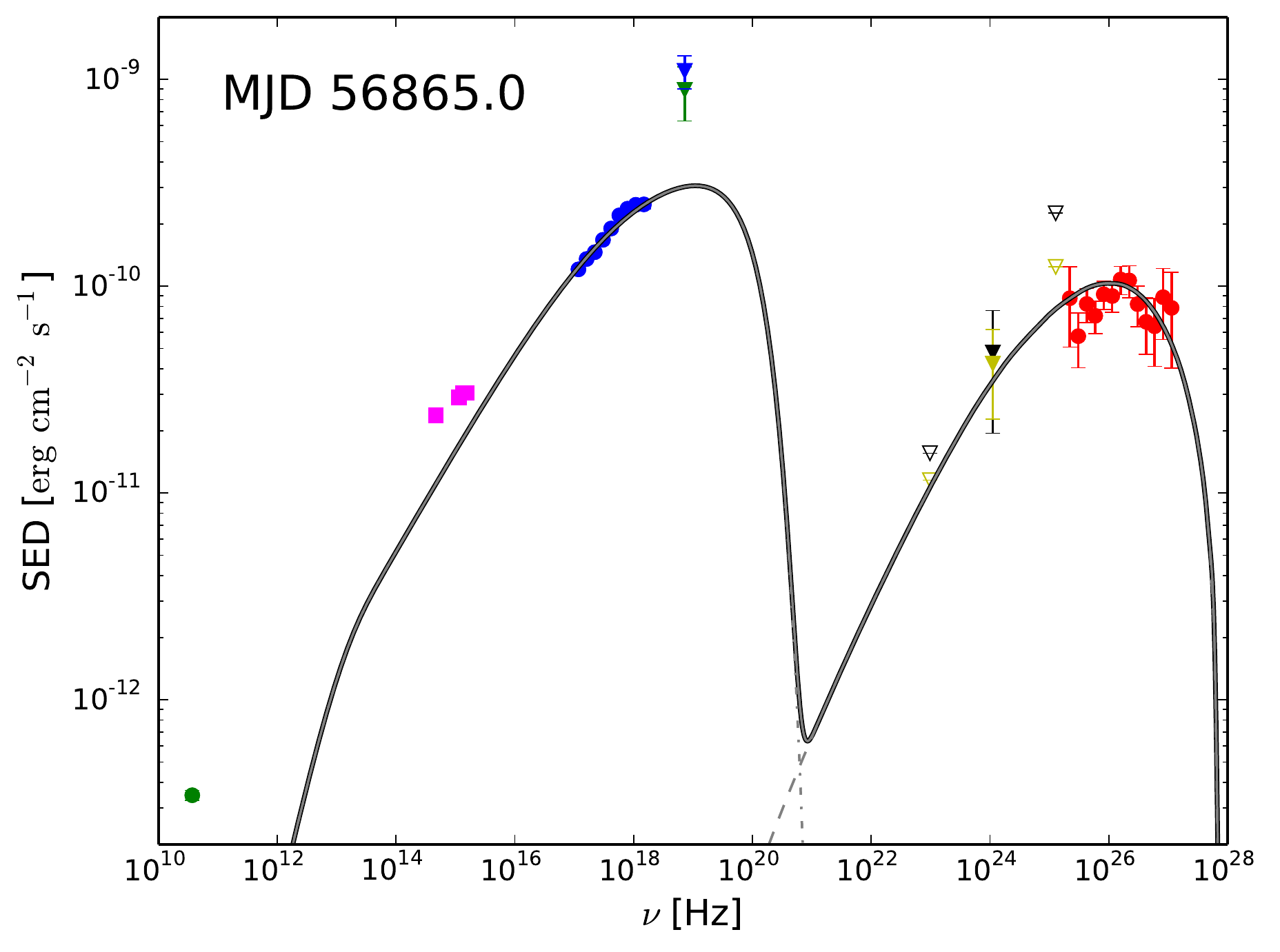}
   \includegraphics[scale=0.4, angle=0]{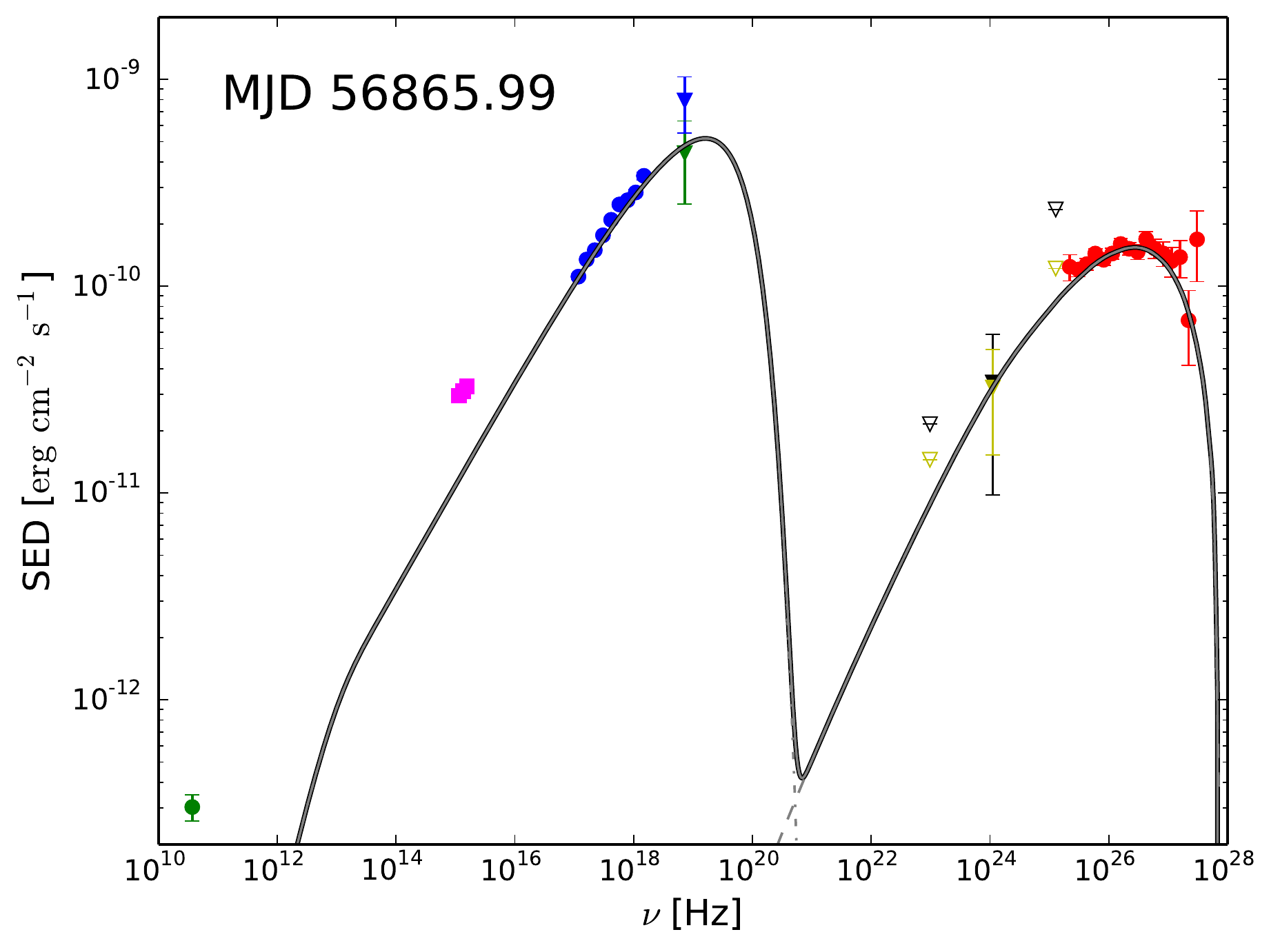}
   \includegraphics[scale=0.4, angle=0]{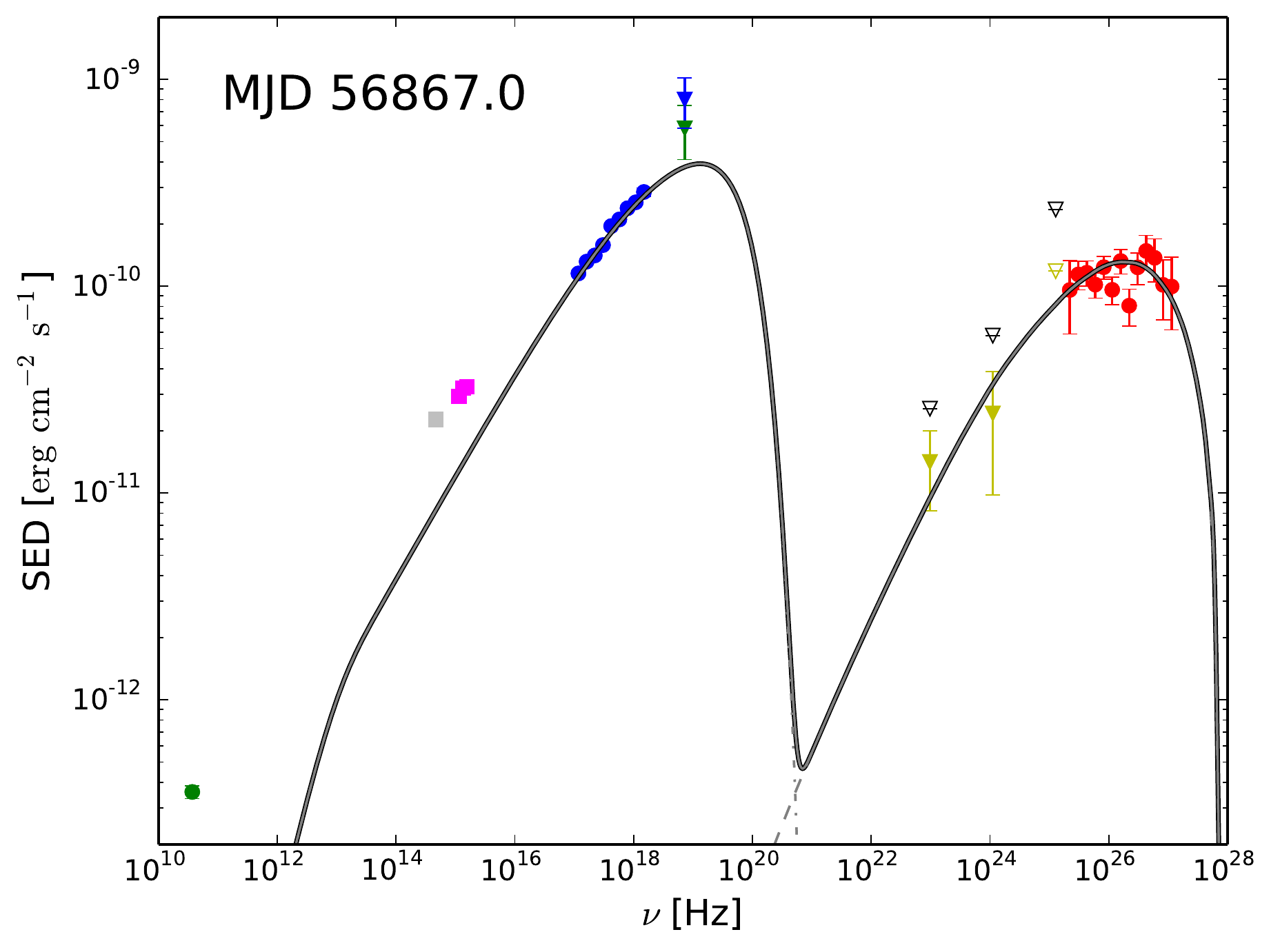}
   \includegraphics[scale=0.4, angle=0]{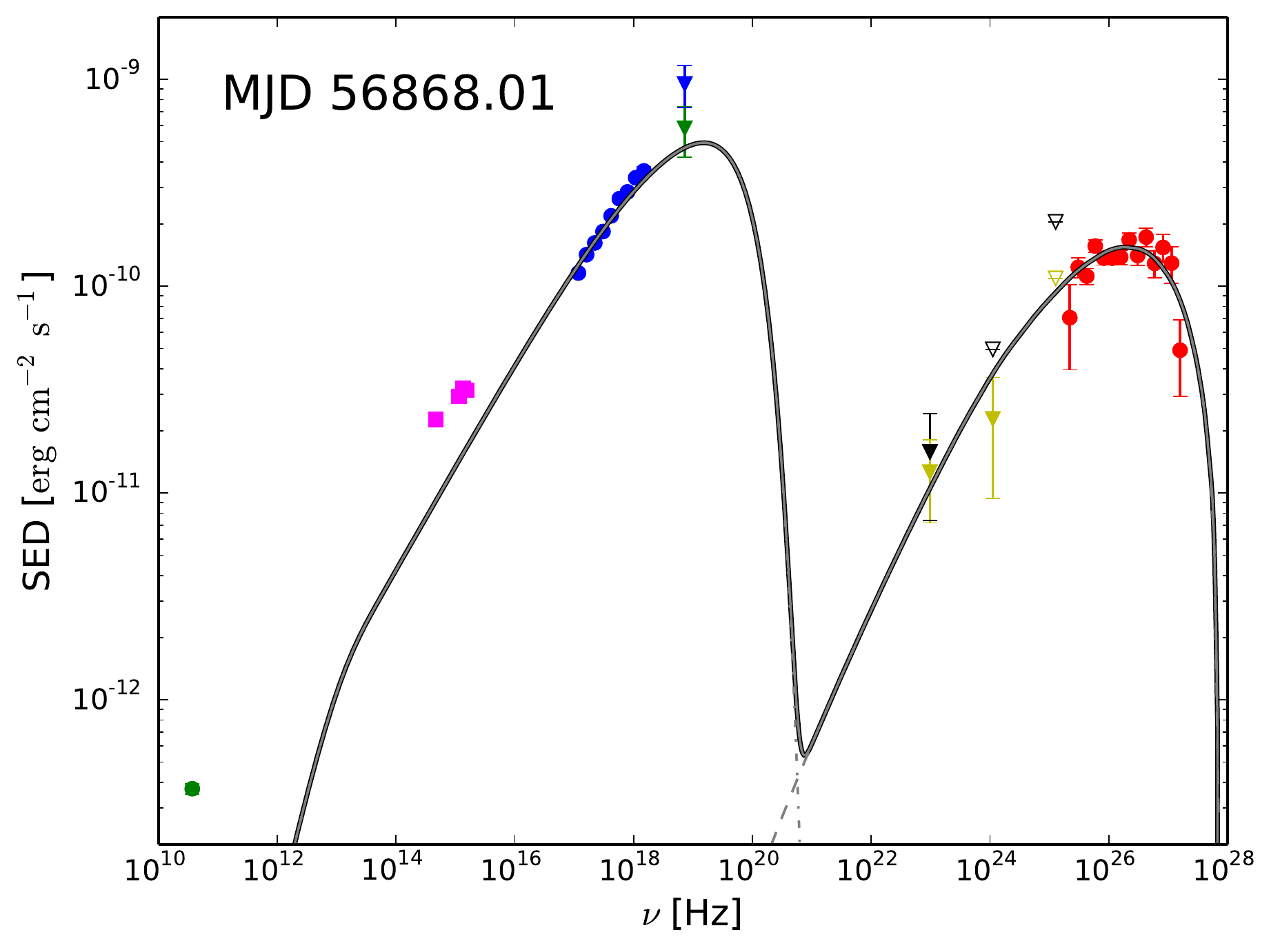}
   \includegraphics[scale=0.4, angle=0]{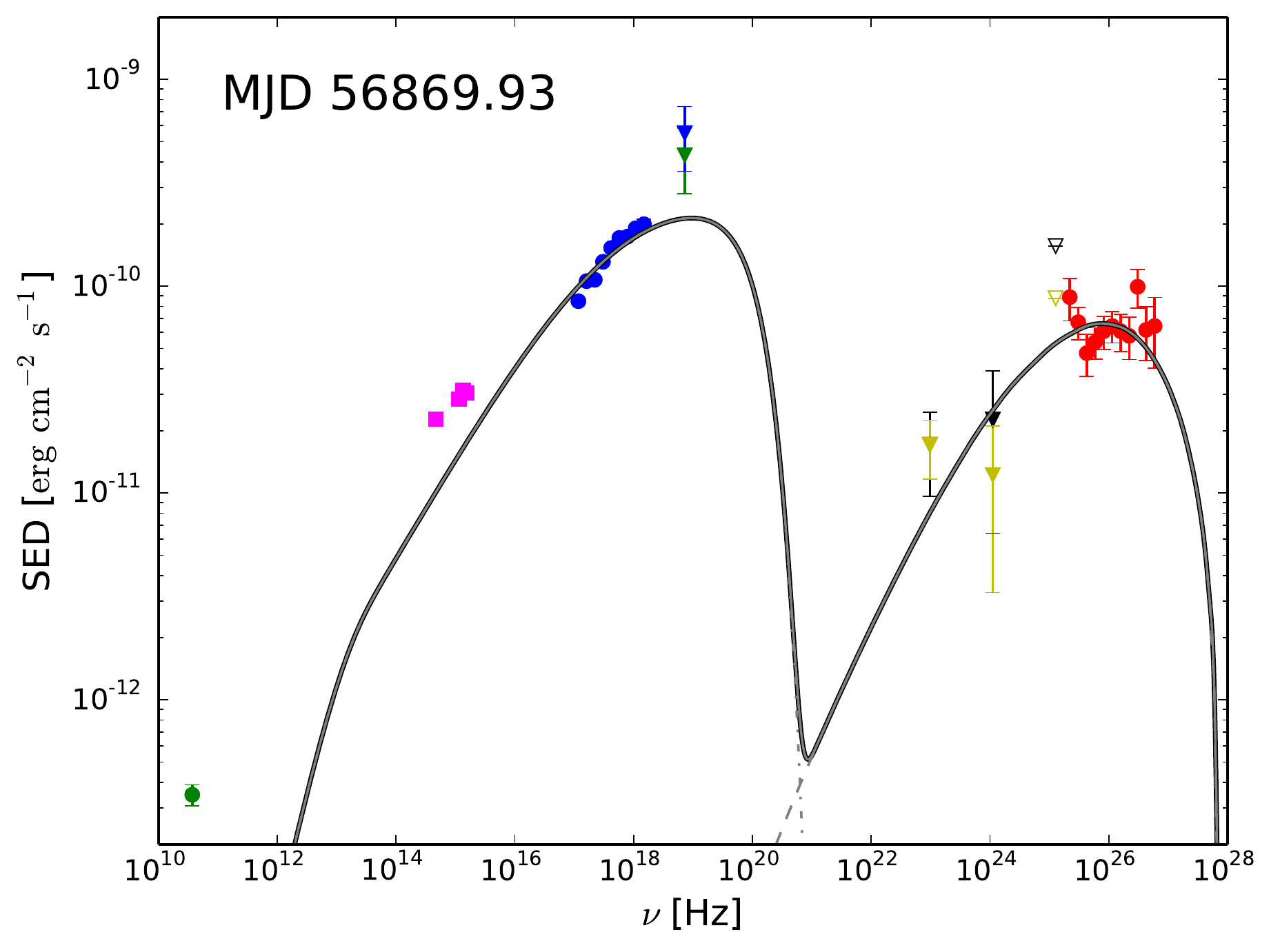}
   \caption{}
\end{figure*}

\renewcommand{\thefigure}{\arabic{figure}}


\section{Characterization of the broadband SED with a narrow TeV component}
\label{sec:ModelNarrow}

As discussed in Sec.~\ref{bump}, an indication of a narrow spectral feature at $\sim$3~TeV was found in the VHE spectrum of Mrk\,501 from 2014 July 19 (observation from MJD~56857.98). This prevents the parameterization of the VHE spectrum with analytic functions typically used to describe the VHE spectra of blazars (e.g., PL, LP). This feature may also be present at some level in the spectra from the day before (MJD~56856.91) and the day after (MJD~56858.98), as shown in Fig.~\ref{fig:VHE_bump}, but these two spectra can be fit well with simple power-law functions. It is during these three days when \textit{Swift}/XRT measured the highest count rates from Mrk\,501, which can be seen as the highest fluxes reported in Table~\ref{tab:xrt_fits} for the energy band \mbox{0.3-2 keV}\footnote{Owing to the rapidly falling flux with the energy, the X-ray count rate measured with \textit{Swift}/XRT is dominated by the emission at the lowest X-ray energies.}, with the highest X-ray flux observed for the night of July 19-20 (\textit{Swift} observation  from MJD~56858.04), when the VHE spectrum shows the indication for a narrow feature at 3~TeV.  However, if we also consider  the energy flux emitted in the X-ray band 2-10 keV,  the X-ray emission from these three days is comparable to that measured nine days later, during the three consecutive days from July 28 to July 30 (MJD~56866.0, MJD~56867.0,  and MJD~56868.0). The VHE gamma-ray activity measured with MAGIC during the three days from July 28 to July 30 is also comparable to that from the three days from July 18 to July 20, with fluxes above 1 C.U. in the energy band 0.15–1~TeV, and fluxes well above 1 C.U. at energies above 1~TeV (see Fig.~\ref{fig:mwl_lc}); however,  the VHE spectra from July 28-30 do not show any indication of narrow features at TeV energies. Neglecting the potential presence of a feature in the VHE spectra from July 18 and July 20 (which is not significant), out of six VHE spectra with extremely high X-ray emission and (relatively) high VHE gamma-ray emission, the narrow feature is observed in only one VHE spectrum, the one from July 19. If we treat these six observations as similar in terms of X-ray and VHE activity, and assume that the probability of finding a narrow feature in all them is the same, the significance of the narrow feature in the measured VHE spectrum from July 19 should be corrected for six trials. In that case, the $\sim$3.6$\sigma$  derived with the Monte Carlo tests reported in  Sec.~\ref{bump} (see Table~\ref{tab:MC_Results_LRT}) would decrease to $\sim 3.1\sigma$. We could adopt a more conservative scenario, and treat all   15 observations performed with MAGIC (during this period of enhanced X-ray activity in July 2014) as 15 independent trial factors, which would further decrease  the significance  to $\sim 2.8\sigma$. In the most conservative approach, we could consider it equally probable for the many hundreds of Mrk501 VHE spectra obtained during the last two decades to contain a narrow TeV feature, implying several hundreds of independent trial factors (neglecting any correlation with X-ray and/or VHE activity). The latter approach would naturally make the narrow feature observed on July 19 totally insignificant.  Owing to the variable nature of blazars, which show a large diversity of X-ray and VHE spectral behavior over time, often showing outstanding and unexpected behaviors on specific days (i.e., super-large flares, particularly soft or hard spectra), we think it is reasonable to consider the uniqueness of the highest X-ray activity during the July 2014 flare, and hence we think it is reasonable to regard the VHE spectrum from July 19 as special. This would imply that, at most, we should consider only a few trials (instead of tens or hundreds of trials), which would lead to a marginally significant ($\sim$3$\sigma$) indication for the presence of a narrow feature at 3~TeV.

There are no reports in the literature about such narrow features in the VHE spectra, but there are  broadband SEDs with narrow high-energy bumps, such as the ones measured for Mrk~421 on MJD~55265 and MJD~55266 on March 2010 \citep[see Fig.~8 from][]{2015A&A...578A..22A} or the one measured for Mrk~501 on MJD~56087 on June 2012 \citep[see Fig.~7 from][]{2018AhnenSubmitted}. In those cases the broadband SED was better explained when adding an extra component with a relatively narrow EED.  In the literature, we can also find different studies that require extra components to explain broadband SEDs and complex variability patterns. In the case of Mrk\,501, with data from 2009, \citet{2017A&A...603A..31A} showed, using a grid-scan over the model parameters, that a two-zone scenario was statistically preferred over a one-zone scenario. Multiple zones have also been used in non-HBL blazars. For instance, in the case of the flat-spectrum radio quasar (FSRQ) PKS\,1222+21 (also known as 4C\,+21.35), variability on the order of $\sim9$\,min was found in the VHE band with the MAGIC telescopes~\citep{pks1222}. Such short variability time, together with the absorption of VHE gamma rays within the BLR, suggest that a small emitting region or blob located outside of the BLR is needed to reconcile the findings with the canonical emission models for FSRQs~\citep{pks1222_model}.  Different variability patterns have also been found at different wavelengths, as in the case of the gravitationally lensed blazar QSO\,B0218+357 \citep{0218_magic}, suggesting that more than one emitting region is responsible for the MWL emission.   

Under the assumption that the narrow feature in the VHE spectrum of Mrk~501 at $\sim$3~TeV is real, it is legitimate to investigate theoretical scenarios that could produce it. In this work, we present three different frameworks that would produce broadband SEDs compatible with the observations. One possible explanation for the TeV feature could be the formation of a  pileup in the EED due to stochastic acceleration, which would  explain the broadband SED using a single region with a multi-component EED. On the other hand, the TeV spectral feature could be produced by the VHE gamma-ray emission from a completely different region. Two scenarios are considered for the latter: SSC emission from a narrow EED in an additional (small) region within the Mrk~501 jet, and emission from electrons accelerated in a magnetospheric vacuum gap close to the supermassive black hole. In the following paragraphs we describe each of the three theoretical approaches.


\subsection{Pileup in the electron energy distribution due to stochastic acceleration}\label{sec:ModelNarrow-PileUp}

Stochastic acceleration has been invoked to explain curved spectra, described by a log-parabolic law, observed in  blazar SED, and the trends between the corresponding peak energy and the spectral curvature   \citep{Tramacere2009,Tramacere2011}. Moreover, stochastic acceleration can also lead to the formation of a pileup in the high-energy range of the relativistic EED \citep{Virtanen2005,Staw2008,Tramacere2011}. Based on this scenario we interpret the sharp and narrow spectral feature observed in the VHE band, together with the high flux level observed by BAT above 10 keV, as the result of a piled-up EED.

As a first approach, we investigate the case of pileup obtained from a continuous mono-energetic injection, escape, and acceleration, under the condition that the particle escape time ($t_{esc}$) is greater than  the dominant acceleration timescale ($t_{acc}$). Under these circumstances, a pileup will emerge around the equilibrium energy ($\gamma_{eq}$), i.e., the Lorentz factor that satisfies the condition $t_{cool}(\gamma)=t_{acc}(\gamma)$, where $t_{cool}$ is the dominant cooling time. The spectral feature shape is described by a relativistic Maxwellian distribution \citep{Staw2008,Sch1985}

\begin{equation}
n(\gamma)\propto \gamma^2 \exp{\Big[\frac{-1}{f(q,\dot\gamma)} \Big(\frac{\gamma}{\gamma_{eq}}\Big)^{f(q,\dot\gamma)}\Big]},
\label{eq:n_equil}
\end{equation}
where $f(q,\dot\gamma)$ is a function depending on the index of the turbulent magnetic field spectrum and on the cooling process. In particular, when the cooling is quadratic in $\gamma$, $f(q,\dot\gamma)=3-q$. Theoretical scenarios based on multiple blobs with relativistic Maxwellian-type EEDs have been used to explain
the very hard gamma-ray spectrum of Mrk\,501, as measured with
{\it Fermi}-LAT \citep{2011ApJ...743L..19L,2016ApJ...832..177S}. In this paper we use a single relativistic Maxwellian
EED to explain the narrow feature at 3 TeV in the VHE spectrum
measured with MAGIC.\\

In the case of ``hard-sphere'' turbulence (q=2.0) the analytical solution for the steady state solution reads \citep{Staw2008}

\begin{eqnarray}
N(\gamma)\propto\Bigg \{
\begin{array}{rll}
&\frac{1}{(2\sigma+1)}
\gamma_{inj}^{\sigma-2}
\gamma^{\sigma+1} &\gamma\leq \gamma_{inj}  \\
&\frac{1}{(2\sigma+1)}
 \gamma_{inj}^{\sigma-1}
\gamma^{-\sigma} &  \gamma_{inj}<\gamma<<\gamma_{eq}\\
&\frac{\Gamma(\sigma-1) }{\Gamma(2\sigma+2)}
 \gamma_{inj}^{\sigma-1}\gamma_{eq}^{-\sigma-2}
 \gamma^2 \exp  {\Big(-\frac{\gamma}{\gamma_{eq}}\Big)} &
 \gamma \gtrsim \gamma_{eq}
 \label{eq:pileup2}
\end{array}
,\end{eqnarray}where $\gamma_{inj}$ is the injection energy, and $\sigma$ determines the spectral slopes above and below $\gamma_{inj}$ as a  function  on the ratio $\epsilon=t_{acc}/t_{esc}$ and according to  $\sigma=(-1/2)+\sqrt{9/4 +\epsilon}$.

In order to understand whether this scenario can reproduce the observed SED, it is useful to evaluate the relative normalization of the pileup branch in eq.~\ref{eq:n_equil} (defined for $\gamma \gtrsim \gamma_{eq}$) to the power-law branch (defined for  $\gamma_{inj}<\gamma<<\gamma_{eq}$),  at $\gamma=\gamma_{eq}$, given by 

\begin{equation}
\frac{\Gamma(\sigma-1)(2\sigma+1)}{\Gamma(2\sigma+2)e}
,\end{equation}
where $\Gamma$ is the gamma function.
According to eq.~\ref{eq:pileup2},  the pileup shape will be significantly dominant over the high-energy power-law branch, only for  $\sigma \lesssim 1.3$, a value that is too hard to reproduce the IC spectrum below the TeV bump, and the X-ray spectrum observed in the XRT window. Hence, we conclude that this scenario is not easily adaptable to our observed data.

A second possible scenario is given by  two injection episodes of mono-energetic particles with $\gamma_{inj}<<\gamma_{eq}$, occurring within the same acceleration region, with a duration of   $T^1_{inj}$ and $T^2_{inj}$, respectively, and delayed by a time interval  $\Delta T_{inj}$.
As long as $\Delta T_{inj}$  is larger than a few $t_{acc}$,  the first population of particles will ``thermalize'' toward a  relativistic Maxwellian around $\gamma_{eq}$ \citep{katar2006,Tramacere2011},  and these particles will be mostly responsible for the emission in the TeV bump, and in the X-rays above 10 keV. 

If the second  injection of particles occurs with a delay  $\Delta T_{inj}$ of a few  $t_{esc}$,  then a lower energy branch will develop cospatially with the initial relativistic Maxwellian population. The distribution  resulting from the second injection, before and close to the equilibrium, can be described by a power law turning into a log-parabola (LPPL), above  a critical energy $\gamma_{0}$ \citep{Tramacere2009,Tramacere2011}.
The phenomenological representation for this scenario can be provided by the following EED:

\begin{eqnarray}
N(\gamma)\propto\Bigg \{
\begin{array}{rll}
&(\gamma/\gamma_{0})^{-s} & \gamma_{inj}\leq \gamma \leq<\gamma_{0} \\
&(\gamma/\gamma_{0})^{-s+r \log(\gamma/\gamma_{0})}  & \gamma_{0}<\gamma<<\gamma_{eq}\\
& K\gamma^2 \exp{-\Big(\frac{\gamma}{\gamma_{eq}}\Big)} &
\gamma \gtrsim \gamma_{eq}
\end{array}
\end{eqnarray}

The first two terms represent the LPPL branch corresponding to the evolution of the second population, and 
the last term corresponds to the  thermalized Maxwellian obtained from the first injection.
The parameter $s$ correspond to the $\sigma$ parameter in eq.~\ref{eq:pileup2}, and the parameter $r$ describes the curvature of the LPPL distribution that evolves under the effect of the diffusive component of the acceleration, and the parameter $K$ takes into account  the ration between the two injections of particles.

We  note  that we are ignoring the region of the EED below $\gamma_{inj}$ because, given the parameter space adopted for the modeling, this part of the EED does not impact significantly on the model above the UV frequencies, except that for a normalization factor.

We build two models, a slower cooling model with a value of the magnetic field $B=0.1$ G, and a faster cooling model with a higher value of $B=0.3$ G, and we refer to them as ``slow'' and ``fast'' cooling respectively.
We assume a beaming factor of 10, and according to the timescale variability of $ t_{var}\lesssim$ one day, we set the constraint on the  source size to be  $R\leq c t_{var} \delta/(1+z) \approx 9\times 10^{15}$ cm.

If we take into account the synchrotron cooling alone, the condition for the formation of the Maxwellian bump in the first injection, $t_{cool}=t_{acc}$, and the value of  the best fit   $\gamma_{eq}\simeq4\times10^5$, require  values of $t_{acc}$ of   $\simeq 2.21$ days and $\simeq0.25$ days, for the slow and fast cooling model, respectively.
These timescales refer to the rest frame of the emitting-acceleration region, hence  in the observer frame will be shortened by a factor of $(1+z)/\delta\simeq 0.1$.  If we combine these requirements on  $t_{acc}$ with the constraint  that  $\Delta T_{inj}$  is  larger than a few $t_{acc}$ (necessary for the thermalization of the first injection), we conclude that  the derived observed timescales  are compatible with the temporal behavior observed in the MAGIC and {\it Swift} energy range.

\begin{figure}
        \centering
        \includegraphics[scale=0.47, angle =0]{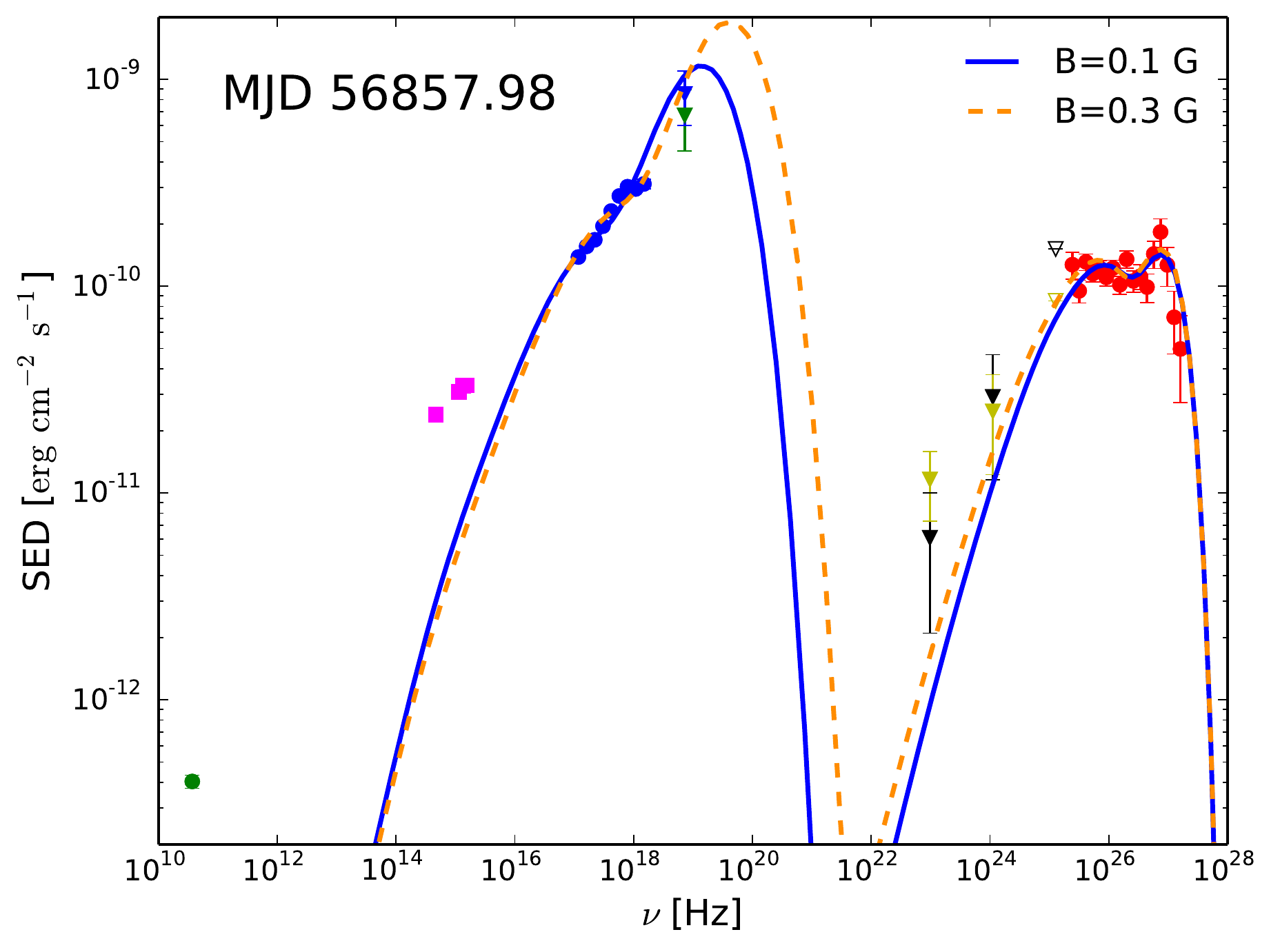}
    \caption{Broadband SED from 2014 July 19 (MJD~56857.98) modeled assuming a pileup in the electron distribution due to stochastic acceleration, and using two different values of the magnetic field: slow cooling with B=0.1 G and fast cooling with B=0.3 G. The color-coding for the data points is the same as in Fig.\ref{fig:SED_modeling}. See text in Sec.~\ref{sec:ModelNarrow-PileUp} for further details.} 
    \label{fig:ModelAndrea}
\end{figure}

The result of our best fit models are shown in Fig.~\ref{fig:ModelAndrea} and the corresponding parameter values are reported in Table~\ref{tab:pileup_model}.
The values of the curvature $r$, for both  models,  is compatible with a distribution that is approaching the equilibrium \citep{Tramacere2011}, hence we might argue that during the second injection episode the acceleration time has decreased compared to the first injection.  
For both scenarios investigated in this section we  used a value of $f(q,\dot{\gamma})=1.0$,  which is compatible with a turbulence index of $q=2$. 
We note that smaller values of $q$  could provide a better description of the narrow bump observed in the TeV spectrum. A  more detailed description of this scenario requires a deeper investigation of the temporal evolution of the emitting plasma under the effect of both acceleration and cooling processes through a numerical solution of the corresponding Fokker-Planck equation, and will be presented in a future publication.
 
\begin{table}
  \centering
  \begin{tabular}{ccc}
         \hline
    Parameter & slow cooling & fast cooling \\
    \hline
    
          $R$ [cm] & $8.8\times 10^{15}$& $3.3\times 10^{15}$    \\
          $B$ [G] &  0.10 & 0.30  \\
          $N$ [cm$^{-3}$] &  0.54 &  2.50  \\
          $\delta$ &  10.00&   10.00 \\
          $\gamma_{max}$ & $1.00\times10^7$&   $1.00\times10^7$   \\
          $\gamma_{inj}$ & $1.00\times10^4$&     $5.00\times10^3$ \\
          $\gamma_{0}$ & $1.50\times10^5$ & $1.30\times10^5$\\
          $s$ & 1.27& 1.28   \\
          $r$ &6.00&   6.10  \\
          $K$  &  $5.30\times10^{-17}$&     $7.00\times10^{-18}$  \\
          $\gamma_{eq}$  &  $4.05\times10^5$ &    $4.0\times10^5$   \\
        \hline
        \hline
        \end{tabular}
    \caption{Parameters used for the stochastic acceleration pileup model applied to the broadband SED from 2014 July 19 (MJD~56857.98), as described in Sec.~\ref{sec:ModelNarrow-PileUp}.}
    \label{tab:pileup_model}
\end{table}

\subsection{Additional SSC model component with a narrow electron energy distribution}
\label{sec:ModelNarrow-2SSC}

\begin{figure*}
   \includegraphics[scale=0.47, angle=0]{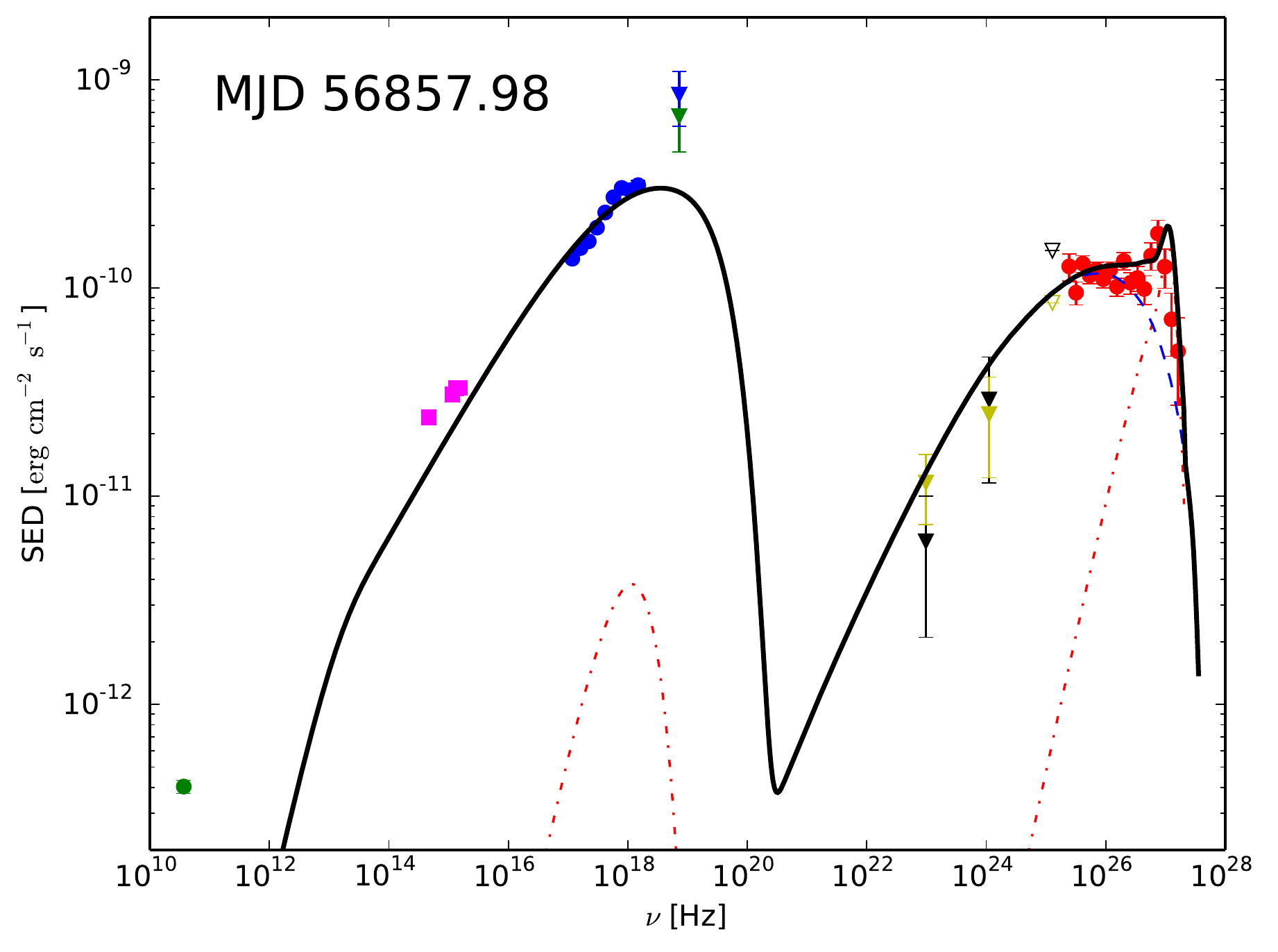}
   \includegraphics[scale=0.47, angle=0]{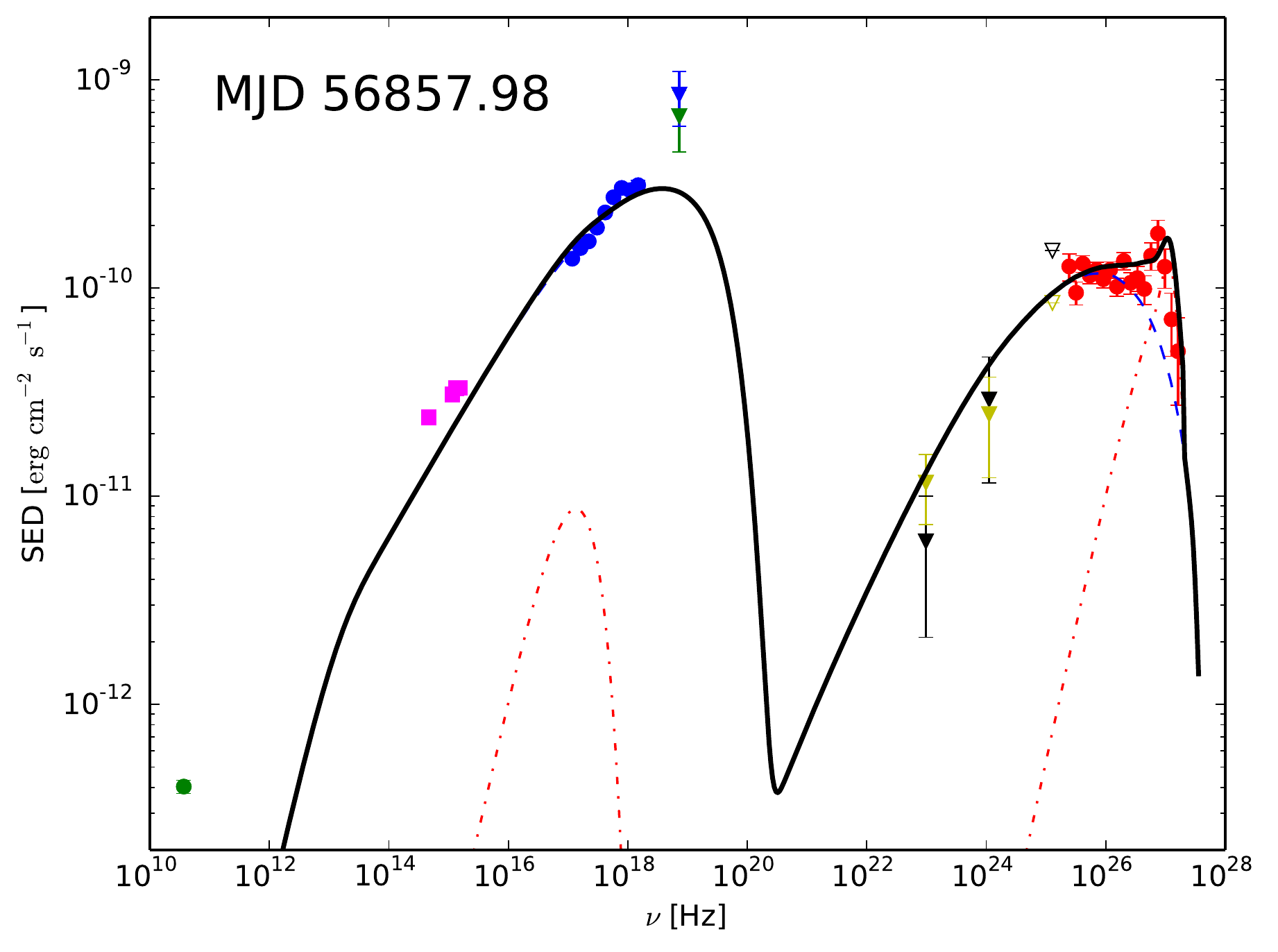}
   \caption{Broadband SED from 2014 July 19 (MJD~56857.98) described with a two-zone SSC model that assumes co-spatial ({\it left panel}) and non co-spatial ({\it right panel}) locations of the emitting regions within the jet. For both panels, the emission from the small region (with narrow EED) is denoted by the red dot-dashed line, while the sum of the emission from the two regions is depicted by the black solid line. The color-coding for the data points is the same as in Fig.\ref{fig:SED_modeling}. See text in Sec.~\ref{sec:ModelNarrow-2SSC} for further details.} 
   \label{fig:SED_modeling_bump}
\end{figure*}

In this theoretical framework, we used a two-zone SSC model to explain the narrow spectral feature at VHE energies. The second (small) emitting region is added to the first (large) one-zone emitting region. Such a scenario can be envisioned as a jet-in-jet model \citep[see, e.g.,][]{giannios2009}, where a small emitting region or blob is embedded within the jet. Two situations are considered: the two emitting regions are co-spatial (i.e., the second blob is embedded within the standard one-zone region) or the   two regions are not co-spatially located. In the case of the co-spatial blob, to avoid a strong interplay between the two regions, the photon density within the small blob needs to be sufficiently high such that the external photon field from the large region is negligible for inverse-Compton scattering and for e$^+$e$^-$ pair creation, otherwise the interaction of the relativistic electrons and the emitted gamma rays from the small blob with the synchrotron emission from the large region would broaden and absorb the spectral TeV feature. For the second scenario, with non co-spatial emitting regions, the conditions can be somewhat relaxed, apart from a very low magnetic field required within the small blob. In the non co-spatial scenario the small emitting region should be located farther  away from the central engine (closer to the observer) than the larger emitting region to prevent the gamma-ray absorption in the low-energy photon field. 
The parameters used to describe the large one-zone emitting region (within the two-zone scenario) were slightly modified to prevent the model from overestimating the measured broadband spectra. The parameters are reported in Table~\ref{tab:modeling_bump} and the models can be found in Fig.~\ref{fig:SED_modeling_bump}. As shown in Table~\ref{tab:modeling_bump}, a large Doppler factor is used, as typically done for jet-in-jet models. Despite the absence of fast (sub-hour) variability, a large Doppler factor is required due to the extremely narrow EED of the small blob. A low (or typical) Doppler factor would require a large (typical) emitting region size, which would imply diffusion, thus making the assumption of a narrow EED unlikely. We note that, due to the large difference in the Doppler factor from the two regions, the co-spatial case would only be possible during a time period on the order of days.

\begin{table}
        \centering
        \begin{tabular}{cccc}
        & Large region &\multicolumn{2}{c}{Small region}\\
                Parameter & & co-spatial & non co-spatial\\
                \hline
        \hline
                        $\gamma_{min}$ [$10^{5}$] & $10^{-2} $&1.2&2.8\\
            $\gamma_{b}$ [$10^{5}$] & $3.7 $&-&-\\
                $\gamma_{max}$ [$10^{5}$] & 17 &1.8&3.0\\
                $n1$& 2 &2.0&2.0\\
            $n2$& 3.3 & - & - \\
                $B$ [G]& 0.125 &0.1&0.005\\
            Density [$\mathrm{cm}^{-3}$] &$2.1\times10^4$ &$1.0\times10^{11}$ & $2.8\times10^8$ \\
                $R$ [cm] & $2.9\times10^{15}$&$1.08\times10^{12}$ & $2.1\times10^{14}$ \\
                $\delta$ & 20 &100 & 60 \\            
                \hline
        \hline          
        \end{tabular}
    \caption{Two-zone SSC model used to describe the broadband SED from 2014 July 19 (MJD~56857.98), as described in Sec.~\ref{sec:ModelNarrow-2SSC}.}
    \label{tab:modeling_bump}
\end{table}

\subsection{IC pair cascade induced by electrons accelerated in a magnetospheric vacuum gap}
\label{sec:ModelNarrow-Gap}

An alternative way to explain the narrow spectral feature at VHE is through the emission resulting from an electromagnetic cascade initiated by electrons accelerated to energies of about 3~TeV in a magnetospheric vacuum gap. In this scenario the electromagnetic cascade, which develops via the interaction of the high-energy electrons with emission line photons from photo-ionized  gas clouds, is responsible for the creation of a narrow component of high-energy photons which, after escaping from the interaction region, get superimposed on the SSC emission from a distinct (large) region. 
 Below we show that such a cascade can develop in the central region of Mrk\,501, and embody the observed broadband SED from 2014 July 19.

Inverse-Compton (IC) pair cascades were first discussed by \citet{Zdz} and were recently treated numerically by \citet{Wendel17}. We adopt a refinement of the scenario in the later work for modeling the emission of an electromagnetic cascade \citep{Wendel19}.
There, the interaction of electrons and positrons (hereafter called electrons) and high-energy photons (HEPs) with a background field of low-energy photons (LEPs) is considered. The LEPs assumed for this scenario are those from the emission of recombination lines from photo-ionized clouds in the inner portion of the host galaxy. We consider only two interaction processes: Breit-Wheeler pair production (PP) and IC-scattering. PP happens solely via collisions of the HEPs with the LEPs, creating electrons that are again available for IC-scattering, and removing the HEPs from their distribution. IC-scattering happens via collisions of relativistic electrons with the LEPs, creating new HEPs that are available for PP, and reducing the energy of the electrons. The interplay of PP and IC-scattering initiates a cascade that evolves the HEP and the electron distributions.

\citet{Zdz} and \citet{Wendel17}  neglected both electron escape and HEP escape from the interaction region (which would imply an infinitely large interaction volume); in contrast we follow \citet{Wendel19} and include these two additional processes into the scenario. Effectively, this means that the IC pair cascade develops only inside a spherical region of radial size $R$. The observer detects the HEPs that escape from this interaction region and arrive to Earth. For the mean escape time, we use $t_{\rm{esc}} = R/c$ with $c$ being the speed of light.

 It has been proposed \citep{LR11,2012A&A...541A..31N,Ptitsyna} that there are charge-depleted regions near the poles of the magnetospheres of spinning black holes. These so-called vacuum gaps exhibit a strong electric field component, which is directed along the magnetic field. Thus, if charged particles enter the gap from the accretion disk, or are created there via PP by photons from an accretion flow, then these charged particles can be accelerated to ultra-relativistic energies and can initiate an IC pair cascade. It is thus justified to approximate the injected relativistic electron distribution ($\dot N_{\rm i}(\gamma)$) per unit space volume by a Gaussian distribution: 

\begin{equation}
\dot N_{\rm i}(\gamma) = \left\{
\begin{array}{ll}
\frac{K_{\rm{G}}}{\sigma \sqrt(2 \pi)} \cdot \exp \left( -\frac{(\gamma-\gamma_{\mathrm{mean}})^2}{2 \, \sigma^2} \right) & \mathrm{if} \; \gamma_{{\mathrm{i}},\,1} \leq \gamma \leq \gamma_{{\mathrm{i}},\,0} \mathrm{,} \\
0 & \mathrm{otherwise}
\end{array}
\right.
\label{EquationDistributionElectrons}
\end{equation}
Here the normalization of the Gaussian ($K_{\rm{G}}$) describes the total number of electrons per unit space volume and per unit time interval that are accelerated in the vacuum gap, propagate away from the vacuum gap along the Mrk\,501 jet axis, and penetrate into the cascade interaction region. The cutoff in $\dot N_{\rm i}$ below $\gamma_{{\mathrm{i}},\,1}$ and above $\gamma_{{\mathrm{i}},\,0}$ was introduced to satisfy the condition $\gamma \cdot x > 1$, where $x$ is the LEP energy divided by the electron rest energy \citep{Zdz, Wendel17}. We chose $\gamma_{{\mathrm{i}},\,1} =\gamma_{\mathrm{mean}} - 3.0 \, \sigma$ and $\gamma_{{\mathrm{i}},\,0} =\gamma_{\mathrm{mean}} + 3.0 \, \sigma$.

Even though Mrk\,501 is classified as a BL Lac-type object and has no pronounced BLR, it is probable that gas clouds from the inner portion of the host galaxy intrude into the AGN. These gas clouds, which stem from the interstellar medium and thus consist mainly of hydrogen and helium \citep{Wilms}, are photo-ionized by the energetic radiation from hot stars and/or the accretion flow. Emission of recombination lines by the photo-ionized gas clouds is thus inevitable. This leads to abundant emission line photon fields in the AGN. Hence, the spectral number density of the LEPs ($n_0(x)$) can be described by a sum of Delta functions

\begin{equation}
n_{0}(x) = K_{\rm{lines}} \cdot \sum _{i=1}^4 \frac{K_{{\rm{line}},i}}{x_{0,i}} \cdot \delta_{\mathrm{Dirac}} \left( x - x_{0,i} \right)
\label{EquationDistributionLEPs}
\end{equation}
The Delta functions are situated at the energy $x_{0,i} = h / (\lambda_{0,i} \, m_{\rm{e}} \, c)$, where $h$, $\lambda_{0,i}$, and $m_{\rm{e}}$ are the Planck constant, the wavelength of line $i$, and the electron rest mass, respectively. The parameter $K_{{\rm{line}},i}$ describes the relative flux of the $i$-th emission line. Dividing $K_{{\rm{line}},i}$ by the energy $x_{0,i}$ of the respective line, gives the relative contribution to the number density. The parameter $K_{\rm{lines}}$ determines the total number density of LEPs. We include these four lines here, which are generally the most prominent ones in broad-line region spectra \citep{Pian} and synthetic photo-ionization spectra \citep{Abolmasov}), and list them in Table \ref{TableLines}. In the following model we do not pay attention to photons from the accretion flow because the demand to synthesize a sharp feature can be met best by usage of a sharp distribution of LEPs. If electrons from the vacuum gap penetrate into the field of emission line photons, they interact with the LEPs and initiate an IC pair cascade, which will create HEPs and secondary electrons.

\begin{table}
\caption{Emission lines used as LEPs in the model producing IC pair cascades reported in Sec.~\ref{sec:ModelNarrow-Gap}.}
\label{TableLines}
\centering 
\begin{tabular}{l l l l} 
$i$ & Line designation                                  & Wavelength & Relative flux  \\
 & & $\lambda_{0,i}$    &  contribution  \\
  & &   [nm]&  $K_{{\rm{line}},i}$ \\
\hline
\hline
1       & Helium II Lyman $\alpha$      & $30.5$                                                                                                                                & 2.00 \\
2       & Hydrogen Lyman series                 & $93.0$                                                                                                                                & 0.17 \\
3       & Hydrogen Lyman $\beta$                & $102.6$                                                                                                                               & 0.57 \\
4       & Hydrogen Lyman $\alpha$               & $121.5$                                                                                                                               & 5.40 \\
\hline
\hline
\end{tabular}
\tablefoot{The fluxes are normalized to the flux of a hypothetical hydrogen Balmer $\beta$ line. The coefficients $K_{\rm{line},3}$ and $K_{\rm{line},4}$ are based on the relative flux ratios given by \citet{Pian}. Plausible values for $K_{\rm{line},1}$ and $K_{\rm{line},2}$ were adopted, cf. \citet{Abolmasov}.}
\end{table}

\begin{table}
\caption{Model parameters used to describe the narrow SED feature from 2014 July 19 (MJD~56857.98) with the emission from an IC pair cascade induced by electrons accelerated in a magnetospheric vacuum gap, as described in Sec.~\ref{sec:ModelNarrow-Gap}.}
\label{TableParametersWendel}
\centering 
\begin{tabular}{l l} 
Parameters & Used value\\                    
\hline
\hline
$\phi$ & $1.8 \, \degr$ \\                                
$R$ [cm]  & $3.0 \cdot 10^{13}$    \\                                              
$K_{\rm{G}}$ [$\rm{s^{-1} cm^{-3}}$]    & $3.3 \cdot 10^{-2}$  \\      
$K_{\rm{lines}}$ [$\rm{cm^{-3}}$] & $9.7 \cdot 10^{6}$   \\
$\gamma_{\mathrm{mean}}$ [${\rm{eV}} / (m_{\rm{e}} c^2)$]  & $3.4 \cdot 10^{12}$  \\
$\sigma$  & $0.23 \, \gamma_{\mathrm{mean}}$ \\
\hline
\end{tabular}
\end{table}

\begin{figure}
 \includegraphics[scale=0.47]{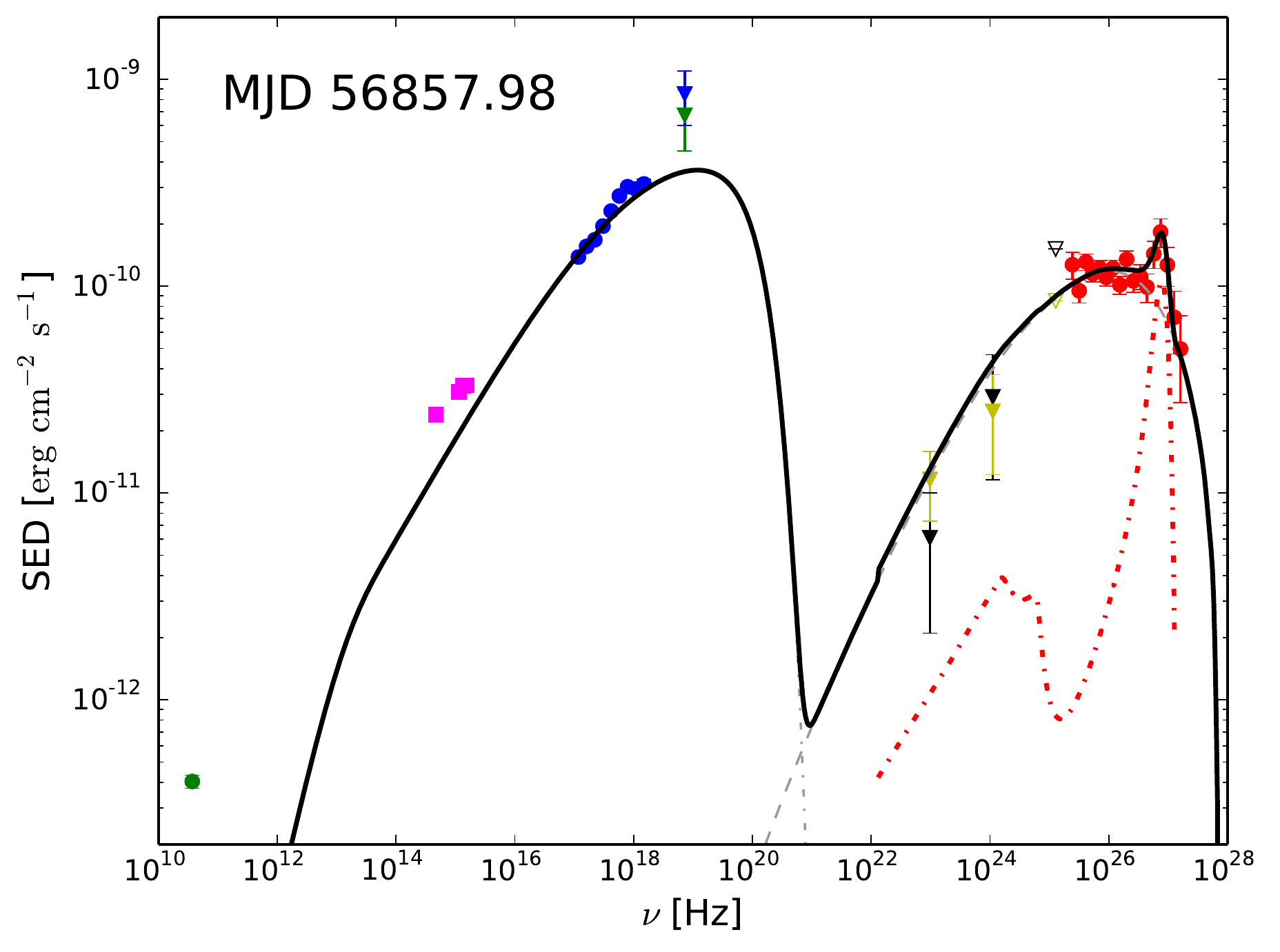} 
 \caption{Broadband SED from 2014 July 19 (MJD~56857.98) modeled with one-zone SSC emission (gray dashed and gray dot-dashed lines) and the emission from an IC pair cascade (red dot-dashed line). The sum of the two components is depicted by the black solid line. The color-coding for the data points is the same as in Fig.~\ref{fig:SED_modeling}. See text in Sec.~\ref{sec:ModelNarrow-Gap} for further details.}
 \label{FigureSSCPlusCascadedEmission}
\end{figure}

The kinetic equation that describes this type of cascade, and the numerical scheme to solve it iteratively to obtain electron and HEP spectral number densities $N(\gamma)$ and  $n_{\gamma}(x_{\gamma})$, with $x_{\gamma}$ being the HEP energy divided by the electron rest energy, is described in \citet{Wendel19}. The HEP spectral number density is determined as the ratio of the IC production rate of HEPs to the loss rate of HEPs, which is the sum of the escape rate and the attenuation rate due to pair absorption. Because of the scattering kinematics, the HEPs leave the volume within a beam of opening angle $\phi$ in the direction the electrons entered the interaction region. The spectral number of photons that stream through a unit area per unit time interval and can be detected at Earth is
\begin{equation}
F(x_\gamma) = n_{\gamma}(x_\gamma) \cdot \frac{4 \pi \, R^2}{\Omega(\phi) \, D^2 \ m_{\rm{e}} c}
\label{EquationCascadedFlux}
,\end{equation}
where $n_{\gamma}$ is measured in units of $m_{\rm{e}} c^2$. The luminosity distance $D$ = 149.4\,Mpc, and $\Omega(\phi)$ is the solid angle of the HEP beam with conical shape and opening angle $\phi$.

The resulting spectrum is, after adding the one-zone SSC model shown in Fig.~\ref{fig:SED_modeling}, adjusted to the narrow SED peak from 2014 July 19. With the parameters listed in Table~\ref{TableParametersWendel}, the narrow feature can be theoretically explained, as is shown in Fig.~\ref{FigureSSCPlusCascadedEmission}, by emission line photons that have been IC upscattered by the electrons from the gap. 

The large dip in the cascaded spectrum above $10^{25} \rm{Hz}$, and the small dip above $10^{24} \rm{Hz}$, are due to the absorption of HEPs due to PP with LEPs from the hydrogen Lyman $\alpha$ line and helium II Lyman $\alpha$ line, respectively\footnote{Absorption troughs due to the hydrogen Lyman $\beta$ line and series are hardly discernible because of the small $K_{{\rm{line}},2}$ and $K_{{\rm{line}},3}$.}. Consequently, the bump at $10^{24} \rm{Hz}$ is due to cascaded emission that is just below the PP threshold, and thus  not pair absorbed.
The cascaded HE radiation   stems from a region that is located within the typical extent of a BLR. Mrk\,501 has no detectable BLR, and hence the density of LEPs must be low, and probably dominated by emission lines from  photo-ionized hydrogen and helium gas clouds that stem from the interstellar medium. The density of LEPs is such that the cascade is well sustained, but the HE photons are not entirely pair absorbed, and can escape and be detected by the MAGIC telescopes. This is in contrast to the case of LBLs and \mbox{FSRQs}, where it is usually considered that the density of LEPs is large, and the HE radiation originating from inside the BLR is completely pair absorbed, implying that the HE radiation sometimes (e.g., during large flares) detected for some of these objects has to originate from outside the BLR~\citep[see, e.g.,][]{pks1222,0218_magic}. 
A discussion on the implications of the used parameters on the physical state of Mrk~501 can be found in \citet{Wendel19}.

Within the theoretical framework presented here, the narrow feature detected with MAGIC in the VHE spectrum of Mrk\,501 is interpreted as a signature of electron acceleration in a magnetospheric vacuum gap, close to the supermassive black hole. Similar theoretical scenarios were also used to explain the fast variability in radio galaxies \citep[e.g.,][]{2014Sci...346.1080A,2018arXiv180601559M}, and to test the stability of a gap and the resulting radiation on a theoretical basis \citep{Ptitsyna, 2016ApJ...818...50H}. The main difference with respect to those scenarios is that, in the study presented here, the inverse-Compton scattering occurs on emission line photons from BLR-like clouds, 
and dominate the broadband gamma-ray emission only in a narrow range of energies. In  other published works, the inverse-Compton scattering occurs on  seed photons emitted by the accretion disk, and   describes a large fraction of the entire gamma-ray emission. From the technical perspective, another difference is that in this work we neglect curvature radiation due to its minor importance for the electron energy loss rate \citep{LR11}, and that we use the electron energy distribution $N(\gamma)$ as a fitting function, whereas in the model by \citet{2016ApJ...818...50H} the electron energy distribution is an inherent feature of the existence and stability of the gap.


\section{Summary and concluding remarks}
\label{sec:Conclusions}

We  presented observational and theoretical results derived with multi-instrument data from Mrk~501 collected during a \mbox{$\sim$two-week} period in July 2014, when the X-ray activity was at its highest among the $\sim$14 years of operation of the \textit{Neil Gehrels Swift Gamma-ray Burst Observatory}. During this outburst, the X-ray spectra measured with XRT (and with BAT) were very hard, and somewhat similar to the large historical flare from 1997 that was measured with BeppoSAX beyond 100~keV \citep{1998ApJ...492L..17P}. 

During this short time interval, the flux variations in the radio, optical, and GeV bands were rather mild ($F_{var}\sim0.05$), but quite substantial in the X-ray bands ($F_{var}>0.15$) and especially substantial in the VHE bands ($F_{var}>0.3$). No intra-night variability was observed on any of the nights. There is a general increase in the fractional variability with energy, with the highest variability occurring at VHE. This variability pattern is similar to that from other multi-wavelength campaigns targeting Mrk~501 \citep[][]{2015A&A...573A..50A,2016A&A...594A..76A,2017A&A...603A..31A,2018AhnenSubmitted}, but very different from the behavior observed in Mrk~421, which shows a clear double-bump structure with the highest variability often observed at X-ray energies \citep{2015A&A...576A.126A,2015A&A...578A..22A,2016ApJ...819..156B}. The correlation between the X-ray and VHE bands (the most variable segments of the electromagnetic spectrum), was investigated using two energy ranges for each, namely 0.3—2~keV and 2—10~keV for X-rays, and 0.15—1~TeV and $>$1~TeV for VHE. This study shows, for the first time for Mrk~501, a significant correlation ($>3 \sigma$)  between these two bands during a relatively short time interval ($\sim$2 weeks)  with a persistent elevated activity. Moreover, we observed that the strength and the significance of this correlation increases with   increasing energy in X-rays, similarly to what was reported by \citet{2018AhnenSubmitted} using the dataset of a  few months in length from 2012. 

During the X-ray flux peak, we observed a narrow feature at about 3 TeV in the VHE gamma-ray spectrum measured with the MAGIC telescopes on 2014 July 19 (observation performed at MJD~56857.98). This TeV~feature cannot be described with the analytic functions typically used for the VHE spectra of blazars, such as power laws, log-parabolas, and log-parabolas with exponential cutoffs: the inconsistencies are larger than $\sim$3~$\sigma$. A fit with a log-parabola below 1.5~TeV and
extrapolated to higher energies shows deviations with the data of 4--5$\sigma$. A likelihood ratio test shows that a log-parabola with an additional narrow component (modeled with another log-parabola with a strong curvature) is preferred with respect to the single log-parabola at more than 4$\sigma$. In addition,  a dedicated Monte Carlo simulation  indicates the presence of the narrow component at a significance larger than 3~$\sigma$. This narrow TeV feature may also be present at some level in the spectra from the earlier (July~18) and later (July~20) nights, but at much lower significance. While the VHE spectra of Mrk~501 have previously shown a prominent peaky structure \citep[see, e.g.,  Fig.~7 in][]{2018AhnenSubmitted}, this is the first time that such a narrow feature has been observed, even if only at a marginally significant level of $\sim$3--4$\sigma$.

A detailed study on the temporal evolution of the broadband SEDs from 2014 July 16 (MJD~56854.9) to 2014 July 31 (MJD~56869.9), resolved on a day-by-day basis, was performed. The time difference between the X-ray and VHE data is mostly below 3 hours which, given the lack of variability on hour timescales, can be considered as simultaneous observations. The daily evolution of the most variable segments of the SED, namely the X-ray and the gamma-ray bands, which is where the most energy is emitted, could be successfully parameterized with a one-zone SSC model, where the main variations are produced by changes in the break energy $\gamma_b$, with some adjustments in the parameters $B$, $n1$, and $n2$. Within this theoretical framework, these results suggest that the flux variations on timescales of days are produced by the acceleration and the cooling of the high-energy electrons.

The SED from 2014 July 19 shows the largest disagreement with the one-zone SSC, which is due to the narrow feature at $\sim$3~TeV observed in the MAGIC spectrum. Under the assumption that this spectral feature is real, we investigated three theoretical scenarios that could reproduce it: a) pileup in the electron energy distribution due to stochastic acceleration; b) a structured jet with two-SSC emitting regions (related or not related), with one region dominated by an extremely narrow electron energy distribution; and c) an emission produced via an IC pair cascade induced by electrons accelerated in a magnetospheric vacuum gap, in addition to the SSC emission from a more conventional region along the jet of Mrk~501.  The three frameworks could reproduce the narrow spectral component reasonably well, given its relatively large uncertainties. Future observations of the gamma-ray emission of Mrk~501 and other bright VHE blazars will help investigate the reliability and potential recurrence of narrow spectral components. 
 
Moreover, these spectral features may also occur at hard X-rays, as predicted by the theoretical scenario from Sect.~\ref{sec:ModelNarrow-PileUp}. Therefore, observations with high-resolution hard X-ray instruments like NuSTAR, together with current and future Cherenkov telescopes such as CTA, would allow for a better characterization of narrow spectral features in both the low- and high-energy bumps, which could have important implications for the understanding of particle acceleration and radiation in Mrk\,501, and in blazars in general.

\begin{acknowledgements}

%
%
The authors thank the anonymous referee for providing a constructive list of remarks that allowed us to clarify and improve some of the results reported in the manuscript.

We would like to thank the Instituto de Astrof\'{\i}sica de Canarias for the excellent working conditions at the Observatorio del Roque de los Muchachos in La Palma. The financial support of the German BMBF and MPG, the Italian INFN and INAF, the Swiss National Fund SNF, the ERDF under the Spanish MINECO (FPA2015-69818-P, FPA2012-36668, FPA2015-68378-P, FPA2015-69210-C6-2-R, FPA2015-69210-C6-4-R, FPA2015-69210-C6-6-R, AYA2015-71042-P, AYA2016-76012-C3-1-P, ESP2015-71662-C2-2-P, FPA2017‐90566‐REDC), and the Japanese JSPS and MEXT is gratefully acknowledged. This work was also supported by the Spanish Centro de Excelencia ``Severo Ochoa'' SEV-2016-0588 and SEV-2015-0548, and Unidad de Excelencia ``Mar\'{\i}a de Maeztu'' MDM-2014-0369, by the Croatian Science Foundation (HrZZ) Project IP-2016-06-9782 and the University of Rijeka Project 13.12.1.3.02, by the DFG Collaborative Research Centers SFB823/C4 and SFB876/C3, the Polish National Research Centre grant UMO-2016/22/M/ST9/00382 and by the Brazilian MCTIC, CNPq and FAPERJ.\\

The \textit{Fermi} LAT Collaboration acknowledges generous ongoing support
from a number of agencies and institutes that have supported both the
development and the operation of the LAT as well as scientific data analysis.
These include the National Aeronautics and Space Administration and the
Department of Energy in the United States, the Commissariat \`a l'Energie Atomique
and the Centre National de la Recherche Scientifique / Institut National de Physique
Nucl\'eaire et de Physique des Particules in France, the Agenzia Spaziale Italiana
and the Istituto Nazionale di Fisica Nucleare in Italy, the Ministry of Education,
Culture, Sports, Science and Technology (MEXT), High Energy Accelerator Research
Organization (KEK) and Japan Aerospace Exploration Agency (JAXA) in Japan, and
the K.~A.~Wallenberg Foundation, the Swedish Research Council and the
Swedish National Space Board in Sweden.
 
Additional support for science analysis during the operations phase is gratefully acknowledged from the Instituto Nazionale di Astrofisica in Italy and the Centre National d'\'Etudes Spatiales in France.\\

The important contributions from ETH Zurich grants ETH-10.08-2 and
ETH-27.12-1 as well as the funding by the Swiss SNF and the German BMBF
(Verbundforschung Astro- und Astroteilchenphysik) and HAP
(Helmoltz Alliance for Astroparticle Physics) are gratefully acknowledged.
Part of this work is supported by Deutsche Forschungsgemeinschaft (DFG)
within the Collaborative Research Center SFB 876, project C3. We thank the
Instituto de Astrofisica de Canarias for allowing us to operate the telescope
at the Observatorio del Roque de los Muchachos in La Palma, the Max-Planck-Institut
fur Physik for providing us with the mount of the former HEGRA CT3 telescope,
and the MAGIC collaboration for their support.\\

JBG acknowledges the support of the Viera y Clavijo program funded by ACIISI and ULL. C. W. is grateful for the support by the project "Promotion inklusive" of the Universit\"{a}t zu K\"{o}ln and the German Bundesministerium f\"{u}r Arbeit und Soziales.\\

This publication makes use of data obtained at Mets\"{a}hovi Radio Observatory, operated by Aalto University, Finland.\\

\end{acknowledgements}

%
%


\appendix
\section{Tables with the X-ray and VHE gamma-ray spectral results}
\label{App:SpectralResults}

The X-ray spectral fits to the {\it Swift}/XRT data from 0.3~keV to 10~keV with a PL and an LP function are reported in Table~\ref{tab:xrt_fits}. The results from {\it Swift}/BAT analysis for the one-day time intervals are reported in  Table~\ref{tab:bat}.

The VHE gamma-ray spectral fits to the MAGIC data from 0.1~TeV to 10~TeV (for both observed and EBL-corrected) with a PL and an LP function are reported in Table~\ref{tab:vhe_fits}. 

\begin{table*}
   \centering
   \begin{tabular}{c c c c c c c c c} 
  MJD& Swift ObsID & Fit Func.& Flux (0.3--2~keV) & Flux (2--10~keV) & $\Gamma$  &   b  & $\mathrm{\chi}^2$/df & p-value \\   
   & & & $[10^{-10}~\mathrm{erg}~\mathrm{cm}^{-2}\mathrm{s}^{-1}]$  & $[10^{-10}~\mathrm{erg}~\mathrm{cm}^{-2}\mathrm{s}^{-1}]$ & & & & \\
        \hline
        \hline
  56855.04& 35023059 & PL &  1.89 $\pm$ 0.03  &  2.33 $\pm$ 0.08 & -1.81 $\pm$ 0.02& - &    205.6/216 & 0.7\\
  56856.04& 35023060 & PL &  2.50 $\pm$ 0.03  &  3.17 $\pm$ 0.08  & -1.77 $\pm$ 0.01& -&    320.9/300 & 0.2\\
  56856.86& 35023061 & PL &  2.67 $\pm$ 0.03  &  3.63$ \pm$ 0.08  & -1.74 $\pm$ 0.01& - &    298.7/316 & 0.8\\
  56858.04& 35023062 & PL &  3.11 $\pm$ 0.03  &  4.64 $\pm$ 0.11  & -1.69 $\pm$ 0.01& -&     308.8/304 & 0.4\\
  56858.93& 35023063 & PL &  3.09 $\pm$ 0.03  &  4.43 $\pm$ 0.09  & -1.69 $\pm$ 0.01& -&     369.5/341 & 0.1\\
  56860.04& 35023064 & LP &  2.74 $\pm$ 0.03  &  3.77 $\pm$ 0.08   & -1.58 $\pm$ 0.02& 0.25 $\pm$ 0.04& 393.5/355 & 0.1\\
  56861.99& 35023065 & PL &  2.25 $\pm$ 0.03  &  2.84 $\pm$ 0.10   & -1.77 $\pm$ 0.02& - &    213.5/223 & 0.7 \\
  56863.99& 35023067 & LP &  2.52 $\pm$ 0.03 &  3.48 $\pm$ 0.09  & -1.65 $\pm$ 0.03& 0.12 $\pm$   0.04& 308.5/301 & 0.4\\
  56865.04& 35023068 & PL &  2.68 $\pm$ 0.03 &  3.69 $\pm$ 0.08   & -1.73 $\pm$ 0.01& -&    344.6/327 & 0.2\\
  56865.99& 35023069 & PL &  2.71 $\pm$ 0.03 &  4.43 $\pm$ 0.09  & -1.63 $\pm$ 0.01& -&    356.3/334 & 0.2\\
  56866.92& 35023070 & PL &  2.59 $\pm$ 0.03 &  3.89 $\pm$ 0.08   & -1.69 $\pm$ 0.01& - &     310.1/329 & 0.8\\
  56867.99& 35023071 & PL &  2.87 $\pm$ 0.03 &  4.89 $\pm$ 0.09  & -1.60 $\pm$ 0.01& - &     308.6/348 & 0.9\\
  56868.92& 35023072 & PL &  2.23 $\pm$ 0.02 &  3.06 $\pm$ 0.09  & -1.74 $\pm$ 0.01& - &     273.0/297 & 0.8\\
  56869.92& 35023073 & PL &  2.00 $\pm$ 0.02 &  2.80 $\pm$ 0.07   & -1.71 $\pm$ 0.01& - &    310.3/292 & 0.2\\
        \hline
        \hline
   \end{tabular}
   \caption{Spectral parameters for the {\it Swift}-XRT data from Mrk~501 in the energy range 0.3--10~keV during the high activity in July 2014. A PL function and an LP function were used. For those cases where the PL function is rejected   at the   2$\sigma$ confidence level (at least), the LP spectral parameters are given in this table.}   
   \label{tab:xrt_fits}
\end{table*}

\begin{table*}
   \centering
   \begin{tabular}{c c c c c c c} 
  MJD& Exposure & SNR & $\mathrm{Flux}_1$  & $\chi^2$/df & $\mathrm{Flux}_2$ & $\chi^2$/df \\  
     & [s] & & [$10^{-10}\,\mathrm{erg}\,\mathrm{s}^{-1}\,\mathrm{cm}^{-2}$]  & & [$10^{-10}\,\mathrm{erg}\,\mathrm{s}^{-1}\,\mathrm{cm}^{-2}$] &  \\
        \hline
        \hline
  56855 & 905 & 2.4 & 4.5$\pm$1.6 & 4.4/7 & 5.2$\pm$2.0 & 5.5/7\\
  56856 & 961 & 1.7 & 2.5$\pm$1.4 & 8.6/7 & 3.8$\pm$1.8 & 7.9/7\\
  56857 & 983 & 3.0 & 4.2$\pm$1.5 & 4.6/7 & 5.4$\pm$2.1 & 4.8/7\\
  56858 & 787 & 3.0 & 6.7$\pm$2.2 & 5.5/7 & 8.5$\pm$2.5 & 5.9/7\\
  56859 & 536 & 2.3 & 4.7$\pm$1.8 & 6.3/7 & 8.2$\pm$2.8 & 3.9/7\\
  56860 & 1320 & 1.6 & 1.9$\pm$1.2 & 1.5/7 & 1.7$\pm$1.1 & 1.5/7\\
  56862 & 640 & -0.2 & $<4.5$ & - & - & - \\
  56864 & 895 & 3.0 & 4.6$\pm$1.6 & 6.0/7 & 5.4$\pm$1.7 & 5.4/7\\
  56865 & 762 & 3.5 & 8.9$\pm$2.6 & 1.7/7 & 11.0$\pm$2.0 & 2.8/7\\
  56866 & 662 & 3.0 & 4.4$\pm$1.9 & 6.6/7 & 7.9$\pm$2.4 & 4.1/7\\
  56867 & 1004 & 3.2 & 5.8$\pm$1.7 & 3.8/7 & 8.0$\pm$2.2 & 3.5/7\\
  56868 & 900 & 4.0 & 5.8$\pm$1.6 & 8.3/7 & 9.5$\pm$2.2 & 4.5/7\\
  56869 & 992 & 2.4 & 3.5$\pm$1.6 & 1.5/7 & 4.8$\pm$2.0 & 1.3/7\\
  56870 & 1001 & 2.7 & 4.3$\pm$1.5 & 10.9/7 & 5.5$\pm$1.9 & 12.0/7\\
  56871 & 1029 & 4.0 & 6.7$\pm$1.5 & 3.0/7 & 8.7$\pm$1.8 & 2.9/7\\
  56872 & 1027 & 1.4 & 2.0$\pm$1.5 & 2.6/7 & 2.1$\pm$1.9 & 3.3/7\\
        \hline
        \hline
   \end{tabular}
   \caption{{\it Swit}-BAT results from the analysis of Mrk\,501 in the energy band 14--195~keV for one-day integration bins. $\mathrm{Flux}_1$ was calculated using the spectral shape from the integrated BAT spectrum during the time interval MJD~56854.5--MJD~56872.5 (see section \ref{SwiftBAT}). $\mathrm{Flux}_2$ was calculated using the spectral shape reported in Table~\ref{tab:xrt_fits}, which is   from the XRT data. The uncertainty on both fluxes is calculated at the 68\% confidence level. In the analysis for MJD~56862, Xspec did not converge because of the very low signal, and a $2\sigma$ flux upper limit was calculated.}
   \label{tab:bat}
\end{table*}

\begin{table*}
   \centering
   \begin{tabular}{c c c c c c c c c} 
        Date & MJD & Fit & $f_0$& $\Gamma$ & b & $\mathrm{\chi}^2$/df & p-value & LP preference\\
        & & &$[10^{-10}\mathrm{TeV}^{-1}\mathrm{cm}^{-2}\mathrm{s}^{-1}]$& & & &\\
        \hline
        \hline
        20140716 & 56854.91 & PL & 0.82$\pm$0.10 & -2.42$\pm$0.11 & -&11.6/8& 0.2 & -\\
        &&PL&0.99$\pm$ 0.12& -2.30$\pm$0.12&& 9.8/8& 0.3&-\\
        \hline
        20140717 & 56855.91 &PL &1.24$\pm$0.13& -2.39$\pm$0.10& -& 6.2/9& 0.7& -\\
        &&PL&1.48$\pm$0.16& -2.27$\pm$0.10&&  5.8/9& 0.8&-\\    
        \hline
    20140718 & 56856.91 & PL&2.78$\pm$0.15& -2.15$\pm$0.05 & - & 10.7/11& 0.5& -\\
    &&PL&3.35$\pm$0.18& -2.01$\pm$0.05& &9.1/11 &0.6&-\\
        \hline
        20140720 & 56858.98 & PL&2.94$\pm$0.12& -2.19$\pm$0.04 & -&17.3/12 &0.1& -\\
        &&PL&3.53$\pm$0.14& -2.04$\pm$0.04&& 14.3/12& 0.3&-\\
        \hline
        20140721 & 56859.97 & LP&2.29$\pm$0.10 & -2.21$\pm$0.04 & 0.15$\pm$0.03 & 7.7/11 &0.7 & 4.5~$\sigma$\\
        &&LP&2.66$\pm$0.11& -2.05$\pm$0.04& 0.11$\pm$0.03& 6.6/11& 0.8 &3.4~$\sigma$\\
        \hline
        20140723 & 56861.01 & PL&1.42$\pm$0.10 & -2.22$\pm$0.08& -& 10.1/10 &0.4& -\\
        &&PL&1.69$\pm$0.11& -2.06$\pm$0.08& -&9.3/10&  0.5& -\\
        \hline
        20140724 & 56862.02 & PL&1.03$\pm$0.05 & -2.24$\pm$0.06& -& 11.8/11 &0.4& -\\
        &&PL&1.23$\pm$0.06 &-2.08$\pm$0.06&-&  9.1/11&  0.6& -\\
        \hline
        20140725 & 56863.00 & PL&1.16$\pm$0.09 & -2.12$\pm$0.09& -& 15.7/10 &0.1& -\\
        &&PL&1.38$\pm$0.11& -1.97$\pm$0.09&-& 14.8/10& 0.1& -\\
        \hline
        20140726 & 56864.02 & PL&1.24$\pm$0.05 & -2.24$\pm$0.04 & -&15.2/13 &0.3& -\\
        &&PL&1.48$\pm$0.06& -2.08$\pm$0.04&-& 12.5/13& 0.5& -\\
        \hline
        20140727 & 56865.00 & PL&1.75$\pm$0.10 & -2.14$\pm$0.06& -& 11.3/12 &0.5& -\\
        &&PL&2.09$\pm$0.12& -1.97$\pm$0.06 & - & 9.6/12 &  0.6& -\\
        \hline
        20140728 & 56866.00 & LP&3.19$\pm$0.08  & -2.10$\pm$0.02 & 0.09$\pm$0.02 & 10.8/14 &  0.7 &5.3 $\sigma$\\
        &&PL&2.50$\pm$0.07& -1.97$\pm$0.02&-& 17.6/15& 0.3& 2.7 $\sigma$\\
        \hline
        20140729 & 56867.00 & PL&2.33$\pm$0.11 & -2.15$\pm$0.05& - &10.2/12 &0.6& -\\
        &&PL&2.79$\pm$0.13& -1.99$\pm$0.05&-&9.8/12&0.6& -\\
        \hline
        20140730 & 56868.01 &LP& 3.17$\pm$0.11   & -2.09$\pm$0.03 & 0.11$\pm$0.03 & 22.8/12 & 0.03& 4.5~$\sigma$\\
        &&LP&3.70$\pm$0.12 & -1.93$\pm$0.03& 0.09$\pm$0.03& 22.2/12& 0.04& 3.2~$\sigma$\\
        \hline
        20140731 & 56869.92 & PL&1.30$\pm$0.08 & -2.11$\pm$0.07 & -&8.2/10& 0.6& -\\
        &&PL&1.55$\pm$0.10& -1.96$\pm$0.07&-& 9.0/10 &0.5& -\\
        \hline
        \hline
   \end{tabular}
   \caption{Spectral parameters for the MAGIC data from Mrk~501 in the energy range 0.1--10~TeV during the high activity in July 2014. Both a PL function and a LP function were used. For each night the observed spectral
fits (first row) and EBL-corrected spectral
fits using the EBL model from \cite{dominguez} (second row)  are provided. The parameters resulting from the fit with an LP  (eq.~\ref{eq:logparabola}) are provided when the LP function is preferred with respect to the PL function (eq.~\ref{eq:power-law}) with a significance higher than 3$\sigma$. The fit parameters for 2014 July 19 are reported in Table~\ref{tab:bump_fit}.}   
   \label{tab:vhe_fits}
\end{table*}

\clearpage

\section{X-ray and VHE gamma-ray spectral index vs flux }
\label{App:SpectrumVSFlux}

\begin{figure}[!b]
   \centering
   \includegraphics[scale=0.6]{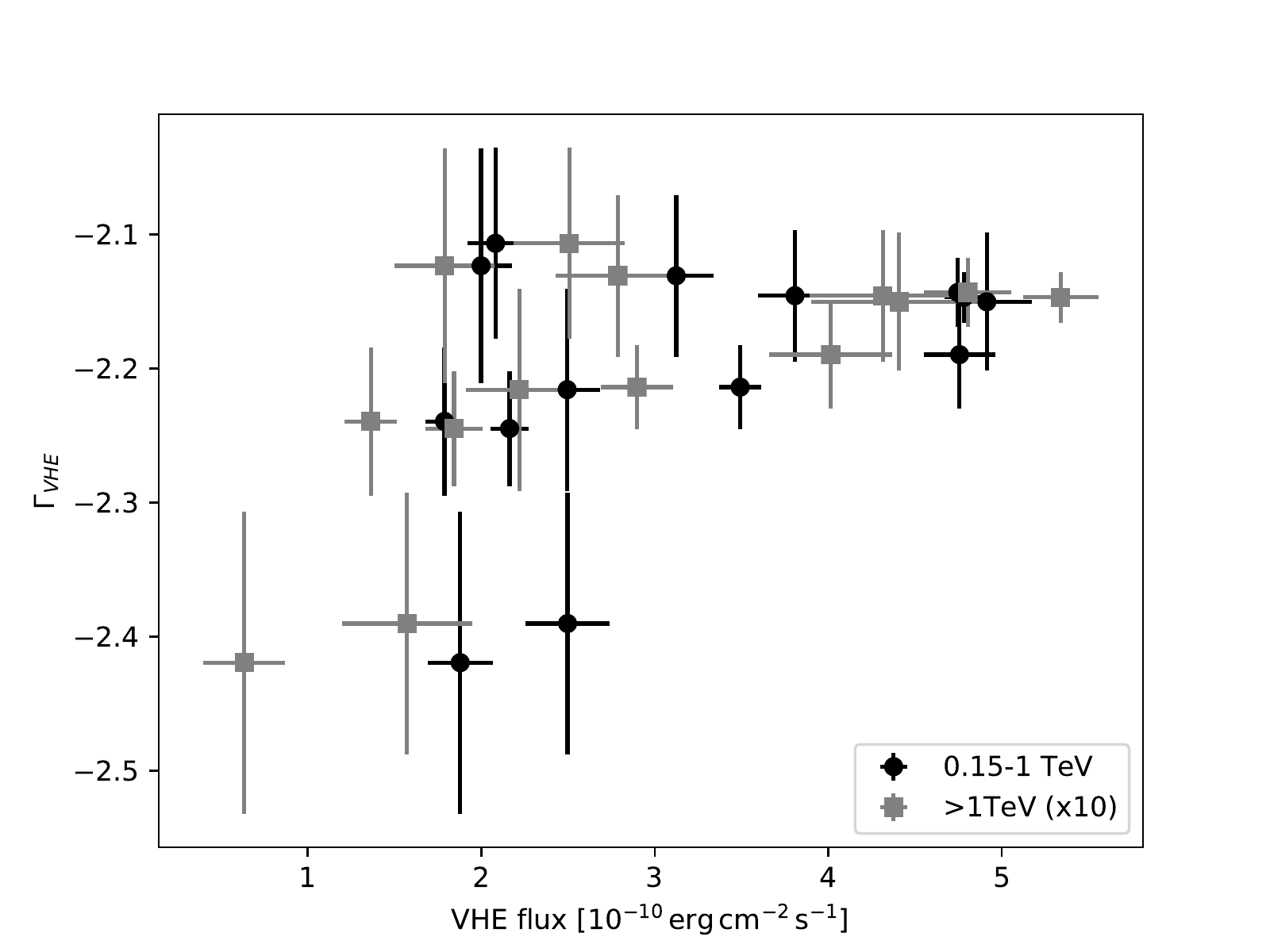} 
   \caption{PL spectral index as a function of the integral flux, as observed by MAGIC in the energy bands 0.15--1~TeV and $>$1~TeV.} 
   \label{fig:magic_flux_vs_index}
\end{figure}

\begin{figure}[!b]
   \centering
   \includegraphics[scale=0.6]{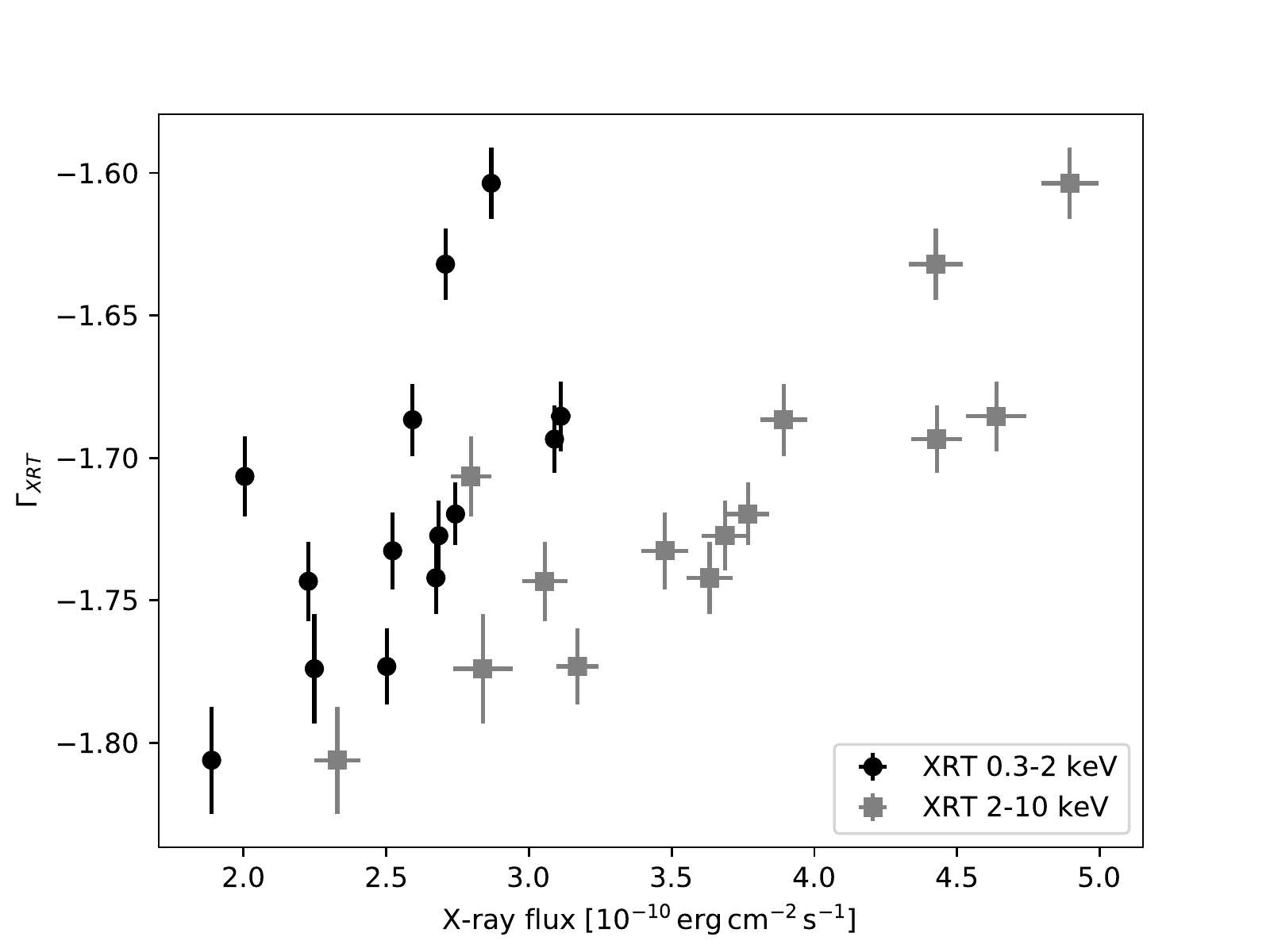} 
   \caption{PL spectral index as a function of the integral flux, as observed by XRT in the energy bands from 0.3-2 keV and 2-10 keV.}
   \label{fig:xrt_flux_vs_index}
\end{figure}

The PL spectral index as a function of the integral fluxes in two energy bands in X-rays and VHE are reported in Fig.~\ref{fig:magic_flux_vs_index} and Fig.~\ref{fig:xrt_flux_vs_index}. The LP function is more suitable than the PL function in a few X-ray spectra and VHE spectra, but the difference is small (see Appendix A). For the sake of simplicity, we decided to use the PL index for the study presented here. The only spectrum that was not considered in this study (for the VHE gamma-ray band) is that from 2014 July 19, which is the one showing a narrow spectral feature at about 3~TeV (see Sec.~\ref{bump}). 

No correlation is found between the PL spectral index and the VHE fluxes, with Pearson coefficients of 0.30 (1.0\,$\sigma$) and 0.50 (1.7\,$\sigma$) for the energy bands  0.15-1\,TeV and $>1$\,TeV, respectively.  On the contrary, the X-ray band shows evidence for the {harder-when-brighter} behavior,  with Pearson coefficients of 0.61 ($2.4\,\sigma$) and 0.86 ($4.3\,\sigma$), and DCF=$0.6\pm0.3$ and DCF=$0.8\pm0.3$, for the soft (0.3--2~keV) and the hard (2--10~keV) X-ray bands, respectively.

\clearpage

\section{Evaluation of extrapolation of LP spectral fit beyond 1.5 TeV }
\label{App:FitExtrapolation}

In order to test the behavior of the low- and high-energy part of the VHE spectrum during the two-week  period with outstanding X-ray activity, we performed the following test. The spectral data were fit with an LP function up to 1.5~TeV. The value of 1.5~TeV is right below the energy of the narrow spectral feature at $\sim$3~TeV in the spectrum from 2014 July 19.  Afterwards, the data-model agreement was quantified above 1.5~TeV for the extrapolation (to higher energies) of the best-fit function up to 1.5~TeV. The results from these spectral fits, and the quantification of the data-model agreement above 1.5~TeV, are reported in Table~\ref{tab:vhe_partialfit}. 

As shown in the table, in general, the extrapolation of the fit up to 1.5\,TeV provides a good description of the spectral shape at energies above 1.5~TeV. The only notable exception is the spectrum from 2014 July 19 (MJD~56857.98), where there is a significant deviation (from the LP function) at energies above 1.5 TeV. The VHE spectrum from July 20 (MJD~56858.98) also shows a marginally significant deviation at high energies with respect to the fit at low energies. 

If we consider the 1$\sigma$ uncertainty in the best fit up to 1.5 TeV (uncertainty in the spectral parameters reported in Table~\ref{tab:vhe_partialfit}), the significance of the deviation of the data points with respect to the extrapolation of the spectral fit below 1.5~TeV decreases to a 3$\sigma$ level for 2014 July 19. For the observed spectrum we find that $\mathrm{\chi^2_{above\,1.5\,TeV}}$/df=24.4/7 (3.3 $\sigma$), while for the EBL-corrected spectrum the values are $\mathrm{\chi^2_{above\,1.5\,TeV}}$/df=19.5/7 (2.7~$\sigma$).

\begin{table*}
   \centering
   \begin{tabular}{c c c c c c} 
        MJD & $f_0$& $\Gamma$ & b & $\mathrm{\chi}^2$/df & $\mathrm{\chi}^2$/df   \\
        &$[10^{-10}\mathrm{TeV}^{-1}\mathrm{cm}^{-2}\mathrm{s}^{-1}]$& & & up to 1.5 TeV & above 1.5 TeV \\
        \hline
        \hline
        \hline
        56857.98 & 2.54 $\pm$0.10& -2.26$\pm$0.06 &0.15$\pm$0.06 &10.3/7 & 54.5/7 (6.0$\sigma$) \\
                &2.92$\pm$0.12& -2.08$\pm$0.06& 0.10$\pm$0.06 &10.6/7& 41.4/7 (5.0$\sigma$)\\
        \hline
        56858.98 & 3.26$\pm$0.19& -2.32$\pm$0.06& 0.18$\pm$0.06 &7.4/8& 18.5/3 (3.6 $\sigma$) \\
        &3.75$\pm$0.21& -2.15$\pm$ 0.06 &0.13$\pm$ 0.06 &8.0/8& 14.8/3 (3.1 $\sigma$) \\
        \hline
        56859.97 & 2.33$\pm$ 0.10& -2.26$\pm$ 0.05& 0.21$\pm$ 0.05 &4.0/8& 8.9/3 (2.2 $\sigma$) \\
        &2.69$\pm$0.12& -2.10$\pm$ 0.05 &0.16$\pm$ 0.05 &3.9/8& 5.7/3 (1.5 $\sigma$) \\
        \hline
        56864.02 & 1.35$\pm$ 0.08 &-2.26$\pm$0.06 &0.12$\pm$ 0.07 & 11.0/8& 4.7/4 (1.0$\sigma$) \\
        &1.56$\pm$ 0.09 &-2.09$\pm$ 0.06 & 0.08$\pm$ 0.07 &10.8/8& 2.9/4 (0.6$\sigma$)\\
        \hline
        56865.00 & 2.06$\pm$ 0.17 &-2.11$\pm$ 0.09 &0.22$\pm$ 0.12 & 5.3/8& 2.7/3 (0.8$\sigma$)  \\
        &2.38$\pm$ 0.20 &-1.94$\pm$ 0.09 & 0.17$\pm$0.12& 5.6/8& 1.7/3 (0.5$\sigma$) \\
        \hline
        56866.00 & 3.14$\pm$ 0.09& -2.09$\pm$ 0.03  &0.07$\pm$ 0.03 &5.9/8& 6.2/6 (0.8$\sigma$) \\
        &3.63$\pm$0.11 &-1.93$\pm$ 0.03 & 0.02$\pm$0.03& 5.3/8 & 9.8/6 (1.5$\sigma$) \\
        \hline
        56867.0 & 2.35$\pm$0.16 &-2.13$\pm$0.07 & $-6.6\times10^{-3}\pm 0.07$ &8.9/8& 1.6/3 (0.4$\sigma$) \\
        &2.71$\pm$0.19 &-1.96$\pm$0.07& 0.05$\pm$0.07 & 8.5/8& 1.2/3(0.3$\sigma$) \\
        \hline
        56868.01 & 3.11$\pm$ 0.12 &-2.08$\pm$ 0.04 &0.08$\pm$0.05 &14.8/8& 10.1/4 (2.1$\sigma$) \\
        &3.59 $\pm$0.14 & -1.91$\pm$ 0.04  &0.03$\pm$ 0.05& 14.0/8 & 16.6/4 (3.0$\sigma$) \\
        \hline
        \hline
   \end{tabular}
   \caption{Results from the forward-folding fits with an LP up to 1.5~TeV to the single-night MAGIC VHE spectra that contains at least three spectral points beyond 1.5~TeV. For each night the fit to the observed (first line) and the EBL-corrected (second line) using the model from \cite{dominguez} are given. The table reports the data-model agreement, quantified with a $\mathrm{\chi}^2$, for the spectral data below 1.5~TeV (goodness of   fit) and above 1.5~TeV. The reported significances in the last column refer the confidence level at which the data-model agreement above 1.5~TeV can be rejected. }
   \label{tab:vhe_partialfit}
\end{table*}

\clearpage

\section{Additional Monte Carlo tests to estimate the chance probability of obtaining a narrow spectral feature on the top of smooth gamma-ray spectra}
\label{App:toyMC}

\begin{figure}
   \centering
   \includegraphics[scale=0.43]{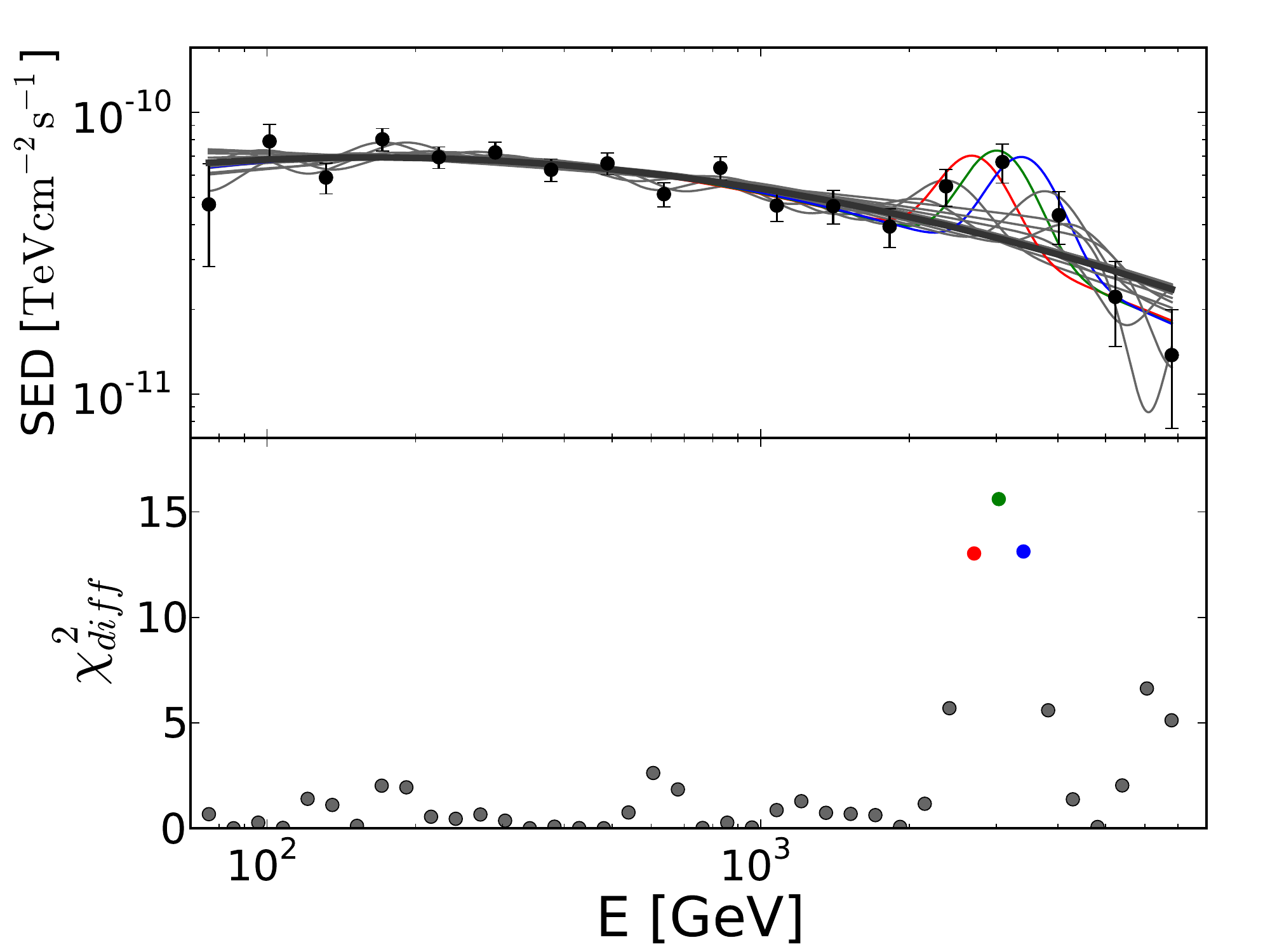} 
   \includegraphics[scale=0.44]{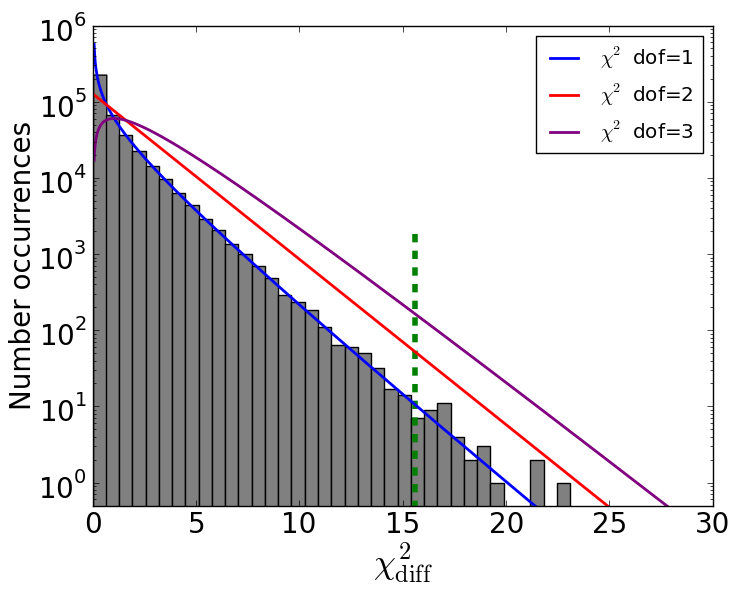}
    \includegraphics[scale=0.44]{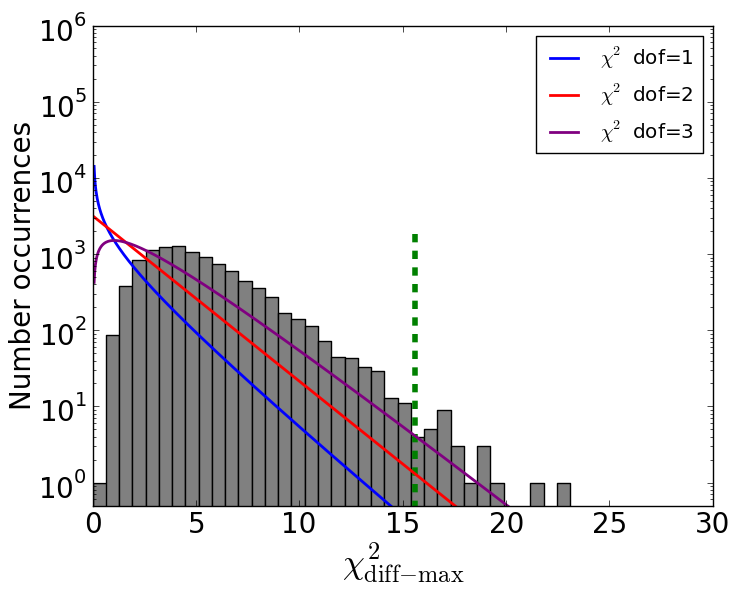}
   \caption{Results from the Monte Carlo simulations for the hypothesis of a narrow feature parameterized with an EP function with fixed curvature. The first (top) panel shows the VHE gamma-ray spectrum from 2014 July 19 (MJD 56857.98), from Fig.~\ref{fig:VHE_bump_extracomponent}, fitted with an LP function (thick dark gray curve), and also fitted with an LP plus an EP with $\beta=9.1$, and centered at various energies  from 0.08~TeV to 6.80~TeV in steps of 0.05 in base-10 logarithmic space (thin light gray lines). The 
   difference in $\chi^2$ values ($\chi^2_{diff}$) is shown below the spectrum, using colors different from gray for cases with $\chi^2_{diff} >8$.  The second and third panels show the resulting $\chi^2_{diff}$ and  $\chi^2_{diff-max}$ distributions from the 10$^4$ simulated spectra. The green dashed line marks the $\chi^2_{diff-data}$ obtained for the measured spectrum and shown in the first panel, while the blue, red, and purple solid lines depict the expected $\chi^2$ distribution for 1, 2, and 3 degrees of freedom. See text in appendix~\ref{App:toyMC} for further details. }
   \label{fig:chi2_tombesi_real_data_LP_Beta9.1}
\end{figure}

\begin{figure}
   \centering
   \includegraphics[scale=0.43]{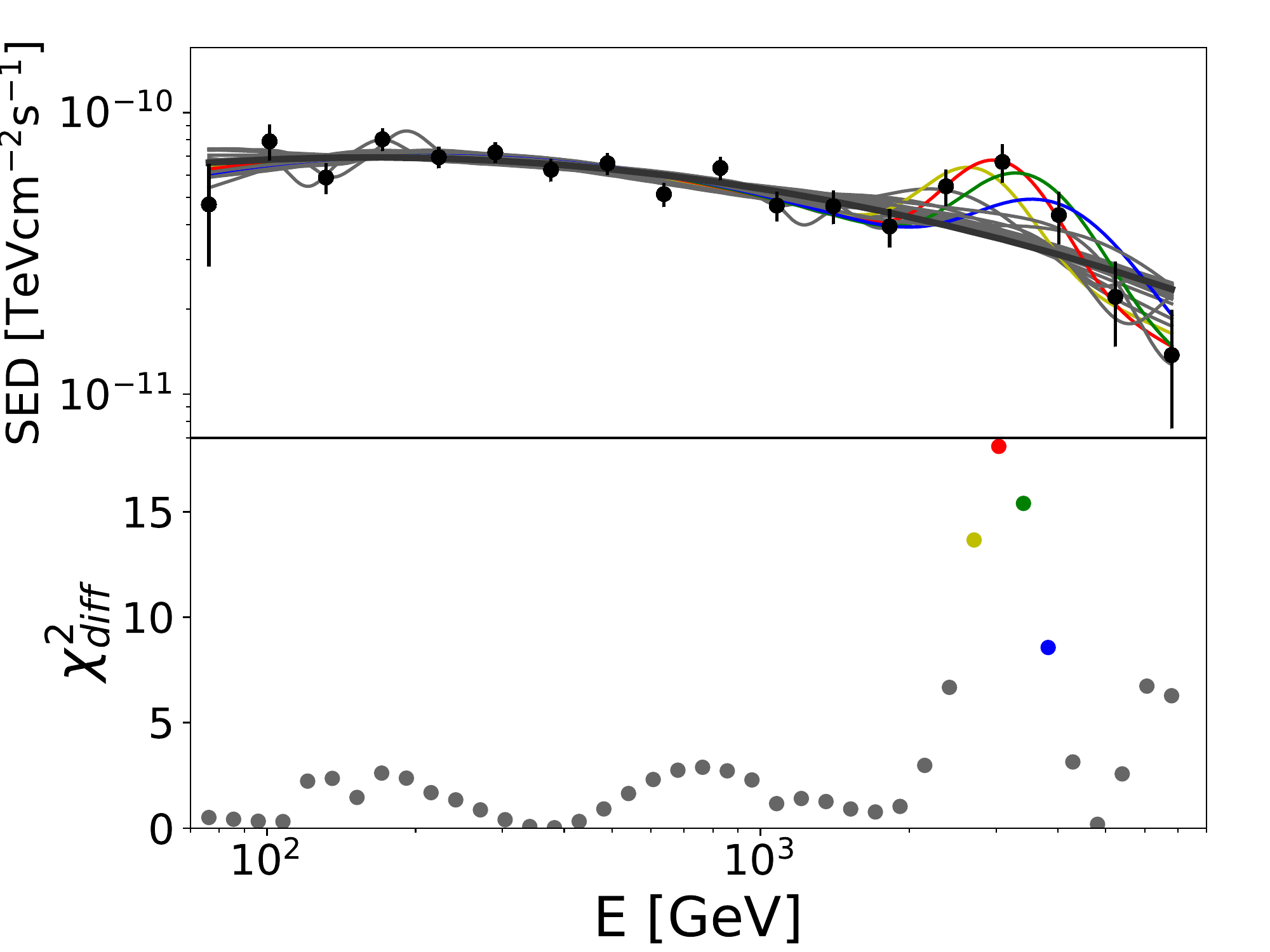} 
   \includegraphics[scale=0.44]{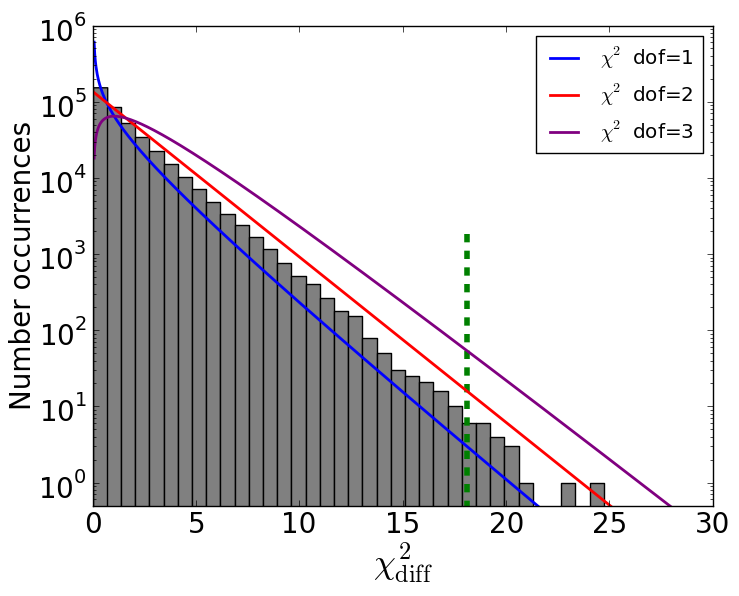}
   \includegraphics[scale=0.44]{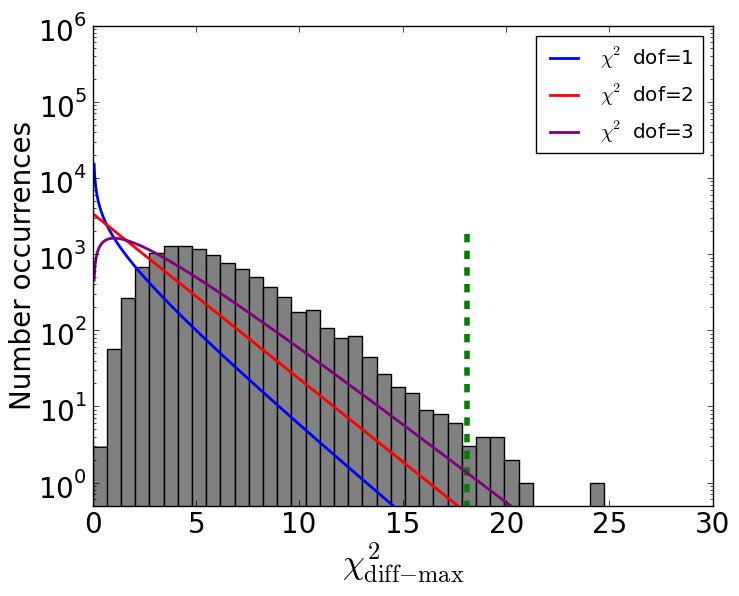}
   \caption{Same as in Fig.~\ref{fig:chi2_tombesi_real_data_LP_Beta9.1}, but for  an EP with a variable curvature (i.e., $\beta$ is left free to vary in the spectral fits) to parameterize the narrow spectral feature. See text in appendix~\ref{App:toyMC} for further details. }
 \label{fig:chi2_tombesi_real_data_LP_VariableCurvature}
\end{figure}

\begin{figure}
   \centering
   \includegraphics[scale=0.43]{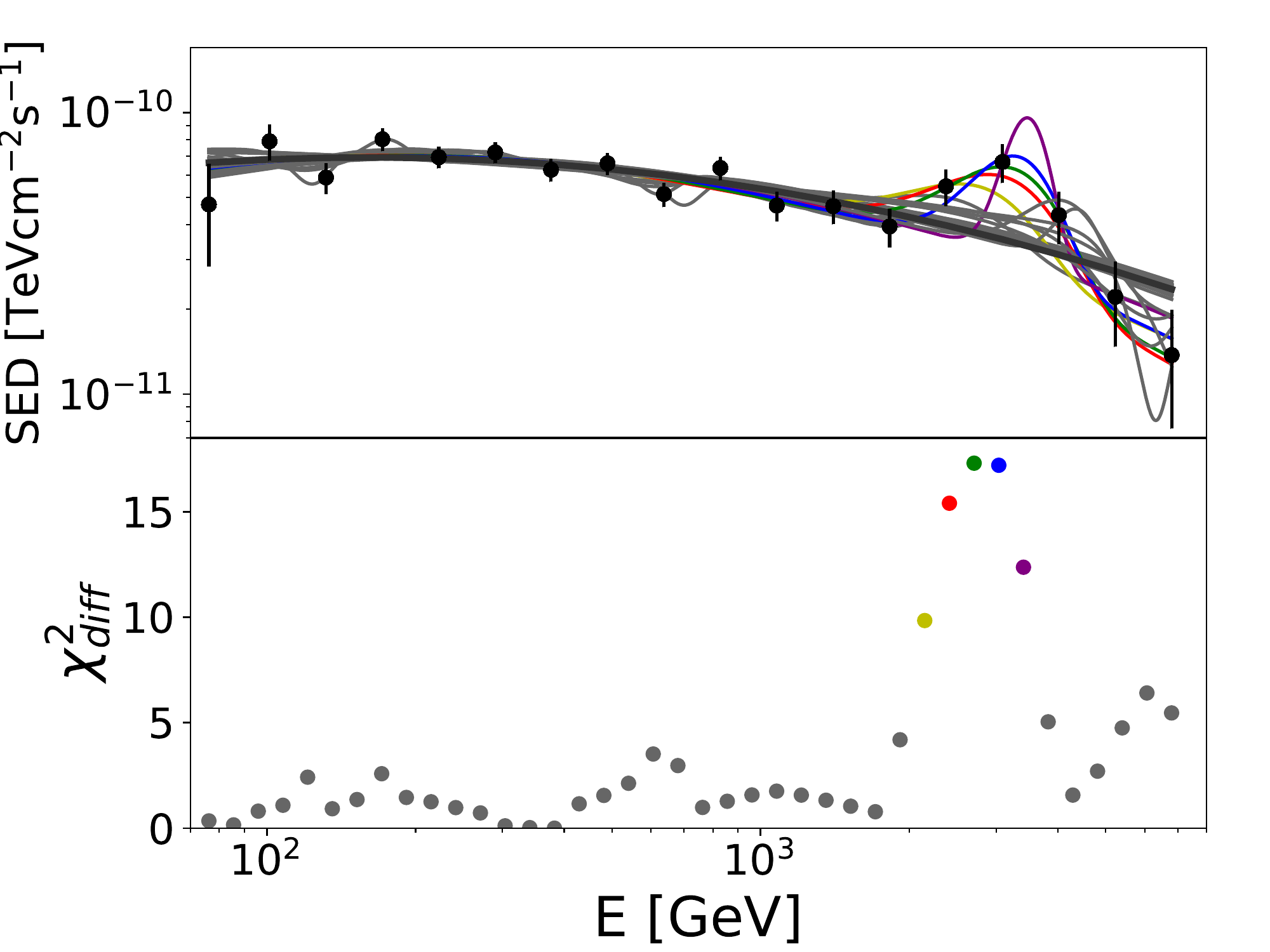} 
   \includegraphics[scale=0.44]{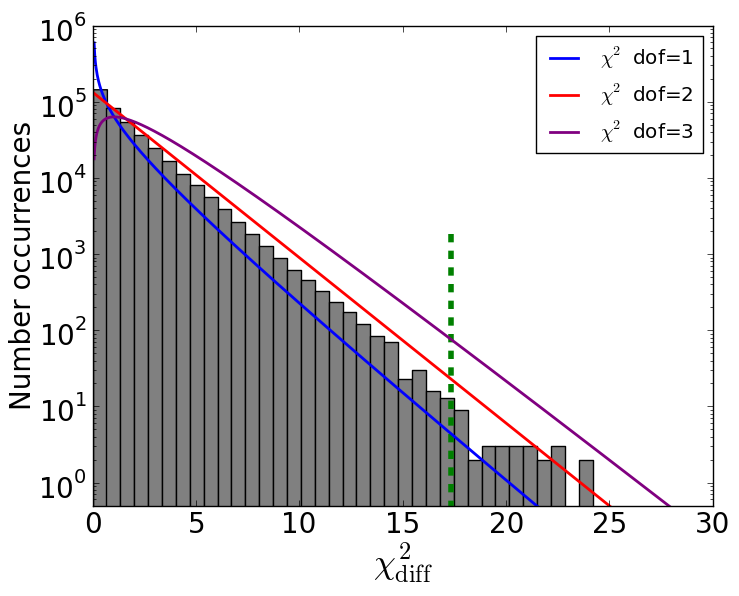}
   \includegraphics[scale=0.44]{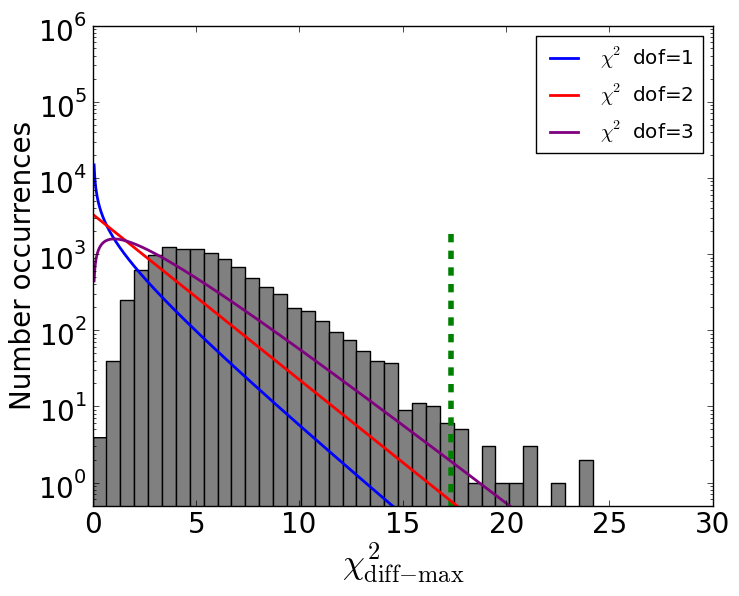}
   \caption{Same as in Fig.~\ref{fig:chi2_tombesi_real_data_LP_Beta9.1}, but for  a Gaussian function with a variable width (i.e., sigma is left free to vary in the spectral fits) to parameterize the narrow spectral feature. See text in appendix~\ref{App:toyMC} for further details. }
 \label{fig:chi2_tombesi_real_data_Gauss_VariableWidth}
\end{figure}

In this section we describe additional Monte Carlo tests that were performed to assess the random chance probability of obtaining a narrow spectral feature like the one observed in the measured VHE gamma-ray spectrum of Mrk\,501 from 2014 July 19 (see section~\ref{bump}). In this case, we followed the prescriptions from \citet{2010A&A...521A..57T}, which had been used to select line-features at $p_{value}<0.05$ in a systematic search over a large number of measured X-ray spectra. The nature of this test is different from  the one described section~\ref{bump}, which relates to the investigation of a feature observed in a single spectrum, but provides an alternative perspective to the evaluation of the random chance probability for the occurrence of narrow features in continuum spectra. The tests are performed on the differential flux spectrum (dN/dE) without applying any correction for the EBL.  The continuum model (which is taken as the null hypothesis) is described with an LP function, and three distinct functions are used to parameterize the narrow feature: a) an EP function (see eq.~\ref{eq:eplogpar}) with curvature fixed to $\beta=9.1$; b) a more generic EP function with variable curvature, where $\beta$ is allowed to change from 1 to 20 in the spectral fits; and c) a Gaussian function with variable width where sigma is allowed to change from 10\% to 40\% of the Gaussian mean. The location of the narrow feature ($E_p$ in the EP function, and the mean of the Gaussian function) is determined from a scan over a 40-bin grid extending from the energy 0.08~TeV (first data point in the spectrum) to the energy 6.80~TeV (last data point in the spectrum), in steps of 0.05 in base-10 logarithmic space.  Each step corresponds to a relative change in the energy of $\sim$12\%, which is comparable to the energy resolution of MAGIC \citep[15\%--20\%; see][]{magic_upgrade2}.  The narrow feature hypothesis {\it a)} is described with one additional free (and unconstrained) parameter,  the normalization $K$, which can take positive and negative values, and follows the prescriptions from \citet{2010A&A...521A..57T}. On the other hand,   hypotheses {\it b)} and {\it c)} relate to a more generic search where the spectral feature hypothesis has a variable shape. In these cases,  the spectral feature is described with two additional free parameters, the normalization $K$ (unconstrained) and the width of the feature, which is parameterized with $\beta$ (for EP) or sigma (for Gauss), and which are constrained to vary within the above-mentioned range of values. The results from these energy scans on the VHE gamma-ray spectrum from July 19 are depicted in the upper panels of Figures~\ref{fig:chi2_tombesi_real_data_LP_Beta9.1}, \ref{fig:chi2_tombesi_real_data_LP_VariableCurvature}, and \ref{fig:chi2_tombesi_real_data_Gauss_VariableWidth}.

Then, in the same way as for the Monte Carlo tests reported in section~\ref{bump}, we use the LP function derived from the fit (thick dark gray curve in Figures~\ref{fig:chi2_tombesi_real_data_LP_Beta9.1}, \ref{fig:chi2_tombesi_real_data_LP_VariableCurvature}, and \ref{fig:chi2_tombesi_real_data_Gauss_VariableWidth}) to generate 10000 realizations of this spectrum with data points that have the same statistical uncertainty as the measured spectrum.  We then performed a series of fits with a model composed of the baseline (LP) and the three cases for the narrow component: EP with fixed $\beta$, EP with variable $\beta$, and Gauss function with variable sigma. The parameter $E_p$ for the EP function and mean for the Gauss function ranges from 0.08~TeV to 6.80~TeV in steps of 0.05 in log10 scale, as done before with the actual measured VHE spectrum. The distribution of $\chi^2_{diff}$ values obtained for the three distinct hypotheses are depicted in the second and third panels of Figures \ref{fig:chi2_tombesi_real_data_LP_Beta9.1}, \ref{fig:chi2_tombesi_real_data_LP_VariableCurvature}, and \ref{fig:chi2_tombesi_real_data_Gauss_VariableWidth}, and a summary of the resulting numbers are reported in Table~\ref{tab:MC_Results}.

\begin{table*}
   \centering
   \begin{tabular}{c c c} %
   \hline
   \hline
    Measured VHE spectrum 
    & MC: 40$\times$10000 spectral fits 
    & MC: 10000 spectral fits with $\chi^2_{diff-max}$ \\
    \hline
    Functional hypothesis for the feature
    
    & N$> \chi^2_{diff-data}$ ~~|~~ $p_{value}$~ (significance) 
        & N$> \chi^2_{diff-data}$ ~~|~~ $p_{value}$~ (significance)\\
        \hline 
EP with $\beta=9.1$ ($\chi^2_{diff-data}$=15.6) 
& 39 ~~~|~~~ $9.7\times10^{-5}$ (3.9$\sigma$) 
& 26 ~~~|~~~ $2.6\times10^{-3}$  (3.0$\sigma$) \\
EP with variable curvature ($\chi^2_{diff-data}$=18.1) 
& 21 ~~~|~~~ $5.2\times10^{-5}$ (4.0$\sigma$) 
& 15 ~~~|~~~ $1.5\times10^{-3}$  (3.2$\sigma$) \\
Gauss with variable width ($\chi^2_{diff-data}$=17.3) 
& 32 ~~~|~~~ $8.0\times10^{-5}$ (3.9$\sigma$) 
& 18 ~~~|~~~ $1.8\times10^{-3}$  (3.1$\sigma$) \\
\hline 
   \hline
   \end{tabular}
   \caption{Results from the Monte Carlo tests following the prescription from \citet{2010A&A...521A..57T}, that are used to assess the chance probability (and related significance) of observing a spectral feature on  top of the measured VHE gamma-ray spectrum described by an LP. See text in Appendix~\ref{App:toyMC} for further details. }
   \label{tab:MC_Results}
\end{table*}

In \citet{2010A&A...521A..57T}, only the highest $\chi^2_{diff}$, namely $\chi^2_{diff-max}$, is considered. This number relates to the largest fluctuation (with the shape of the narrow feature) in the simulated spectrum.  Here we also report the results obtained when using all the $\chi^2_{diff}$ values obtained from the 40$\times$10000 spectral fits. 
When we consider the hypothesis of a narrow feature of fixed shape, which is the one resembling the test performed in \citet{2010A&A...521A..57T}, and the most simple out of the three hypotheses investigated, 
the second panel of Fig.~\ref{fig:chi2_tombesi_real_data_LP_Beta9.1} shows that the distribution of $\chi^2_{diff}$ follows a $\chi^2$ with 1 degree of freedom. This is expected because, for each grid position, there is only one additional degree of freedom in the fit with the narrow component (the parameter $K$ from the EP function); however,  this shows that this test does not take into account that the location of the feature in the spectrum is arbitrary, and that a search in energy space is needed. When generating a large number of random tests, for a continuum spectrum where we make a search for additional components parameterized with two degrees of freedom (normalization and energy location), we would expect that   distribution of $\chi^2_{diff}$ to follow a nominal $\chi^2$ distribution with 2 degrees of freedom. On the other hand, the third panel of Fig.~\ref{fig:chi2_tombesi_real_data_LP_Beta9.1} shows that the distribution of $\chi^2_{diff-max}$ has a large deficit at low $\chi^2_{diff}$ values, and is shifted to the right with respect to a nominal $\chi^2$ distribution for 2 degrees of freedom.  This occurs by construction of the test because the cases with low $\chi^2_{diff}$ values are systematically rejected. A similar situation occurs for the hypotheses where the shape of the narrow feature is not fixed, and hence  an extra degree of freedom is added in the search (the curvature or width of the feature). The results for these tests are shown in  Figures~\ref{fig:chi2_tombesi_real_data_LP_VariableCurvature} and \ref{fig:chi2_tombesi_real_data_Gauss_VariableWidth}, where the reference nominal $\chi^2$ distribution would be the one for 3 degrees of freedom.

We note that, when allowing for the curvature of the EP function to vary, the chance probability for a random fluctuation decreases slightly (see Table~\ref{tab:MC_Results}). This is caused by the marginally better spectral fit to the measured VHE gamma-ray spectral points when
using two degrees of freedom (see upper panels of Figures~\ref{fig:chi2_tombesi_real_data_LP_Beta9.1} and \ref{fig:chi2_tombesi_real_data_LP_VariableCurvature}): $\chi^2_{diff-data}$ increases from 15.6 to 18.1, which counteracts the
larger freedom in the spectral fits to find narrow features in the simulated spectra.
The numbers obtained with the EP and the Gaussian function with variable width are very similar because
of the relatively large statistical uncertainties in the measured spectral data points:
the results are dominated by the peak of the mathematical function used to
describe the narrow spectral feature, and they are not affected by the tails of such function,
which is where the Gauss and the EP function differ most.

\clearpage

\section{Characterization of the radio-optical emission with another SSC component}
\label{App:SEDRadioOptical}

\begin{figure}
   \includegraphics[scale=0.45, angle=0]{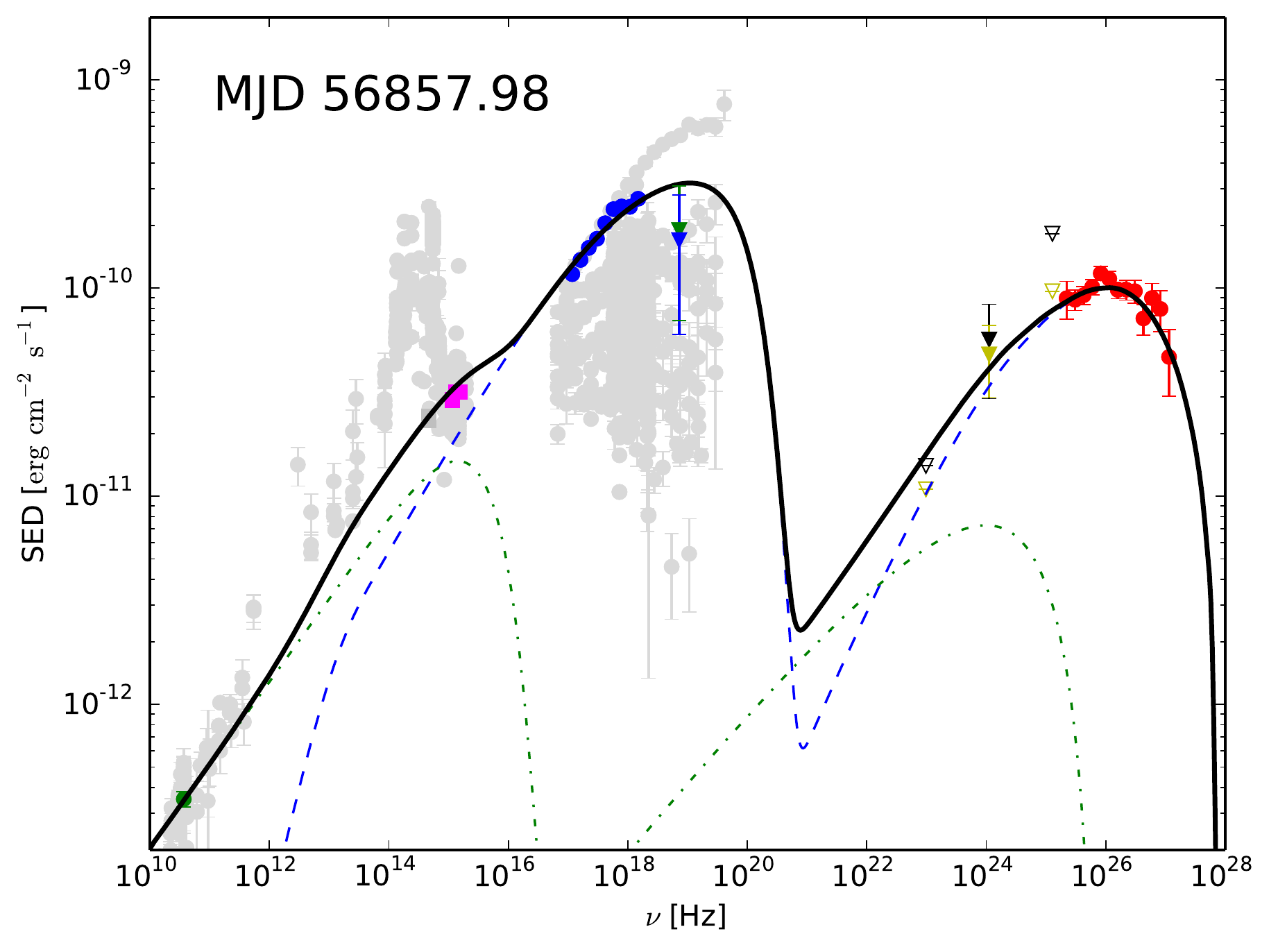}
     \caption{Broadband SED from 2014 July 21 (MJD~56859.97) where a two-zone SSC model has been used to describe the overall emission. One emitting region is responsible for the gamma-ray, X-ray and partially optical emission (dashed blue line). A second component explains the radio emission together with some optical emission (dashed green line).  The colored spectral points represent the data sample from 56859.97. The gray spectral points represent the archival spectral points taken from the SED builder at SSDC. The peak emission at $\sim 10^{14}-10^{15}$ Hz is due to the host galaxy contribution, which is not taken into account in our SSC model of the jet emission.}
   \label{fig:SED_modeling_radio}
\end{figure}

In this section we model the radio to optical UV emission for one of the nights of the sample (see Fig.~\ref{fig:SED_modeling_radio}). For this model a simple PL electron distribution was used instead of the broken PL. The parameters used for this modeling are $\gamma_{min}=1$, $\gamma_{max}=8\times10^4$, $n1$=$n2$=2.2, $B$=0.02 [G], Density$=6\times10^3 [\mathrm{cm}^{-3}]$, R$=4.7\times10^{16}$[cm], $\delta=10$.
With this approach the radio to optical emission could be fitted.

\clearpage

\section{Relation between $\gamma_{b}$ and $B$}
\label{App:RelationBreakAndB}

\begin{figure}[h]
   \includegraphics[scale=0.47, angle=0]{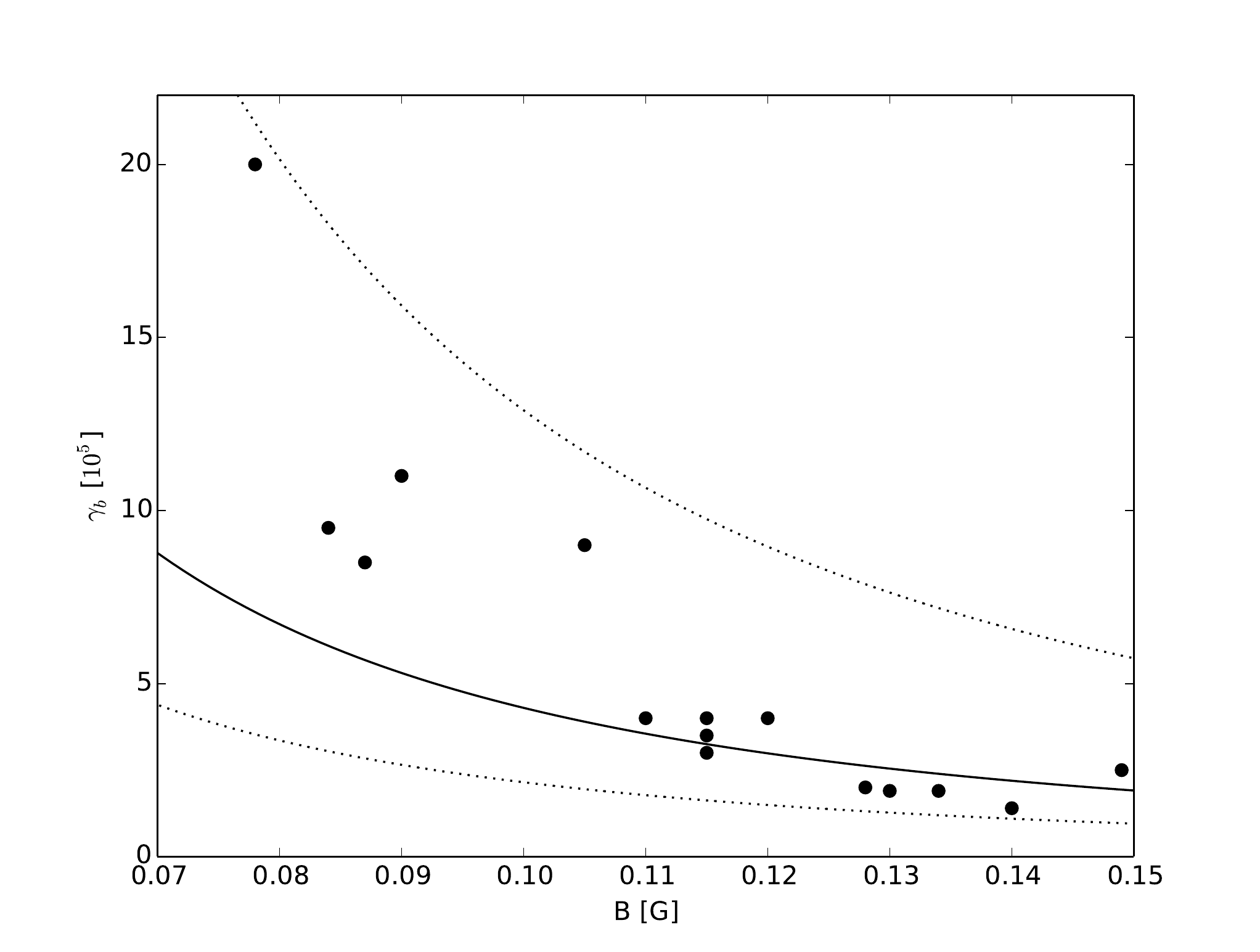}
   \caption{Evolution of the $\gamma_b$ as a function of the magnetic field $B$ for the one-zone SSC model reported in Sec.\ref{sec:Model}. The solid line represents the theoretical expectation assuming that $\gamma_b$ is due to synchrotron and IC cooling (see eq.~\ref{eq_syn_cooling}). The dotted lines depict the region with the energy break located a factor of 3 higher and a 50\% lower than the theoretical expectation.}
   \label{fig:ssc_param}
\end{figure}

In the canonical one-zone SSC framework, we expect a break in the electron energy distribution, where the spectral indices change by one unit. This break occurs at the energy at which the timescale  for energy loss  is equal to the dynamical timescale.  Given that the synchrotron bump and the inverse-Compton bump appear quite similar in the Mrk~501 SEDs from July 2014 (see Fig.\ref{fig:SED_modeling}), we can assume that the electrons lose energy roughly equally through synchrotron and inverse-Compton emission. In this case, the theoretical expectation for the location of the break would be given by the  relation 

\begin{equation}
\gamma_b=\frac{3\, \pi \,m_e\,c^2}{(\sigma_T\,B^2\,R)}
\label{eq_syn_cooling}
,\end{equation}
where $m_e$ is the electron mass, $\sigma_T$ the Thompson cross section, and R the radius of the emitting region. Figure \ref{fig:ssc_param} shows the evolution of $\gamma_b$ as a function of $B$ for the theoretical exercise reported in Sec.\ref{sec:Model}. The values used to parameterize the broadband SEDs agree typically within a factor of $\sim$2 with the theoretical expectations. Given that the one-zone SSC is a relatively simple theoretical scenario (e.g., the emission region may not be perfectly spherical and homogeneous), these differences between the employed values and the theoretical expectations in the canonical one-zone SSC can be considered in reasonable agreement. 

\end{document}